\documentclass[apj,appendixfloats,numberedappendix]{emulateapj}

\usepackage{wasysym}
\usepackage{rotating}
\usepackage[dvips]{color}
\usepackage{verbatim}
\usepackage{soul}
\usepackage[hypertex]{hyperref}
\usepackage{subfigure}
\usepackage{lineno}

\slugcomment{Accepted for publication in ApJ Main Journal.}

\shorttitle{Exploring the Variability of 1633+382. Light Curves}

\shortauthors{Algaba et al.}

\begin{document}


\title{Exploring the Variability of the Flat Spectrum Radio Source 1633+382.\\ I. Phenomenology of the Light Curves}

\author{
Juan-Carlos Algaba$^{1,2}$, Sang-Sung Lee$^{1,3}$, Dae-Won Kim$^2$, Bindu Rani$^{4\dagger}$, Jeffrey Hodgson$^1$, Motoki Kino$^{5,6}$, Sascha Trippe$^2$, Jong-Ho Park$^2$, Guang-Yao Zhao$^1$, Do-Young Byun$^1$, Mark Gurwell$^7$, Sin-Cheol Kang$^{1,3}$, Jae-Young Kim$^2$, Jeong-Sook Kim$^6$, Soon-Wook Kim$^{1,3}$,  Benoit Lott$^8$, Atsushi Miyazaki$^9$, Kiyoaki Wajima$^1$
} 
\thanks{$^{\dagger}$NASA Postdoctoral Program (NPP) Fellow}

\affil{
$^1$Korea Astronomy \& Space Science Institute, 776, Daedeokdae-ro, Yuseong-gu, Daejeon, Republic of Korea 305-348\\
$^2$Department of Physics and Astronomy, Seoul National University, 1 Gwanak-ro, Gwanak-gu, Seoul 08826, Korea\\
$^3$Korea University of Science and Technology, 217 Gajeong-ro, Yuseong-gu, Daejeon 34113, Korea\\
$^4$NASA Goddard Space Flight Center, Greenbelt, MD 20771, USA\\
$^5$National Astronomical Observatory of Japan, 2211 Osawa, Mitaka, Tokyo 1818588, Japan\\
$^6$Kogakuin University,  Academic Support Center, 2665-1 Nakano, Hachioji, Tokyo 192-0015, Japan\\
$^7$Harvard-Smithsonian Center for Astrophysics, Cambridge, MA USA\\
$^8$Universit\'e Bordeaux 1, CNRS/IN2P3, Centre d'\'Etudes Nucl\'eaires de Bordeaux Gradignan, 33175 Gradignan, France\\
$^9$Faculty of Science and Engineering, Hosei University, 372 Kajino-cho, Koganei, Tokyo 1848584, Japan\\
}

\begin{abstract}
We present multi--frequency simultaneous VLBI radio observations of the flat spectrum radio quasar 1633+382 (4C~38.41) as part of the interferometric monitoring of gamma-ray bright active galactic nuclei (iMOGABA) program combined with additional observations in radio, optical, X--rays and $\gamma-$rays carried out between the period 2012 March -- 2015 August. The monitoring of this source reveals a significant long-lived increase in its activity since approximately two years in the radio bands, which correlates with a similar increase on all other bands from sub--millimeter to $\gamma-$rays. A significant correlation is also found between radio fluxes and simultaneous spectral indices during this period. The study of the discrete correlation function (DCF) indicates time lags smaller than the $\sim40$~days uncertainties among both radio bands and also high-energy bands, and a time lag of $\sim$70 days, with $\gamma-$rays leading radio. We interpret that the high-energy and radio fluxes are arising from different emitting regions, located at $1\pm12$ and $40\pm13$~pc from the central engine respectively.

\end{abstract}


\section{Introduction}
The source 1633+382 (4C~38.41) is a flat spectrum radio quasar (FSRQ) at a redshift $z=1.813$ \citep{Hewett10}. Very Large Array (VLA) observations have found that its kilo--parsec scale morphology shows a core--dominated triple structure  in the north-south direction with an extension of about 12 arcseconds  \citep{Murphy93}, whereas  parsec scales show a misalignment of about $90\degr$ with a single jetted structure detected by the Very Long Baseline Array (VLBA) towards the west. Superluminal motion with jet velocities up to $373\pm16~\mu$as~yr$^{-1}$ ($29.2\pm1.3~c$) has been detected based on 10 moving features \citep{Lister13}. It has been estimated that the parsec scale jet is aligned at $\sim1-3\degr$ to our line of sight \citep{Hovatta09,Liu10}. Owing to its jet trajectories, it has been suggested that its central engine may host a binary supermassive black hole \citep{Liu10} with a combined black hole mass of $M\sim1.32\times10^9M_{\odot}$ \citep{Zamaninasab14}.

The source 1633+382 is one of the most powerful extragalactic objects detected by the Energetic Gamma Ray Experiment Telescope (EGRET) \citep{Fichtel94,Thompson95,Hartman99}. This object has also been included in the {\
{\it Fermi}-LAT (Large Area Telescope) bright active galactic nuclei (AGN) list \citep{Abdo09} and its $\gamma-$ray flux is consistent with those reported in the Second \citep[2FGL; ][]{Nolan12} and the Third {\it Fermi}-LAT source catalogs \citep[3FGL; ][]{Acero15}. In optical bands, it has been classified as an optically violent variable (OVV) blazar \citep{Mattox93} and strong variability in radio flux has also been observed \citep{SpanglerCotton81,Kuhr81,Seielstad85,Aller92}. Multifrequency observations of the $\gamma-$ray flares observed by \emph{Fermi}-LAT in 2009--2010 suggested that their origin was connected to an interaction between a component emerging from the core downstream of the jet and the 43~GHz VLBI core \citep{Jorstad11}. Similarly, a large outburst observed in 2011 could be explained geometrically, ascribed to the variations of the Doppler factor owing to changes in the viewing angle \citep{Raiteri12}.

Being one of the bright $\gamma-$ray blazars, 1633+382 is a typical target of several other monitoring programs, such as the Steward Observatory blazar monitoring program \citep{Smith09}, the VLBA-BU-BLAZAR program \citep{Jorstad16}, the OVRO 15~GHz 40m monitoring program \citep{Richards11}, the Interferometric Monitoring of Gamma-ray Bright AGN (iMOGABA) \citep{Lee13,Algaba15,Lee16}, and {\it Swift} X-Ray Telescope (XRT) \citep{StrohFalcone13}, to cite some. These programs aim to answer two questions: where are the $\gamma-$ray flares located and what are the emission mechanisms responsible for their origin. Under this context, iMOGABA observed this source and detected a significant increase of its radio--mm flux density between 2013 and 2015, based on the simultaneous 22 to 129~GHz VLBI observations using the Korean VLBI Network (KVN). This provides an excellent opportunity to study the flux and spectral evolution of radio flares in great detail. 

In this paper, we study broadband flaring activity of 1633+382 observed from 2012 to 2015. A detailed broadband spectral analysis and estimation of physical parameters of the emission regions will be given in a forthcoming paper. The organization of the paper is as follows. In Section 2 we introduce the observations performed at the various frequencies; in Section 3 we investigate the light curves and analyze their phenomenological properties, such as variability, spectral indices and time delays; and in Section 4 we discuss our results. Summary and Conclusions can be found in Section 5.

\section{Observations and Data Reduction}

To explore the broadband  variability properties of 1633+386 from radio to $\gamma-$rays, we monitored the source and collected data from various ground-- and space--based instruments between March 2012 and August 2015 (MJD~56000 to 57250). In Table \ref{obs-summary} we summarize the various bands and instruments discussed here, including the mean cadence (i.e, number of observations over the given interval) in days.

\begin{table}
\begin{center}
\caption{List of Observations}
\label{obs-summary}
\begin{tabular}{cccc}
\tableline
\hline
Band &  Instrument & Frequency (Hz) & Cadence (d)\\
(1)&(2)&(3)&(4)\\
\tableline
Radio & OVRO& $1.50\times10^{10}$&7\\
Radio & KVN & $2.20\times10^{10}$&36\\
Radio & KVN/VLBA & $4.30\times10^{10}$&23\\
Radio & KVN & $8.60\times10^{10}$&36\\
Radio & KVN & $1.29\times10^{11}$&76\\
Radio & SMA & $2.25\times10^{11}$&22\\
Optical & Steward Obs.& $5.44\times10^{14}$&8\\
X--rays & {\it Swift}--XRT & $4.84{\times}10^{16}-2.42{\times}10^{18}$&18\\
$\gamma-$rays & {\it Fermi}--LAT & $2.42{\times}10^{22}-7.25{\times}10^{25}$&7\\	
\tableline
\end{tabular}
\end{center}
\end{table}

\subsection{OVRO 40m}
Beginning in late 2007, approximately a year before the start of LAT science operations, \cite{Richards11} began a large-scale, fast--cadence 15~GHz radio monitoring program with the 40~m telescope at the Owens Valley Radio Observatory. Reduced data for their core sample, the 1158 CGRaBS \citep{Healey08} north of $-20\degr$ declination, are available to the public\footnote{\url{http://www.astro.caltech.edu/ovroblazars}}. Public data available spanning from 2008 to 2016 were used. 

\subsection{iMOGABA}
The iMOGABA program observed this source with multi-frequency VLBI at 22, 43, 86 and 129~GHz using the KVN \citep{Lee14}\footnote{\url{http://radio.kasi.re.kr/sslee/}}. Observations were done quasi--continuously starting in January 2013 with a cadence of roughly one epoch per month, except during KVN maintenance season, between June--August 2013 and July--August 2014, when VLBI observations could not be performed. An amplitude correction factor of 1.1 due to requantization loss from the digital filter \citep{Lee15} was applied to the data. Frequency phase transfer technique was used after typical data reduction in order to recover fringe solutions at the highest frequencies \citep{Algaba15}. Reliable maps and flux densities were obtained for all epochs at 22, 43 and 86~GHz but not for 129~GHz.  For further details on the iMOGABA calibration process see e.g. \cite{Algaba15,HodgsonJKAS16,Lee16}.

\subsection{43~GHz VLBA--BU--BLAZAR Program}
The VLBA Boston University (BU) Blazar program observes a sample of 33 $\gamma-$ray bright blazars and 3 radio--galaxies at 43~GHz about once per month with the Very Long Baseline Array (VLBA). With their large VLBA project, they obtain total and polarized intensity images of such objects. Data in the form of total and polarized intensity peaks, electric vector polarization angle (EVPA), images and \emph{fits} files with the UV-data, models and maps are public\footnote{\url{http://www.bu.edu/blazars/VLBAproject.html}}. Information for 1633+382 is available from June 14, 2007 up to the most recent observation.

We obtained data for 1633+382 and combined them with the iMOGABA observations at 43~GHz. To compare the data from the two programs, we have to take into account the limitations on both resolution and sensitivity limits of the KVN compared with those of the VLBA. In order to do this, we convolved the total intensity map with a beam similar in size and angle to the one from iMOGABA observations at 43~GHz. We then compared both maps limiting the sensitivity to that on the iMOGABA map. 

We found that iMOGABA and VLBA--BU--BLAZAR maps are reasonably similar both in terms of observed morphology and intensity. In particular, regarding their structure, they both appear as a point--like source. With respect to their total flux, there is unfortunately no simultaneous observations with the two arrays, but some observations within only few days difference indicate a peak flux density difference between the two arrays smaller than the uncertainty. For example, peak flux densities of iMOGABA at MJD~56310 and BU at MJD~56308 data at 43~GHz are $3.2\pm0.4$ and $3.37\pm0.05$~Jy, respectively. Similar discussion can be derived for the total flux, with $3.5\pm0.6$ and $3.48\pm0.07$~Jy, respectively.

We thus combined both data sets to have a better time coverage of the light curve. The BU data conveniently provide us information at 43~GHz prior to the start of iMOGABA observations, before the flux of 1633+386 increased, and fills the gaps produced by the maintenance seasons of iMOGABA (see Figure \ref{lightcurve}). Additionally, inspection of this figure clearly shows that the 43~GHz total flux densities obtained with BU are mostly within the uncertainties of the iMOGABA flux densities.

\subsection{SMA}

The 225 GHz flux density data for 1633+382 were obtained at the Submillimeter Array (SMA) near the summit of Mauna Kea (Hawaii). 1633+382 is included in an ongoing monitoring program at the SMA to determine the fluxes of compact extragalactic radio sources that can be used as calibrators at mm wavelengths \citep{Gurwell07}.  Observations of available potential calibrators are from time to time performed for 3 to 5 minutes, and the measured source signal strength calibrated against known standards, typically solar system objects (Titan, Uranus, Neptune, or Callisto).  Data from this program are updated regularly and are available at the SMA website\footnote{\url{http://sma1.sma.hawaii.edu/callist/callist.html}}.

\subsection{Optical Program at University of Arizona}
Steward Observatory of the University of Arizona contributes optical data for the LAT-monitored blazars and {\it Fermi} Targets of Opportunity. The optical program utilizes either the 2.3~m Bok Telescope on Kitt Peak, AZ  or the 1.54~m Kuiper Telescope on Mt. Bigelow, AZ.   All observations are performed using the SPOL CCD Imaging/spectropolarimeter. Calibration of the data is performed via differential aperture photometry \citep[see][for more details]{Smith09}. Reduced V-band photometry data are publicly available\footnote{\url{http://james.as.arizona.edu/~psmith/Fermi}}. 

\subsection{Swift}
{\it Swift}--XRT monitoring of {\it Fermi}--LAT sources of interest data are publicly available thanks to support from the {\it Fermi} GI program and the {\it Swift} Team\footnote{\url{http://swift.gsfc.nasa.gov/archive/}}. Data analysis to produce the public data is described in their monitoring website\footnote{\url{http://www.swift.psu.edu/monitoring/readme.php}}. Full details of the procedure can be found in \cite{StrohFalcone13}. A brief summary is provided here.

The {\it Swift}--XRT data were processed using the most recent versions of the standard {\it Swift} tools: {\it Swift} Software version 3.9, FTOOLS version 6.12 and XSPEC version 12.7.1. Light curves are generated using xrtgrblc version 1.6. Circular and annular regions are used to describe the source and background areas respectively and the radii of both depend on the current count rate. In order to handle both piled-up observations and cases where the sources land on bad columns, PSF correction is handled using xrtlccorr. Single observation light curves use a step binning procedure so that if the count rate is less than 0.4, 1, 10, 100 and 10000 cts/s, then 20 cts/bin, 50 cts/bin, 200 cts/bin, 1000 cts/bin and 2000 cts/bin are used respectively. All error bars are reported at the 1-sigma level. 

We obtained data for 1633+382 in the energy range 0.2--10~keV. We approximated the X--ray flux from the count rates with the HEARSAC WebPIMMS tool\footnote{\url{http://heasarc.gsfc.nasa.gov/cgi-bin/Tools/w3pimms/w3pimms.pl}} assuming a power law spectra with photon index between 1.5 and 2.0 and Galactic column density at the coordinates of the source provided by the web version of the nH~FTOOL\footnote{\url{http://heasarc.gsfc.nasa.gov/cgi-bin/Tools/w3nh/w3nh.pl}}.

\subsection{Fermi}
To investigate the flux variations at GeV energies, we used the {\it Fermi}-LAT \citep{Atwood09} data observed in survey mode\footnote{\url{https://fermi.gsfc.nasa.gov/cgi-bin/ssc/LAT/LATDataQuery.cgi}}. Photons in the source event class  were analyzed using the standard ScienceTools (software version v10.r0.p5) and instrument response functions P8R2$\_$SOURCE$\_$V6. A  region of interest (ROI) of 20$^{\circ}$  radius centered at the position of the source was analyzed using a maximum-likelihood algorithm \citep{Mattox96}. In the unbinned likelihood analysis\footnote{\url{http://fermi.gsfc.nasa.gov/ssc/data/analysis/scitools/likelihood\_tutorial.html}}, sources from the 3FGL catalog \citep{Acero15} within 20$^{\circ}$ and the recommended Galactic diffuse background ($gll\_iem\_v06.fits$) and the isotropic background ($iso\_P8R2\_SOURCE\_V6\_v06.txt$) emission components \citep{Acero16} were included. We set the model parameters for the sources within 5$^{\circ}$ of the center of the ROI as free. Model parameters for the rest, except for the sources reported as being significantly variable (variability index $\geq$72.44) in the 3FGL catalog, were fixed to their catalog values. For the variable sources, we set all the model parameters free.  The weekly binned light curve of the source at E$>$100~MeV was produced by modeling the spectra over each bin by a simple power law (N(E) = N$_0$ E$^{-\Gamma}$, N$_0$ : prefactor, and $\Gamma$ : power law index). Except during the flaring phases,  variability  at shorter timescales cannot be investigated as the source is not bright enough. 

In order to investigate the photon index variations and their correlation with photon flux variations, we computed $\gamma-$ray flux light curves above the decorrelation energy E$_0$\footnote{De-correlation energy minimizes the correlations between integrated photon flux and photon index}. During our period of interest, we found E$_0 = 167$~MeV. Finally, we generated the constant uncertainty (20\%) light curve above E$_0$   following the adaptive binning method analysis method as described in  \cite{Lott12}.

\section{Analysis and Results}
In this section we present the broadband analysis of the flux variability of 1633+382 observed between March 2012 and August 2015 (MJD 56000--57250) with the data compiled above. 

\subsection{Light curve analysis}
Figure \ref{lightcurve} shows the light curves of 1633+382 for the different bands compiled here. In the case of resolved images (e.g., VLBI), we include the integrated flux, which is more convenient for comparison. Nonetheless, even in this case we may expect a certain flux loss due to some extended structure. Given the compactness of this source, and the fact that any extended emission should have a flux smaller than the typical iMOGABA image rms of 20, 30, 60 and 75 mJy/beam at 22, 43, 86 and 129~GHz respectively, we consider that such emission will not have a significant impact on the measured integrated flux.

\begin{figure*}
\includegraphics[scale=1,trim={1cm 1.5cm 1.5cm 1.3cm},clip]{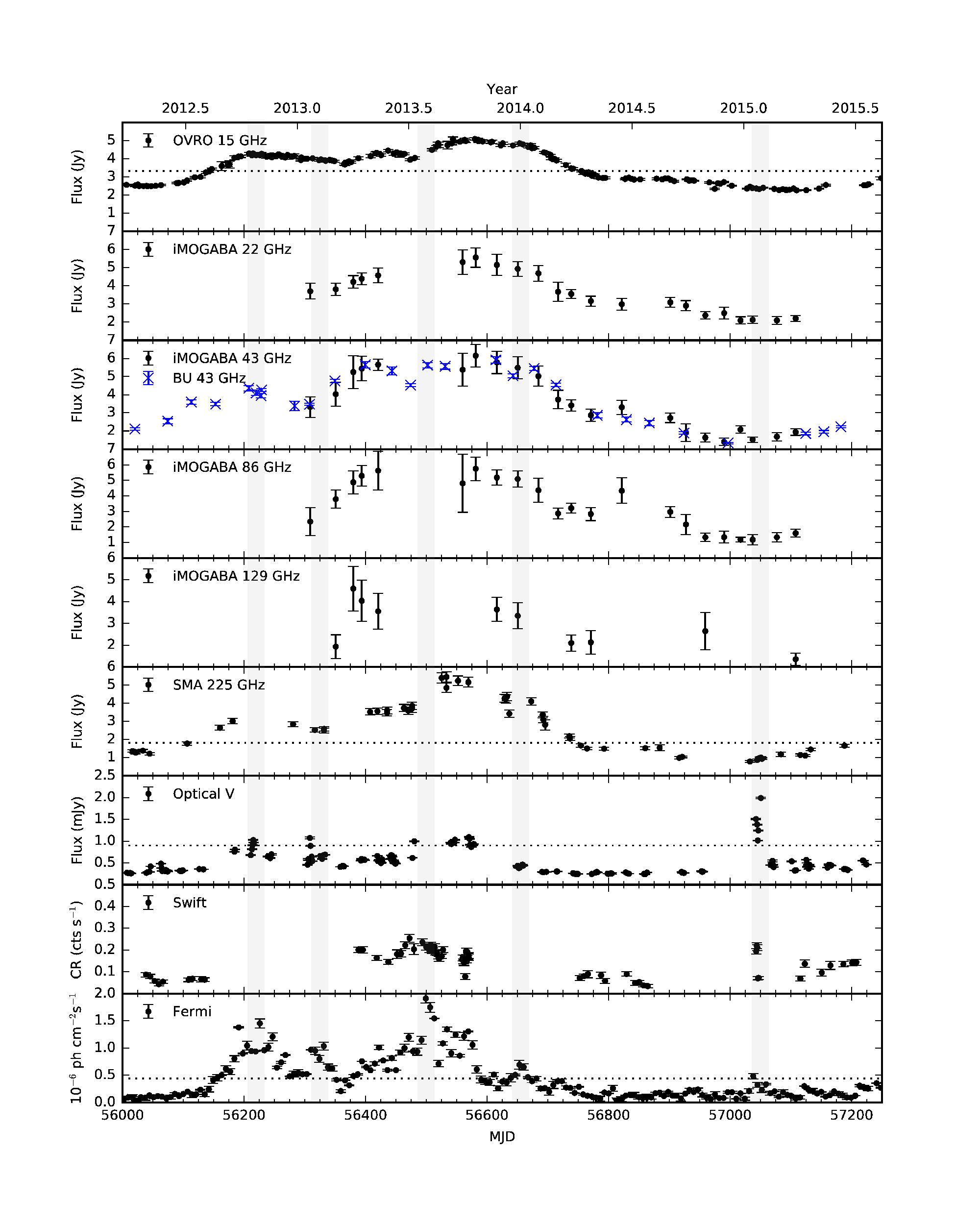}
\caption{Light curve of 1633+382. From top to bottom: OVRO 15~GHz, iMOGABA 22~GHz, iMOGABA 43~GHz, iMOGABA+BU 86~GHz, iMOGABA 129~GHz, SMA 225~GHz, Optical V~band, {\it Swift}--XRT X--ray and {\it Fermi}--LAT $\gamma-$ray. All units are in Jy except for optical (mJy), X--ray (count rate, cts/s) and $\gamma-$ray ($10^{-6}$ photons/cm$^2$/s). Horizontal dotted lines indicate the flux threshold to estimate the periods of flux enhancement, as described in the text. Gray vertical areas show the peak epochs of the $\gamma-$ray local maxima, which roughly coincide with the maximum of the smoothed data using 5 points average.}
\label{lightcurve}
\end{figure*}

The coverage of the light curves is of very different quality, with 15 GHz and gamma-ray being the best and 129~GHz being very poorly sampled. In the case of the {\it Swift}-XRT data, large gaps in the data prevent us from following the overall trend for the flux at 0.2--10~keV. Note that, although iMOGABA data are simultaneous, some data are missing at 129~GHz due to the lack of reliable flux measurements. We were not able to observe the beginning of the high flux state with iMOGABA. Furthermore, the maintenance gaps in iMOGABA, especially that in the summer of 2013, are also apparent in the light curve. These time gaps are mitigated at 43~GHz with the inclusion of the VLBA--BU-Blazar data.

In general, the light curves at various bands follow a very similar trend. Between roughly 56100$<$MJD$<$56800 there seems to be a period of high activity, where the flux density seems to be significantly larger and more variable than outside this time range. In order to discuss this flux density activity, we will divide the state of the source in flaring versus quiescent periods based on a flux density threshold. We define this threshold as the median flux density value plus three times the standard deviation, based on the period of low activity (note that, for the case of the optical band, another period around MJD$\sim57050$ also shows a significant increase (by a factor of up to few) in the flux density, and thus has also been removed for this calculation). The threshold levels are defined for the better sampled light curves to be 3.33~Jy for OVRO 15~GHz data, 1.83~Jy for SMA 225~GHz data, 0.9~mJy for optical data and $0.44\times10^{-6}$~ph~cm$^{-2}$s$^{-1}$ for $\gamma-$ray data, respectively.

The OVRO 15~GHz  light curve of 1633+386 is the most densely sampled at radio frequencies. At the beginning of the observations discussed here, the source appears to be at a quiescent level with a flux $S_{15}<3.3$~Jy. During the high activity period, the flux density increases reaching three clear local maxima  of $S_{15}\sim4.3$~Jy at MJD~56210, $S_{15}\sim4.4$~Jy at MJD~56440 and $S_{15}\sim5.1$~Jy at MJD~56555. After the last local maximum, the flux density smoothly drops and after MJD~56780 it is below $S_{15}<3.3$~Jy characteristic of the quiescent state. We note that at this relatively low frequency, the light curve exhibits the typical characteristics of smoothed variability patterns, and thus we cannot fit the typical exponential models in the literature \citep[see e.g.][]{Valtaoja99,LeonTavares11}.

Examination of the iMOGABA data indicates that the radio flux density at higher frequencies seems to follow a similar variability trend as the 15~GHz radio band, although due to the lack of data before the start of the program and during maintenance the overall comparison is very challenging. Nonetheless, inspection of the 43~GHz data, containing both iMOGABA and VLBA--BU observations, shows that the flux density trend at this frequency qualitatively correlates with that at 15~GHz. This trend is also seen in the SMA data at 225~GHz, although in this case the last local maximum seems to be comparatively brighter. 

Although there are indications for an increase of the optical flux between MJD~56150 and MJD~56600, the flux does not surpass the threshold limit of 0.9~mJy discussed above with the exception of few epochs e.g., at MJD~56213, MJD~56306 and around MJD~56530. These optical maxima are much sharper ($\lesssim$ three weeks) than their radio counterparts and only the first of them seems to correlate with a maximum in radio.
Additionally, there is a very bright and rapid flux enhancement occurring in optical at MJD~57050 which is not seen at any radio frequency. While it is true that the cadence of iMOGABA does not provide enough sampling, the more dense observations with OVRO and possibly SMA should have detected flaring at these timescales if present in the radio data. This seems not to be the case, suggesting that such flux enhancement did not happen at radio frequencies.

The X-ray data, although sparse, show similar activity as that of radio bands except the rapid flux enhancement at MJD$\sim$57050, which is only observed at energies higher than radio (optical to $\gamma-$rays). Unfortunately, due to the lack of X-ray data during large periods, a one-to-one comparison between the flaring activity between X-rays and other bands cannot be performed in a robust way.

The {\it Fermi}--LAT data at $\gamma-$rays show prominent activity very well in agreement with the epoch of radio--enhancement. The $\gamma-$rays weekly binned data exhibit rapid ($\sim$1-2 weeks) large fluctuations (as a note, in fine-binned data, we notice even faster variations on timescales of a few hours.) Inspection of the data suggests local maxima at around MJD~56210, 56330, 56500 and 56650. There are also indications of an enhancement of the activity marginally over the threshold at MJD~57040.  Given the complexity of the $\gamma-$ray activity, we defer the more detailed analysis for a forthcoming paper.

Although the overall $\gamma-$rays trend seems to follow that of the radio-bands, the detailed morphology seem to convey a different behaviour. The first and second $\gamma-$ray local maxima appear to have respective clear counterparts in the optical light curve, and may be associated with the first region of the radio flux density enhancement. For the third and fourth $\gamma-$ray local maxima, the association is not so precise, as there is not a clear overlap with a simple optical maximum, and there seems to be a time lag between the radio local maximum. Interestingly, the $\gamma-$ray enhancement at MJD~57040 has very clear counterparts in X-ray and optical bands, whereas no counterpart is seen in radio bands.

In summary, the gradual (i.e., long--term) activity shows similar morphology at all bands from radio to $\gamma-$rays. The first radio maximum seems to be very well correlated with a flux enhancement at optical and $\gamma-$rays, whereas the second radio maximum seems to involve a certain delay with respect to their high energy counterparts. On the other hand, there appear to be very rapid ($\lesssim$ few weeks) enhancements in high energy bands that lack a radio counterpart.

\subsubsection{Flux Variability}

In order to investigate the variability of the flux amplitude at the different bands we follow the approach of \cite{Chidiac16} and \cite{Vaughan03}, where they compute the fractional variability amplitude $F_{var}$ defined as
\begin{equation}
F_{var}=\sqrt{\frac{V^2-\overline{\sigma^2_{\text{err}}}}{\bar{x}^2}},
\end{equation}
where $V$ is the variance of the light curve,  $\bar{x}^2$ the arithmetic mean on the flux and $\overline{\sigma^2_{\text{err}}}$ is its mean square error. The uncertainty of the fractional variability is given by
\begin{equation}
err(F_{var})=\sqrt{\left(\sqrt{\frac{1}{2N}}\frac{\overline{\sigma^2_{\text{err}}}}{\bar{x}^2 F_{var}}\right)^2+\left(\frac{\overline{\sigma^2_{\text{err}}}}{N}\frac{1}{\bar{x}}\right)^2},
\end{equation}
with $N$ the number of data points in the light curve. We also performed a bootstrap analysis and obtained uncertainties to estimate random errors due to the sampling that we added quadratically to the formal ones. We note that sampling effects may also lead to errors in the formal calculation, specially for these bands with fewer data such as 129~GHz or X-rays. In Table \ref{FvarTable} we summarize the estimates of the fractional variability obtained. A plot of $F_{var}$ showing its trend along the different bands studied here is shown in Figure \ref{FvarPlot}.

\begin{table}
\begin{center}
\caption{Variability Characteristics.}
\label{FvarTable}
\begin{tabular}{cc}
\tableline
\hline
Freq (Hz) & $F_{\mbox{var}}$ \\
(1)&(2)\\
\tableline
$	1.50\times10^{10}	$&$	0.47	\pm	0.01	$	\\
$	2.20\times10^{10}	$&$	0.46	\pm	0.05	$	\\
$	4.30\times10^{10}	$&$	0.58	\pm	0.03	$	\\
$	8.60\times10^{10}	$&$	0.55	\pm	0.06	$	\\
$	1.29\times10^{11}	$&$	0.34	\pm	0.13	$	\\
$	2.55\times10^{11}	$&$	0.69	\pm	0.03	$	\\
$	5.44\times10^{14}	$&$	0.75	\pm	0.03	$	\\
$	4.84\times10^{17}	$&$	0.59	\pm	0.04	$	\\ 
$	3.63\times10^{25}	$&$	0.89	\pm	0.03	$	\\
\tableline
\end{tabular}
\end{center}
\end{table}

\begin{figure}
\includegraphics[scale=0.42,trim={0cm 0cm 0cm 0cm},clip]{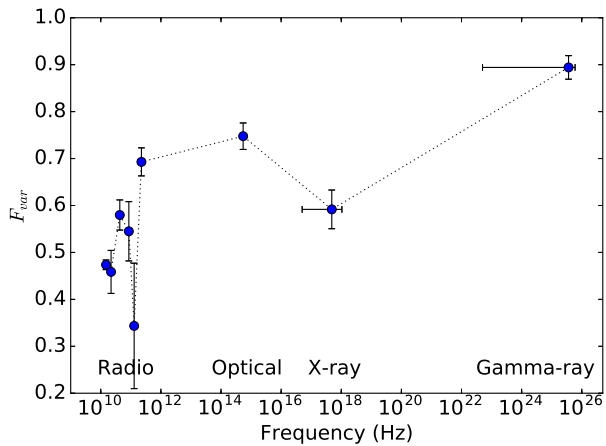}
\caption{Fractional Variability of the 1633+382 light curves analyzed here. Different bands are indicated in the bottom.}
\label{FvarPlot}
\end{figure}

$F_{var}$ increases with frequency at radio wavelengths, except for 129~GHz, for which the lack of data, in combination with the relatively large flux errors, leads to a large uncertainty for this data point. From millimeter to $\gamma-$rays, we notice a steady increase in $F_{var}$ 
except the X-ray point. We note however that both of these may not be intrinsic values, but rather due to observational bias. Sensitivity limits or the lack of sampling due to other diverse effects affecting these data may lead to the lower apparent X-ray $F_{var}$. To examine this possibility, we first dropped samples from the $\gamma-$ray light curve choosing only the samples that coincide within a tolerance with a sample in the X--rays light curve and re--computed $F_{var}$. In this case, we obtained a slightly lower$F^{\gamma-drop}_{var}=0.84\pm0.10$.  Second, we considered the $F_{var}$ of the $\gamma-$rays light curve when only data points with flux larger than $0.44\times10^{-6}$ ph~cm$^{-2}$~s$^{-1}$ (i.e, the above defined threshold limit) were considered. In this case we obtain $F^{\gamma-bias}_{var}=0.57\pm0.14$; i.e., a decrease of 36\% from its to original value. We thus consider that the lower $F_{var}$ values for both 129~GHz and X-rays are not representative of the intrinsic variability of the source at these bands.

At radio frequencies, $F_{var}$ amplitude values typically increase with frequency because at higher frequencies we tend to probe
optically thin regions (which are typically self-absorbed at low-radio frequencies) in the jet. As the source becomes optically thin, the increase of $F_{var}$ becomes less dramatic and, after radio--frequencies, the smoother increase may be rather related with the electron energy probed. The highest $F_{var}$ amplitude is found for the $\gamma-$ray light curve, suggesting that the largest variations are seen for the highest energy electrons. 

\subsection{Spectral Indices}

\subsubsection{Radio--Optical Spectral Indices}

We now investigate the spectral indices $\alpha=\log(S_1/S_2)/\log(\nu_1/\nu_2)$ associated with the 1633+386 light curve. Given that only data from iMOGABA are truly simultaneous, we proceed in the following way: for each pair of frequencies, we compared the nearby observing epochs and, if they were separated by five days or less, we then assumed them to be quasi--simultaneous and calculated the spectral index. The epoch we assign to such index is the average between the epochs of the two frequencies. 

In Figure \ref{spix} we plot spectral indices for all combinations between each pair of frequencies in radio and optical bands. 
The average spectral indices at the various frequency pairs are shown in Table \ref{table:spix}, where the uncertainties refer to the standard deviation. Overall, we see that $\alpha$ is very close to zero for radio-pairs and it becomes slightly negative when we compare low radio-frequencies with SMA 225~GHz. Only in the optical band the source becomes optically thin. This is expected from the classification of this source as a FSRQ \citep{Healey07} and is consistent with previous spectral index observations of this source \citep[e.g.][]{Algaba11,Algaba12}.

\begin{figure}
\includegraphics[scale=1.0,trim={0.6cm 1.3cm 0.7cm 1.1cm},clip]{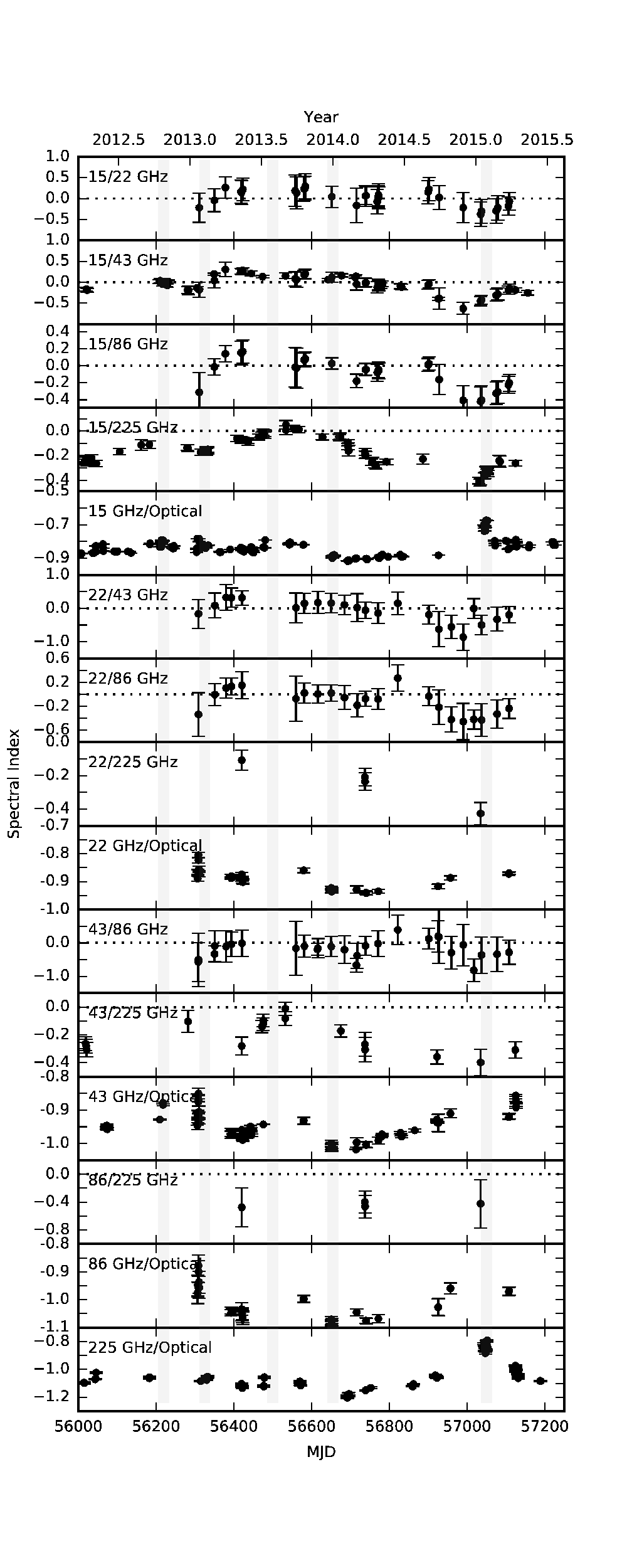}
\caption{Radio to optical spectral indices of 1633+382. Note that the scales for each panel are different. Horizontal dotted lines indicate a zero spectral index, for eye guidance.}
\label{spix}
\end{figure}

\begin{table}
\begin{center}
\caption{Average Spectral Indices.}
\label{table:spix}
\begin{tabular}{cc}
\tableline
\hline
Frequency Pair & Average Spectral Index \\
(1)&(2)\\
\tableline
15--22~GHz&$0.02\pm0.19$\\
15--43~GHz&$-0.04\pm0.19$\\ 
15--86~GHz&$-0.08\pm0.18$\\
15--225~GHz&$-0.18\pm0.12$\\ 
15~GHz--Optical&$-0.84\pm0.04$\\
22--43~GHz&$-0.08\pm0.31$\\
22--86~GHz&$-0.12\pm0.20$\\
22~GHz--SMA&$-0.24\pm0.11$\\ 
22~GHz--Optical&$-0.89\pm0.03$\\
43--86~GHz&$-0.19\pm0.26$\\
43--225~GHz&$-0.22\pm0.11$\\
43~GHz--Optical&$-0.95\pm0.04$\\
86--225~GHz&$-0.44\pm0.03$\\
86~GHz--Optical&$-1.02\pm0.06$\\
225~GHz--Optical&$-1.04\pm0.11$\\
\tableline
\end{tabular}
\end{center}
\end{table}

The time evolution of the spectral indices seems to be smooth except for significant jumps in these involving optical bands. These jumps are coincident with the periods of increase of optical flux at MJD~56306 and MJD~57050 for which a clear radio counterpart is not seen. In general, evolution of the spectral index seems not to be random in nature but to follow a trend during the period studied here, which seems to be similar for, at least, radio bands.

\subsubsection{$\gamma-$ray photon index}
\begin{figure}
\includegraphics[angle=90,scale=0.35,trim={0cm 0cm 0cm 0cm},clip]{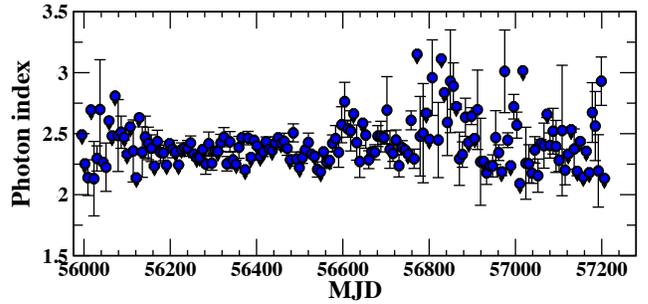}
\caption{Observed $\gamma-$ray photon index variations in 1633+382.  }
\label{photonindex}
\end{figure}

In Figure \ref{photonindex} we plot the $\gamma-$ray photon index curve. 
Unlike the photon flux, the $\gamma-$ray photon index did not show any prominent 
variation during the course of our campaign period; the average photon index value 
is 2.42$\pm$0.21. The scatter in the photon index values around the mean is quite 
large during quiescent state. On the other hand, during the period of high activity, the scatter is comparatively 
smaller, possibly partly due to the higher signal to noise ratio of the measurements.

\begin{figure}
\includegraphics[scale=0.43,trim={0cm 0cm 0cm 0cm},clip]{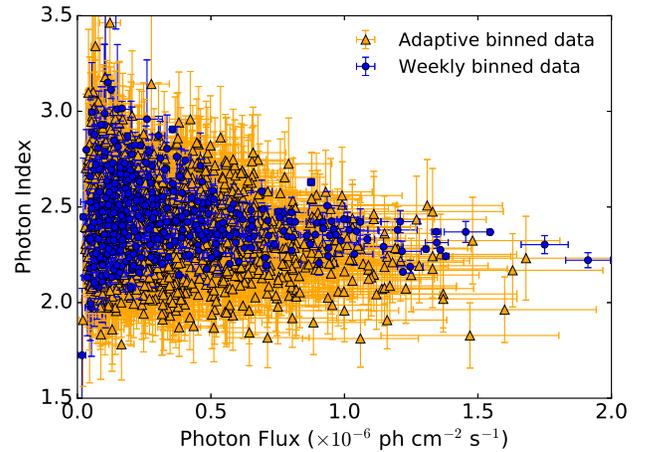}
\caption{Observed $\gamma-$ray photon index versus flux in 1633+382. Blue points: standard weekly binned data; orange triangles: adaptive binned data}
\label{gammafluxvsphotonindex}
\end{figure}

Further insight can be gained studying the photon flux vs. photon index correlation. Given that our data have poor statistics for the faint states, specially in the case of the adaptive data, we discuss the full dataset. Results for the dataset limited to the period studied here and the full data available are qualitatively similar and only differ in the statistical significance. In Figure \ref{gammafluxvsphotonindex} we plot the standard weekly binned data as blue points and the adaptive binned data as orange triangles. We do see a hint of spectral hardening as the source gets brighter in the weekly binned data. Formally the linear-Pearson correlation test gives $r_P = -0.264$, with a  $p-$value=0.004. Significance of such a correlation between photon flux and photon index is thus marginal. The adaptive binning analysis on the other hand suggests that there is a negative correlation between the $\gamma-$ray photon flux and photon index variations. Formally we obtained $r_P = -0.255$, with $p-$value=2$\times10^{-16}$. A bootstrap analysis leads to median Pearson $r_P = -0.16$, and $p-$value=6$\times10^{-10}$. A linear fit conveys similar results, with a slope of $-2.1\pm0.2$. This suggests a harder-when-brighter trend. Spectral hardening during bright $\gamma-$ray flares is a very common feature observed in several {\it Fermi} blazars \citep[see e.g.][]{Abdo10,Abdo11,Rani13b}.

\subsubsection{Connection with the Flux Density}

Simultaneous observations from iMOGABA allow us to make a study of the flux density variability comparing directly flux densities at 22, 43, 86 and 129~GHz during the same epoch and investigating any possible trend as a function of flux density variability. Together with the spectral indices discussed above, this analysis provides additional information. 
Comparison of Figures \ref{lightcurve} and \ref{spix} suggests that a possible correlation between flux density and spectral index exists. 
Higher radio flux densities seem to be correlated with optically thicker spectral indices. We can analyze this if we directly compare the flux densities and the derived spectral index variations. In Figure \ref{iMvsiMspix}, we plot the simultaneous iMOGABA spectral indices as a function of flux density, considering both the lower and higher flux densities used to calculate the spectral index (e.g., if we consider $\alpha^{22}_{43}$ we plot it against $S_{22}$ and $S_{43}$). It is clearly seen that, as discussed above, a larger flux density seems to be typically associated with comparatively larger spectral indices, with a Pearson correlation coefficient $R>0.65$ for  $\alpha^{22}_{43}$ and $\alpha^{22}_{86}$. It is worth noting that the correlation is always stronger when we compare the spectral index with the high frequency flux density.

\begin{figure}
\includegraphics[scale=0.47,trim={0.5cm 0cm 1cm 0cm},clip]{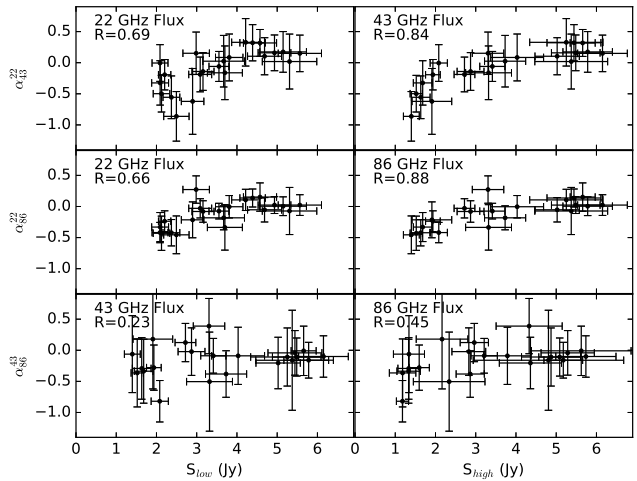}
\caption{Comparison between iMOGABA flux densities and simultaneous spectral indices. For each row, the various spectral indices $\alpha^{22}_{43}$, $\alpha^{22}_{86}$ and $\alpha^{43}_{86}$ are represented. In each column, the flux density of the lower (higher) frequency for each spectral index is plotted. On the top left corner in each figure the flux density and the Pearson correlation coefficient are shown.}
\label{iMvsiMspix}
\end{figure}

This clearly shows the dependence of the source opacity with flux density at radio frequencies. When the source is at its quiescent state, it shows a flat or slightly optically thinner spectrum, whereas when it is showing some increase in the flux activity, the spectrum becomes optically thicker. This effect seems to be more dramatic at lower radio frequencies and can be quantitatively characterized by the trends in Figure \ref{iMvsiMspix}. Further analysis of the multi--band flux time dependence, including SED analysis, will be discussed in Paper II.

Although non--iMOGABA observations are not simultaneous, and spectral indices obtained together with optical bands have been obtained considering a maximum time difference of five days, a similar analysis can be done for radio--optical spectral indices. Figure \ref{RADIOvsOPTspix} shows the radio--optical spectral indices as a function of quasi--simultaneous flux densities. As above, there seems to be a strong correlation, with a larger Pearson correlation coefficient when the spectral index is compared with the higher frequency (i.e, optical in this case). We note that the sign of the coefficient is simply related with the optical depth via the spectral index.

\begin{figure}
\includegraphics[scale=0.47,trim={0.5cm 0cm 1cm 0cm},clip]{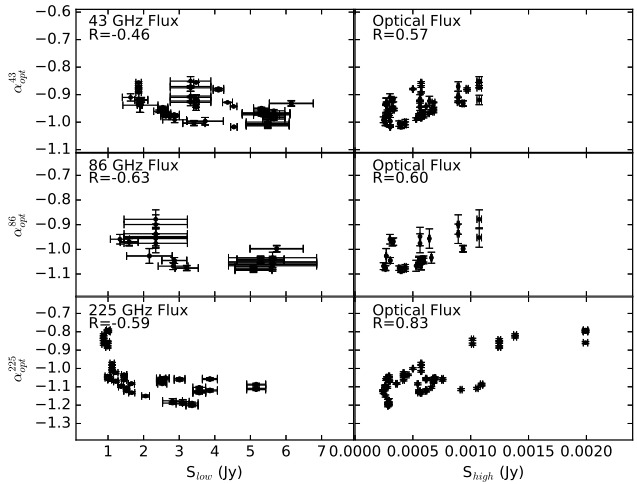}
\caption{Same as Figure \ref{iMvsiMspix} but for comparison between 43, 86 and 225~GHz flux densities and non-simultaneous corresponding optical spectral indices.}
\label{RADIOvsOPTspix}
\end{figure}

\subsection{Light Curves Statistical Properties}

Statistical properties of the light curves such as the probability distribution function (PDF) and power spectral density (PSD)  were estimated, then Monte Carlo simulations were run to test the significance of the correlation function (see below). The model implementation for the simulated light curves is crucial, and the significance of the correlations can have a strong dependence on the model used for the light curves, as demonstrated by  \cite{Emmanoulopoulos13} for the dependence on the PDF and \cite{Max-Moerbeck14} for the dependence on the PSD power law index. Additionally, for unevenly sampled data, the shape of the PSD can be distorted. In this case, linear interpolation and rebinning of the unevenly sampled light curves to a regular grid allows for a more reliable PSD fitting, as shown in \cite{Max-Moerbeck14}.

Following \cite{Emmanoulopoulos13}, we interpolated the data and obtained their PSD and PDF. We note a possible bias associated with the interpolation, which is fully discussed in \cite{Uttley02} and others following the same ideas. In general, given the flaring behaviour of the light curves, their PSD can be fitted with a power law of the form PSD~$\propto1/\nu^{\beta}$. On the other hand, although gaussian-like PDFs have been treated in the literature (\citealp[see e.g.][]{TK95} and subsequent discussion in \citealp{Emmanoulopoulos13}), it is clear that this may not always be the case, specially for high energy light curves. Indeed, inspection of our data (see Figure \ref{fig:lc_stats}) shows that a uniform distribution PDF=Constant between $x$ and $x+dx$ seems to model the data better for radio light curves (and possibly X--rays), whereas a gamma distribution of the form PDF~$\propto x^{\alpha-1} e^{-x}$ seems to model better the optical and $\gamma-$rays light curves.

Several considerations are to be taken into account regarding the PDF. First, despite an uniform model seems to fit radio data better than a gaussian model, it is also true that this may not be either the best model, specially for OVRO 15~GHz data, which shows indications of a bimodal distribution. However, for simplicity and consistency with the rest of the data, we still consider a uniform model for this band. Second, it seems that the data shift from a uniform distribution at low frequencies towards a gamma distribution at higher ones. This is clear specially for SMA data, where a uniform distribution is still a reasonable fit, but some skewness can already be hinted. The case of X-rays is very interesting, as it fits better with a uniform distribution, which is interesting if we consider the progression towards a gamma distribution already clear in optical data. This may be caused by an observational bias due to detectability limits or observational trigger in X-rays.

In order to find the parameters that best fit the models for each PSD and PDF, we performed a bootstrap analysis. This method is preferable to obtain the fitted parameters as well as a more reliable estimation of their uncertainties as it considers stability under variation of the data, which is significantly important given the lack of them for various bands leading to poor statistics. To calculate the PDF, the number of bins considered was calculated as $\sqrt{N}+1$, where $N$ is the number of data points for each light curve. In any case, the uncertainty in the parameters was chosen to be not smaller than one half of the bin width. 
In Table \ref{table:lc_stats} we include the fitted values for all the light curves as follows. In column 1 we indicate the frequency band, in columns 2, 3 and 4 we include the parameters, $x$ and $dx$ for the PDF fitted uniform distribution, and $\alpha$ for a gamma distribution respectively; and in column 5 we include the parameter $\beta$ for the fitted PSD power law model. In Figure \ref{fig:lc_stats} the PSD and PDF are shown with the corresponding fits.

The fitted power law for the PSD has a roughly similar value $\beta\sim-2.3$ for all radio bands within the uncertainties, flattening to about  $\beta\sim-1.7$ for optical, X-rays and $\gamma-$ray values. 
The improvement in the statistics of 43~GHz with respect to other iMOGABA data sets is quite noticeable, as this frequency benefits from the inclusion of Boston University data. The overall robustness of the fitted parameters over the various bands will ensure representative PSD and PDF models that will lead to adequate artificial light curves to estimate the correlation functions confidence intervals.

\begin{table}
\begin{center}
\caption{Light curves statistical properties.}
\label{table:lc_stats}
\begin{tabular}{ccccc}
\tableline
\hline
Freq. Band & PDF $x$ & PDF $dx$ & PDF $\alpha$ & PSD $\beta$ \\
(1)&(2)&(3)&(4)&(5)\\
\tableline
$	1.50\times10^{10}	$&$	2.2	\pm	0.1	$&$	2.8	\pm	0.1	$&$		 - 		$&$	-2.1	\pm	0.1	$\\
$	2.20\times10^{10}	$&$	2.0	\pm	0.4	$&$	3.2	\pm	0.4	$&$		 - 		$&$	-2.4	\pm	0.4	$\\
$	4.30\times10^{10}	$&$	1.5	\pm	0.3	$&$	4.5	\pm	0.3	$&$		 - 		$&$	-2.1	\pm	0.2	$\\
$	8.60\times10^{10}	$&$	1.1	\pm	0.5	$&$	4.3	\pm	0.5	$&$		 - 		$&$	-2.5	\pm	0.4	$\\
$	1.29\times10^{11}	$&$	1.5	\pm	0.4	$&$	2.1	\pm	0.4	$&$		 - 		$&$	-2.3	\pm	0.9	$\\
$	2.55\times10^{11}	$&$	0.9	\pm	0.3	$&$	4.4	\pm	0.3	$&$		 - 		$&$	-2.3	\pm	0.2	$\\
$	5.44\times10^{14}	$&$	 - 			$&$		-		$&$	0.87	\pm	0.2	$&$	-1.9	\pm	0.2	$\\
$	4.84\times10^{17}	$&$	0.05	\pm	0.01	$&$	0.18	\pm	0.01	$&$		 - 		$&$	-1.7	\pm	0.2	$\\
$	3.63\times10^{25}	$&$	 - 			$&$		-		$&$	0.76	\pm	0.2	$&$	-1.7	\pm	0.2	$\\
\tableline
\end{tabular}
\end{center}
\end{table}

\begin{figure*}
\includegraphics[scale=0.9,trim={0cm 0cm 0cm 0cm},clip]{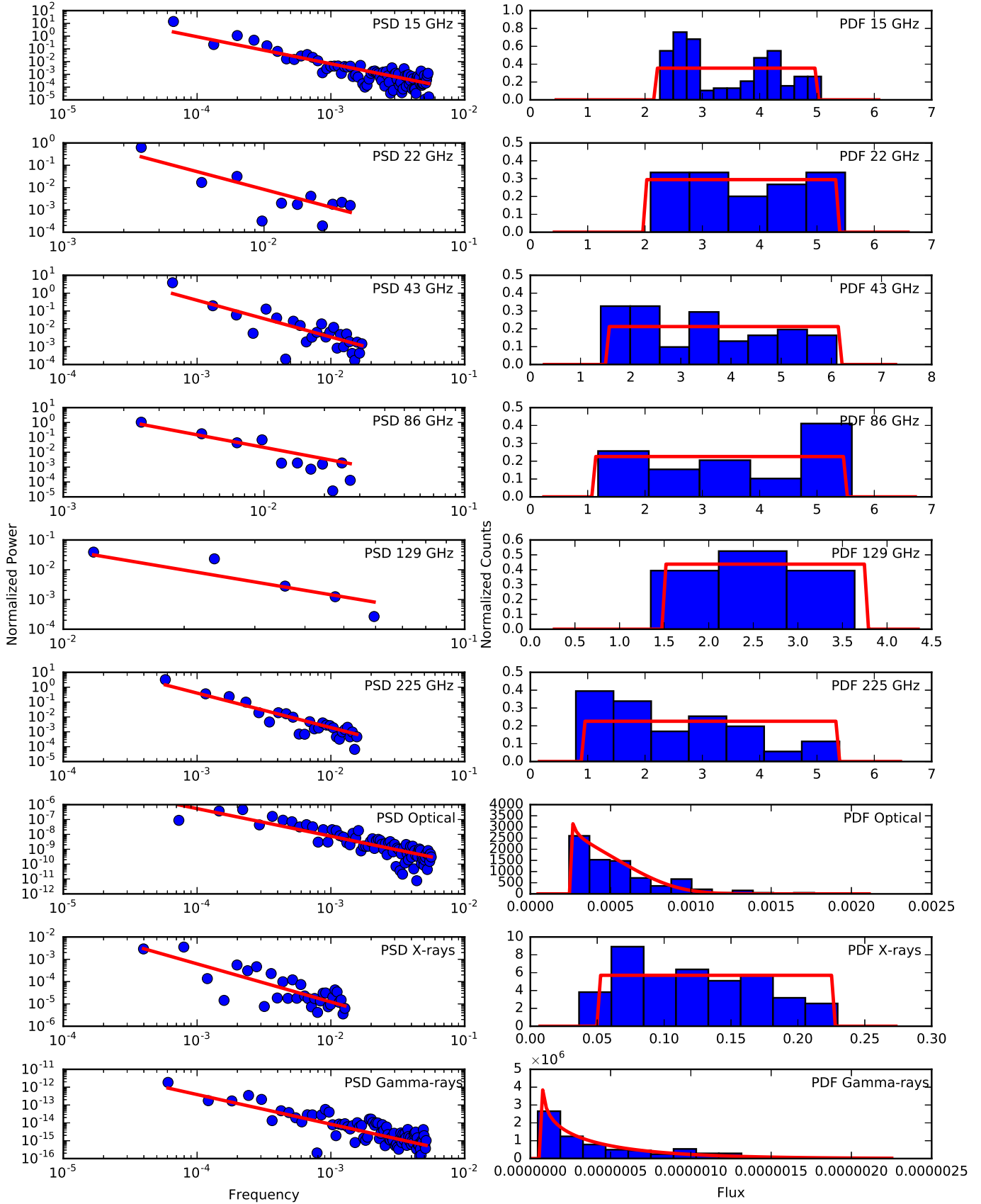}
\caption{Statistical properties of the light curves. Left: PSD. Red lines indicate the fit to a simple power law PSD; Right: PDF. Red line indicates the fit to a uniform or gamma distribution function (see text for details).}
\label{fig:lc_stats}
\end{figure*}

A plot showing the PSD slopes a a function of frequency is shown in Figure \ref{PSDvsFreq}. Within the error bars, the PSD slopes are similar at mm, radio, optical, X-ray and gamma-ray frequencies. Different slope infer dominance of different physical processes but here it seems not to be the case.

\begin{figure}
\includegraphics[scale=0.43,trim={0cm 0cm 0cm 0cm},clip]{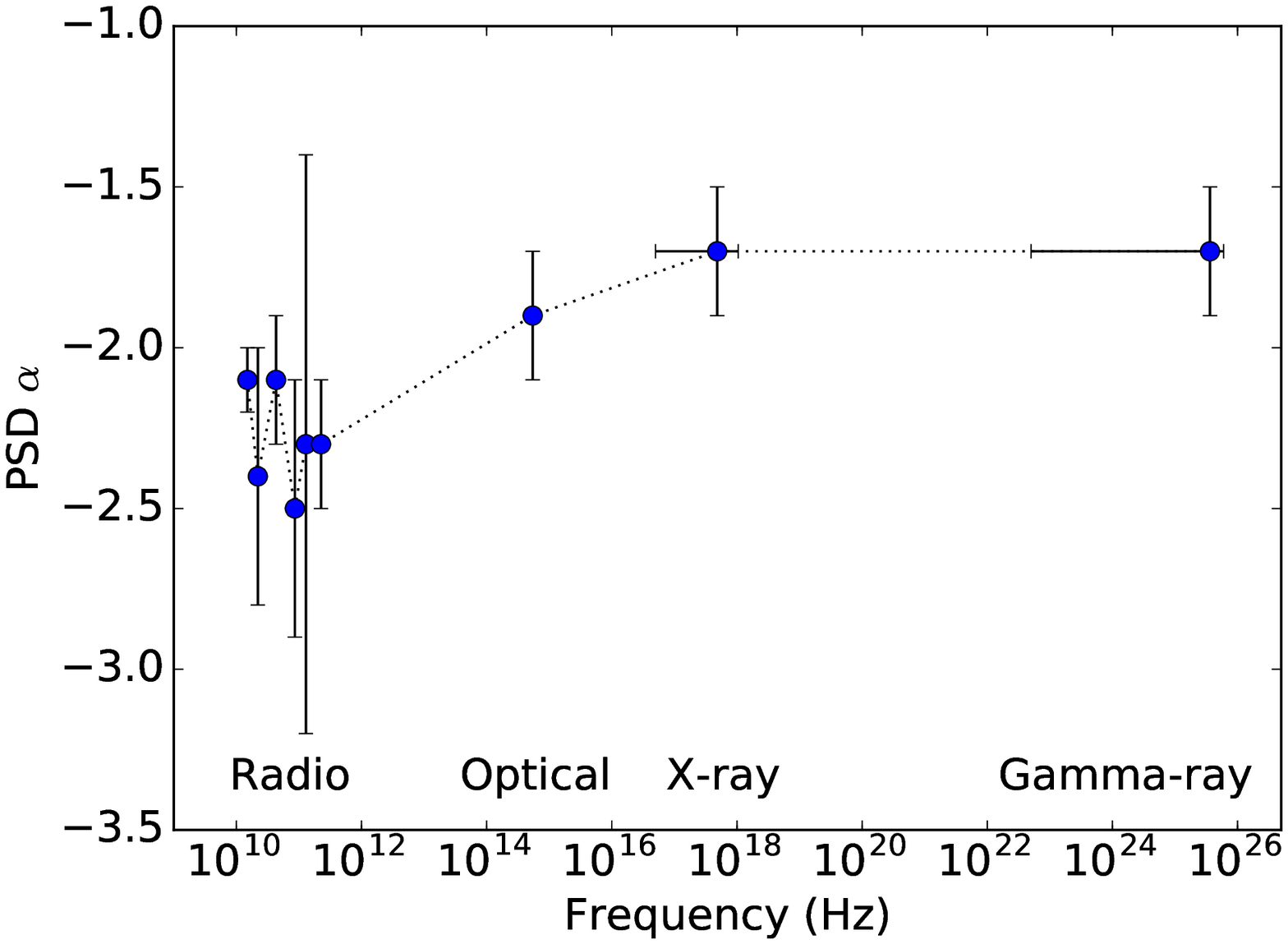}
\caption{PSD power law index as a function of frequency band.}
\label{PSDvsFreq}
\end{figure}

\subsection{Cross--Correlations and Time Delays}
\label{cc&td}

We used the discrete correlation function method first introduced by \cite{EdelsonKrolik88} to investigate the correlation and possible time lags between the observed light curves. This method is very useful for unevenly sampled data, which is our case here. First, we calculate the unbinned discrete correlation function
\begin{equation}
UDCF_{ij}=\frac{(a_i-\bar{a})(b_j-\bar{b})}{\sqrt{\sigma^2_a \sigma^2_b}},
\end{equation}
where $a_i$, $b_i$ are the individual data points for each light curve; $\bar{a}$, $\bar{b}$ are the mean values of the respective time series and $\sigma^2_a$, $\sigma^2_b$ are the variances of the time series. The discrete correlation function is then found via
\begin{equation}
DCF(\tau)=\frac{1}{N}\sum{UDCF_{ij}(\tau)}
\end{equation}
with error
\begin{equation}
err(DCF)=\frac{1}{N-1}\sqrt{\sum{[UDCF_{ij}(\tau)-DCF(\tau)]^2}}.
\end{equation}

Note that the bin size is not implicit in this definition of the DCF, but has to be determined for each particular case. 
For the data presented here, we considered the bin size as the minimum between the typical cadences of the two cross--correlated light curves. As a consistency check, we also considered the DCF with a bin two times larger and smaller, finding our results consistent with our a priori estimation. The profile of the DCF was approximated by a Gaussian distribution, and the time lag was derived by obtaining the peak of the fitted Gaussian. Time lags obtained with different time bins agreed within one day difference at most. The positive time lag indicates that the flaring activity at the most energetic (higher frequency) band is leading.

In order to study the DCF significance, we performed a Monte Carlo analysis. We proceeded in the following manner: first, for each of the bands that we wanted to cross--correlate, we simulated 20000 light curves with the same variability and statistical properties as the original one (i.e., probability distribution functions and power spectral densities). For this purpose, we used the public code by \cite{Connolly15} based on the implementation described in \cite{Emmanoulopoulos13}. We then cross--correlated the simulated light curves and compared their DCF to obtain the 68, 95 and 99.7\% confidence levels (1, 2 and 3$\sigma$ confidence levels, respectively). DCF peaks  above the 95\% confidence level are considered to be significant.

In general, the profile of the DCF can be quite well approximated by a Gaussian distribution. We have thus fitted such a Gaussian around the peak of the DCF. We consider the location of the peak of such Gaussian as a good estimator for the time delay between the two frequency pairs. As an estimate of the error of the time delay, we used  the half width at 90\% height of the fitted Gaussian. This approach is similar, although slightly less conservative, than taking the FWHM, followed by e.g. \cite{Rani13} or \cite{Chidiac16}. We note that, in any case, the results are not significantly affected  by the errors on the time lag. Table \ref{timelag} contains information about the time lags and confidence levels estimated for the various DCFs.

\begin{table}
\begin{center}
\caption{Time Lag.}
\label{timelag}
\begin{tabular}{lcccc}
\tableline
\hline
&\multicolumn{2}{c}{All LC} & \multicolumn{2}{c}{LC MJD$>$56500}\\
\cline{2-3} \cline{4\,-5}
Frequency pair & lag (days) & CL ($\sigma$) & lag (days) & CL ($\sigma$)\\
(1)&(2)&(3)&(4)&(5)\\
\tableline
15-22	GHz			&$	27	\pm	39	$& 	3	&$	3	\pm	31	$& 	3	\\
15-43	GHz			&$	15	\pm	38	$& 	3	&$	5	\pm	32	$& 	3	\\
15-86	GHz			&$	35	\pm	37	$& 	3	&$	-2	\pm	31	$& 	3	\\
15-129	GHz			&$	30	\pm	29	$& 	3	&$	31	\pm	27	$& 	3	\\
15-225	GHz			&$	24	\pm	38	$& 	3	&$	37	\pm	30	$& 	3	\\
\rule{0pt}{4ex}15	GHz	-	Optical	&$	41	\pm	20	$& 	1	&$	101	\pm	21	$& 	1	\\
22	GHz	 - 	Optical	&$	151	\pm	30	$& 	2	&$	138	\pm	21	$& 	2	\\
43	GHz	 - 	Optical	&$	161	\pm	36	$& 	1	&$	192	\pm	32	$& 	1	\\
86	GHz	 -	Optical	&$	204	\pm	28	$& 	2	&$	246	\pm	33	$& 	2	\\
129	GHz	 -	Optical	&$	 -			$& 	1	&$		-		$& 	1	\\
225	GHz	 -	Optical	&$	 - 			$& 	1	&$	135	\pm	30	$& 	1	\\
\rule{0pt}{4ex}15	GHz	-	X--ray		&$	91	\pm	52	$& 	2	&$	97	\pm	25	$& 	2	\\
43	GHz	 -	X--ray		&$	101	\pm	38	$& 	2	&$	124	\pm	43	$& 	2	\\
225	GHz	 -	X--ray		&$	90	\pm	35	$& 	2	&$	82	\pm	28	$& 	2	\\
Optical	 -	X--ray		&$	9	\pm	17	$& 	3	&$	12	\pm	12	$& 	3	\\
\rule{0pt}{4ex}15	GHz	-	$\gamma-$ray	&$	67	\pm	40	$& 	3	&$	73	\pm	30	$& 	2	\\
43	GHz	 -	$\gamma-$ray	&$	89	\pm	41	$& 	3	&$	70	\pm	28	$& 	3	\\
225	GHz	 -	$\gamma-$ray	&$	69	\pm	40	$& 	3	&$	39	\pm	25	$& 	3	\\
Optical	 -	$\gamma-$ray	&$	8	\pm	17	$& 	2	&$	-6	\pm	15	$& 	2	\\
X--ray	 -	$\gamma-$ray	&$	17	\pm	30	$& 	3	&$	-5	\pm	19	$& 	3	\\
\tableline
\end{tabular}
\end{center}
\end{table}

Since the light curves consist of more than one flux density enhancement, the DCF analysis of the entire data  shows an average of these. To cross-check 
the consistency of the estimated DCF results, we choose the data for the last radio flux density enhancement (MJD $>$56500); in both cases, we obtained similar results.  The DCF results for 
different bands for the entire data sets and also for the third radio flux density enhancement are listed in Table \ref{timelag}. The DCF curves are shown in Figs. \ref{DCF} to \ref{DCFgamma} and in Fig.\ \ref{DCFRappendix}. Time lags as a function of frequency are shown in Figure \ref{tlag}. In the following sub-sections, we discuss the cross-correlation analysis results in detail.

\subsubsection{Radio--Radio correlations}
In Figure \ref{DCF} we plot the DCF for 15~GHz and the rest of the radio--frequencies. The DCF at all different radio bands show a correlation with a significance of $\geq3\sigma$. The estimated time lags between radio--radio bands, 
listed in Table \ref{timelag}, are consistent with no lag within uncertainties.  The analysis therefore suggests 
a significantly correlated concurrent flaring activity at radio frequencies. It is however important to note 
that the average sampling of radio data is about a month. 

\begin{figure}
\includegraphics[scale=0.21,trim={0cm 0cm 0cm 0cm},clip]{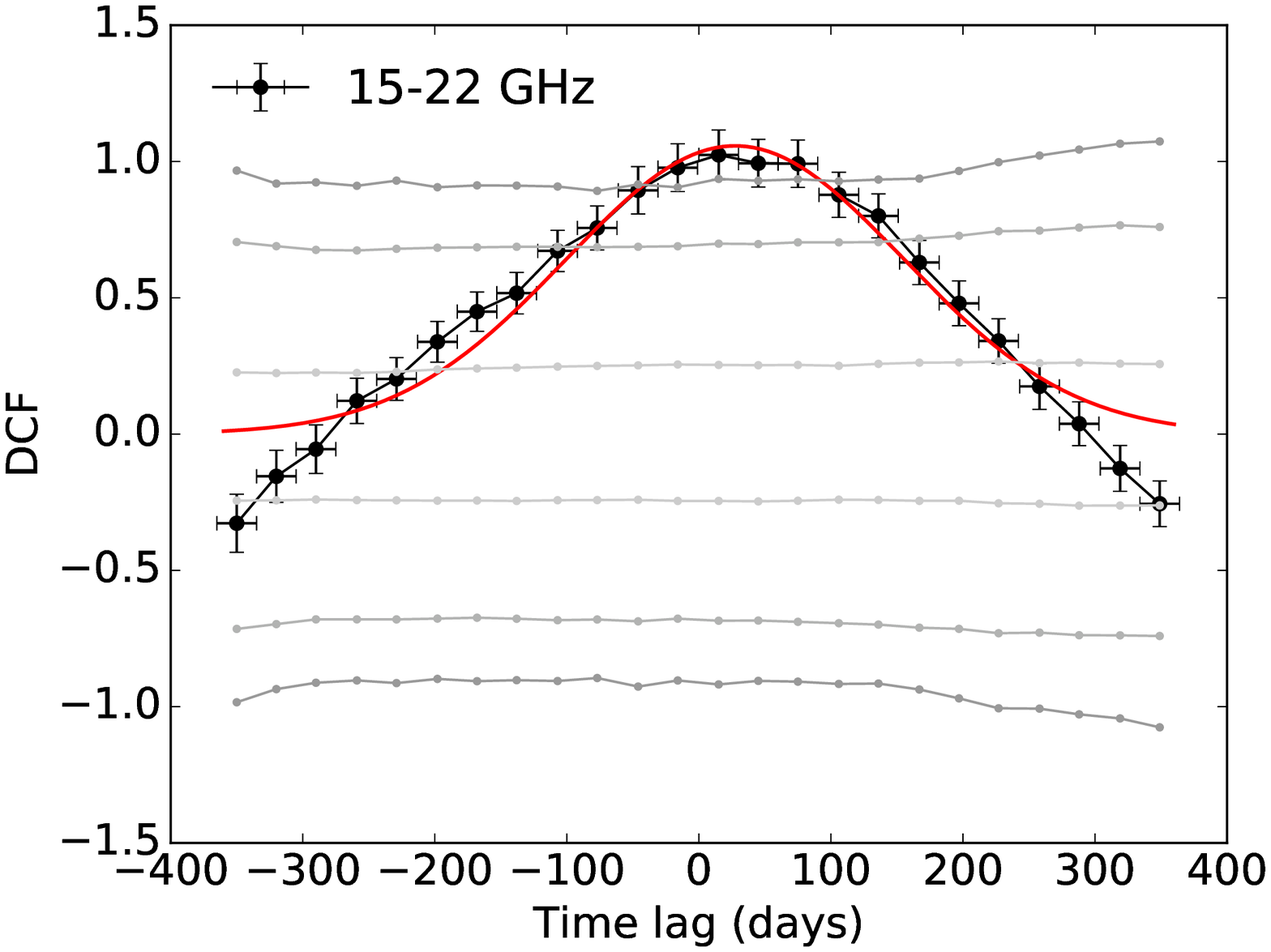}
\includegraphics[scale=0.21,trim={0cm 0cm 0cm 0cm},clip]{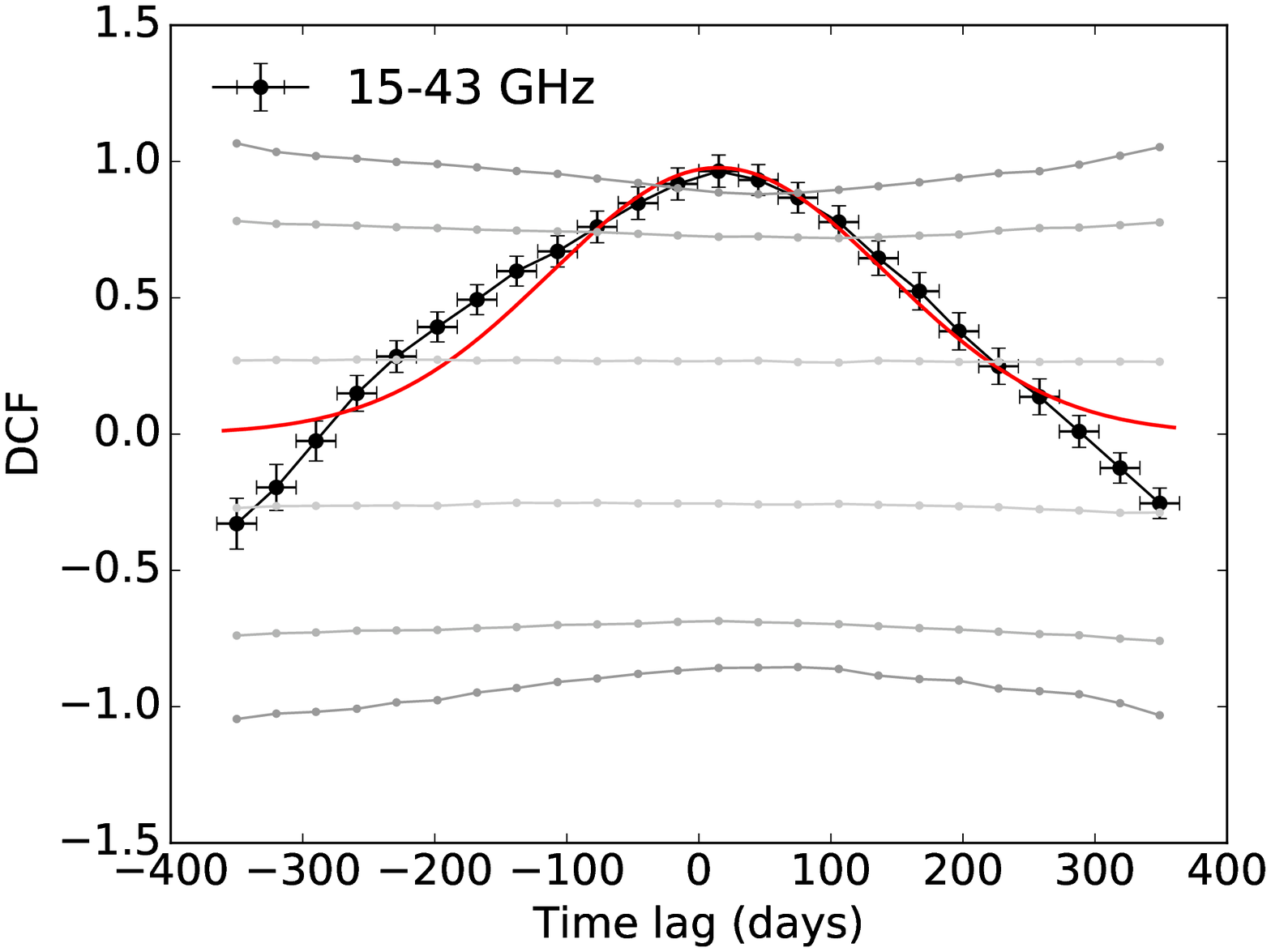}\\
\includegraphics[scale=0.21,trim={0cm 0cm 0cm 0cm},clip]{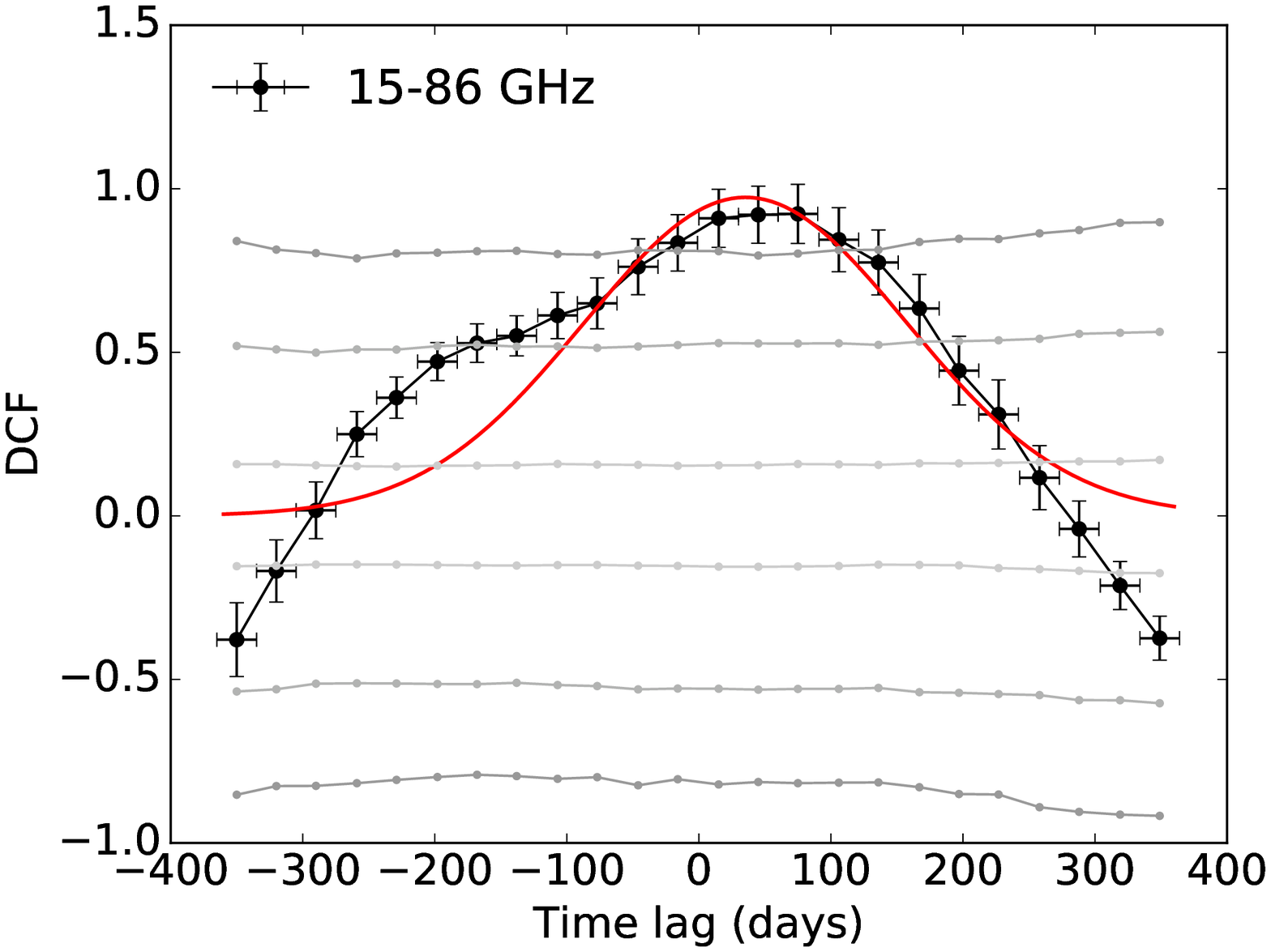}
\includegraphics[scale=0.21,trim={0cm 0cm 0cm 0cm},clip]{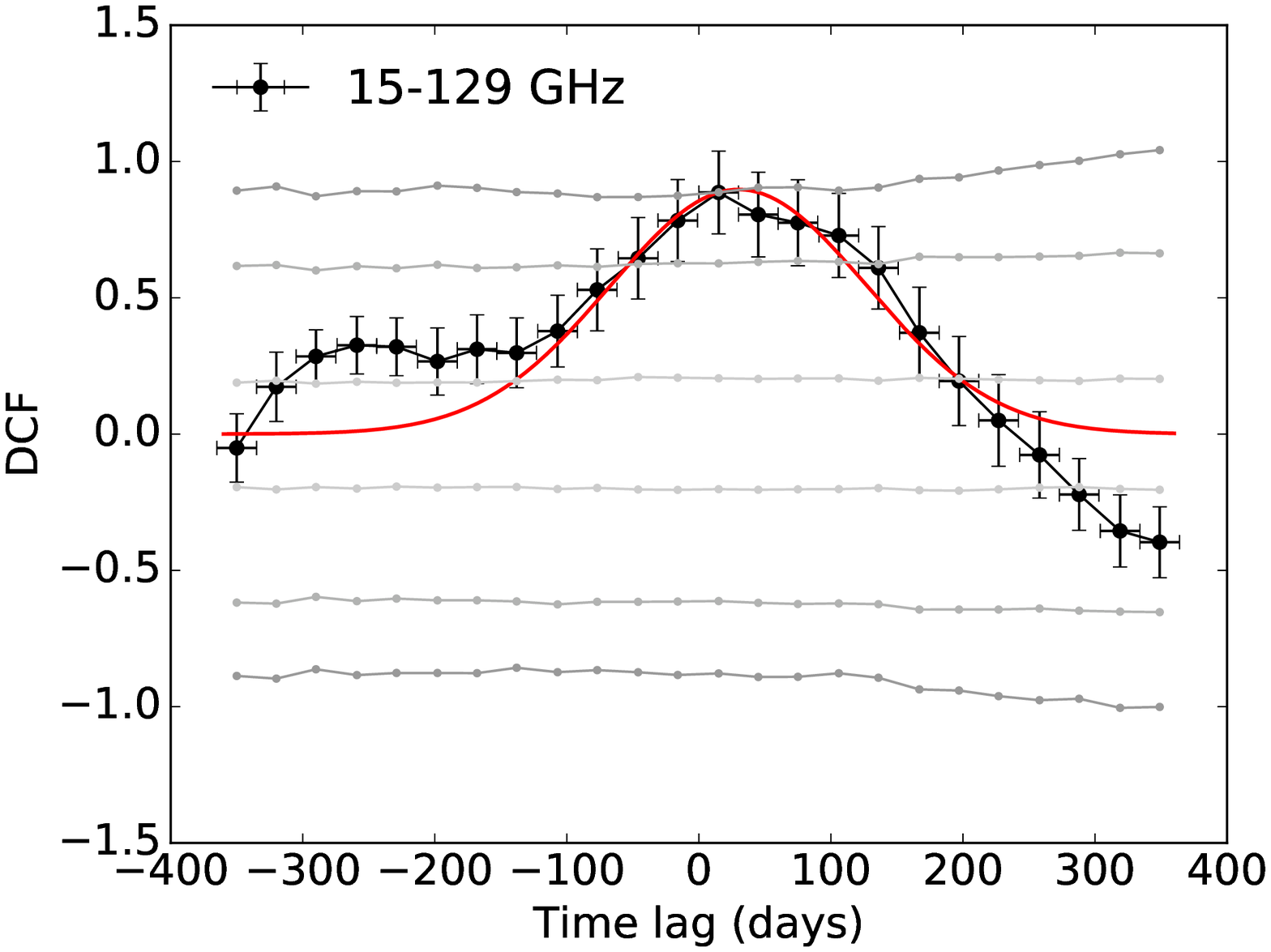}\\
\includegraphics[scale=0.21,trim={0cm 0cm 0cm 0cm},clip]{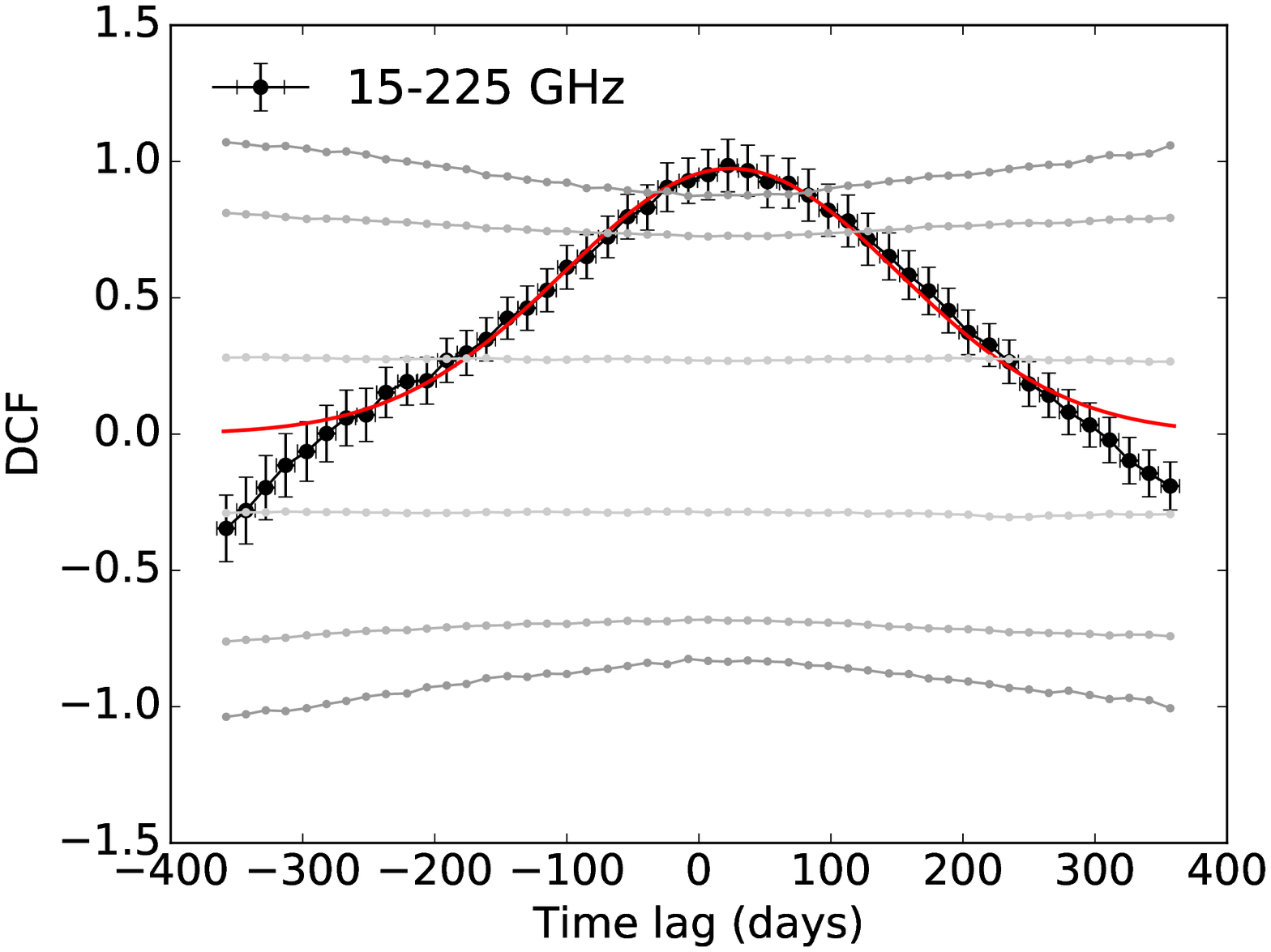}
\caption{Discrete Correlation Function of 1633+382 for 15~GHz OVRO data with the rest of radio--bands. Black connected dots indicate the DCF for the given pair of frequencies indicated in the top left corner. Gray lines indicate the 68, 95 and 99.7\% confidence levels, with darker color denoting higher confidence levels. The red thicker line corresponds to a Gaussian fitting around the peak of the DCF. Discrete correlation functions for MJD$>$56500 are shown in Figure \ref{DCFRappendix} in the appendix.}
\label{DCF}
\end{figure}

\subsubsection{Optical--Radio Correlations}
In Figure \ref{DCFro} we plot the DCF for each of the radio bands with optical data, and the estimated time lags 
are listed in Table \ref{timelag}. Positive time lags suggest that the flux variations at optical 
frequencies lead those at radio  bands by  $\sim$150~days. However, the significance of the DCF is in general below $95\%$. 
The absence of a significant optical--radio correlation can either be due to 
poor data sampling, hence in future could be tested via good cadence radio--optical observations; or it 
could be intrinsic to the source as is seen for 3C 273 \citep{Chidiac16}.

\begin{figure}
\includegraphics[scale=0.21,trim={0cm 0cm 0cm 0cm},clip]{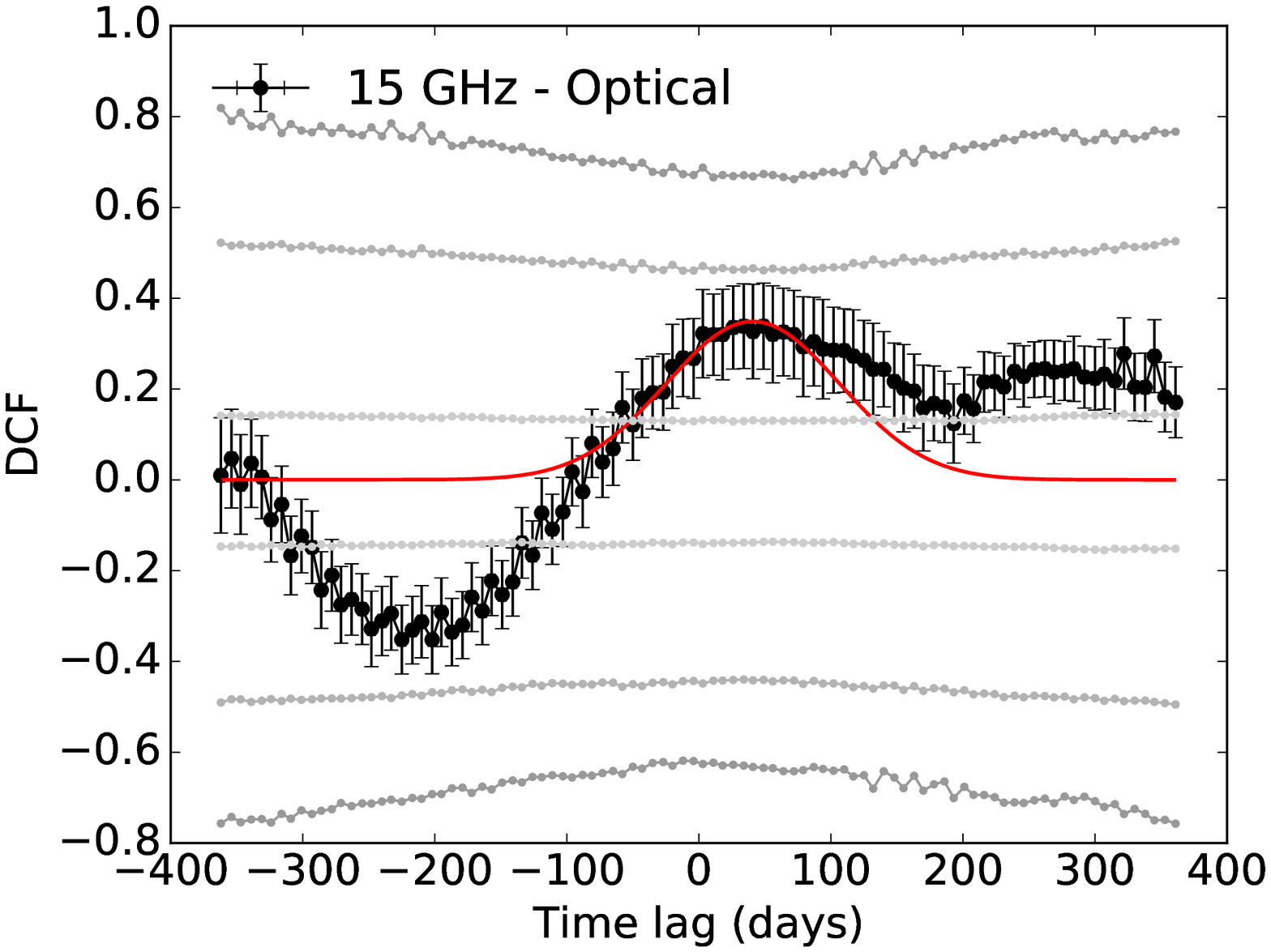}
\includegraphics[scale=0.21,trim={0cm 0cm 0cm 0cm},clip]{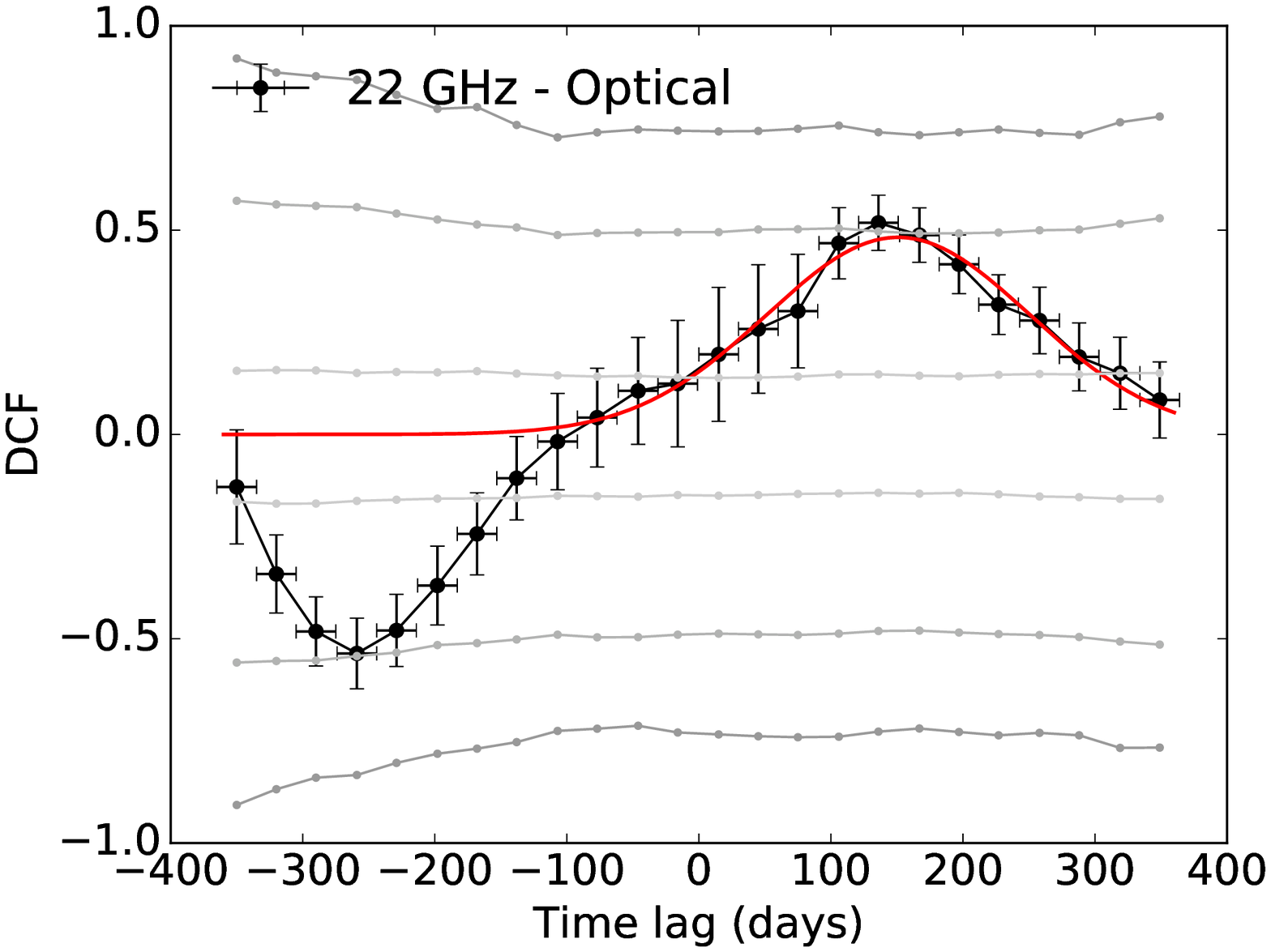}\\
\includegraphics[scale=0.21,trim={0cm 0cm 0cm 0cm},clip]{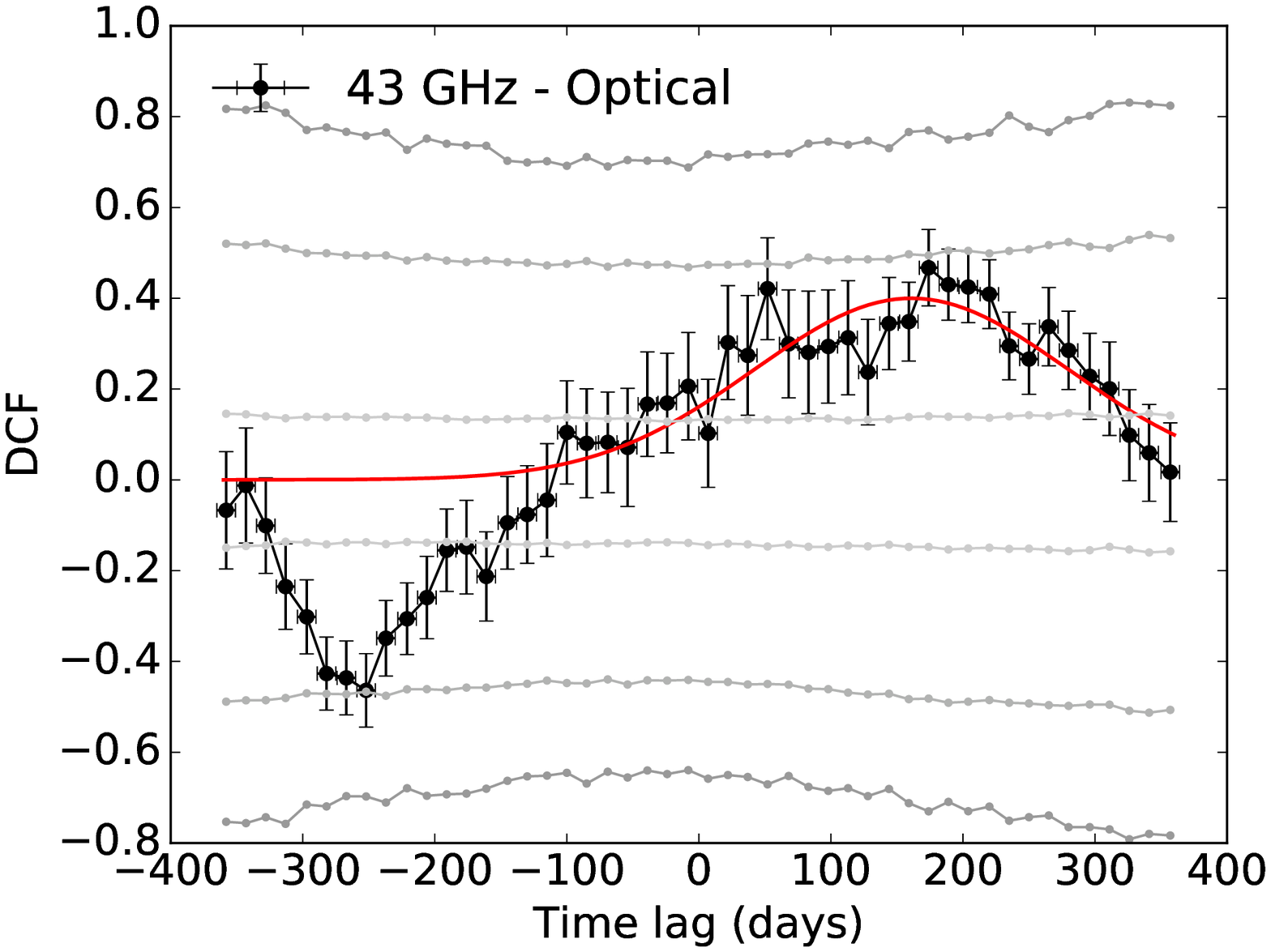}
\includegraphics[scale=0.21,trim={0cm 0cm 0cm 0cm},clip]{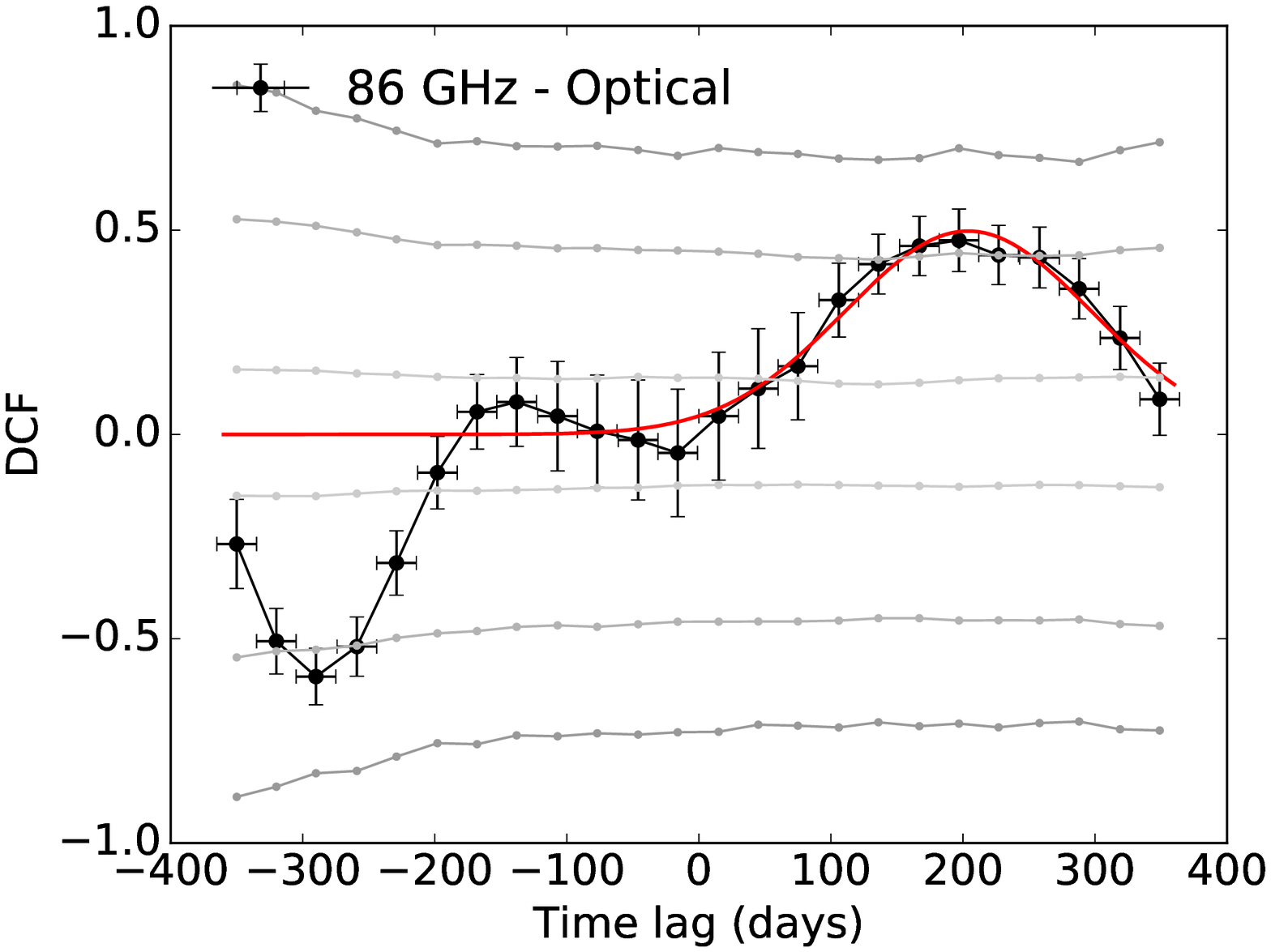}\\
\includegraphics[scale=0.21,trim={0cm 0cm 0cm 0cm},clip]{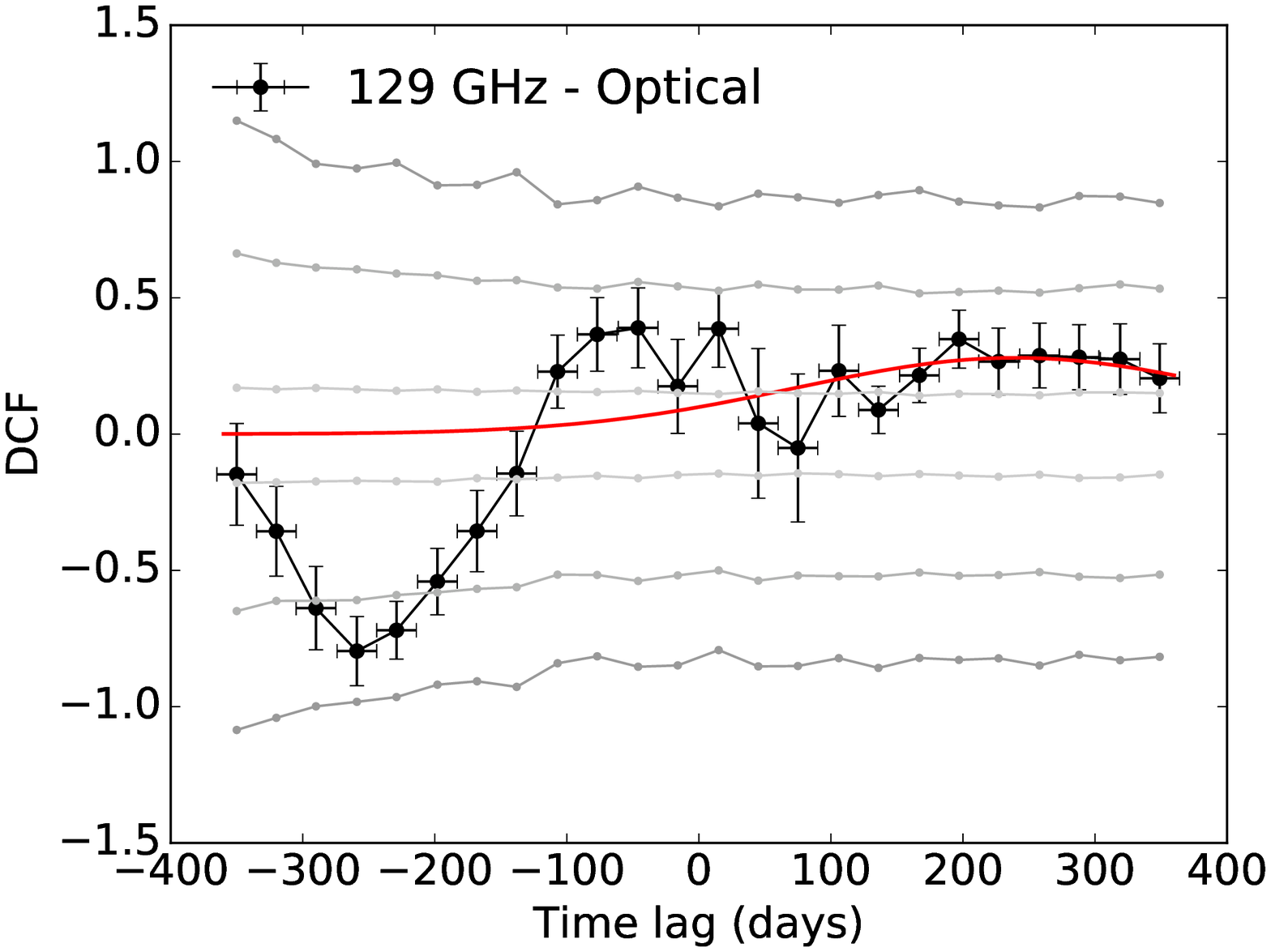}
\includegraphics[scale=0.21,trim={0cm 0cm 0cm 0cm},clip]{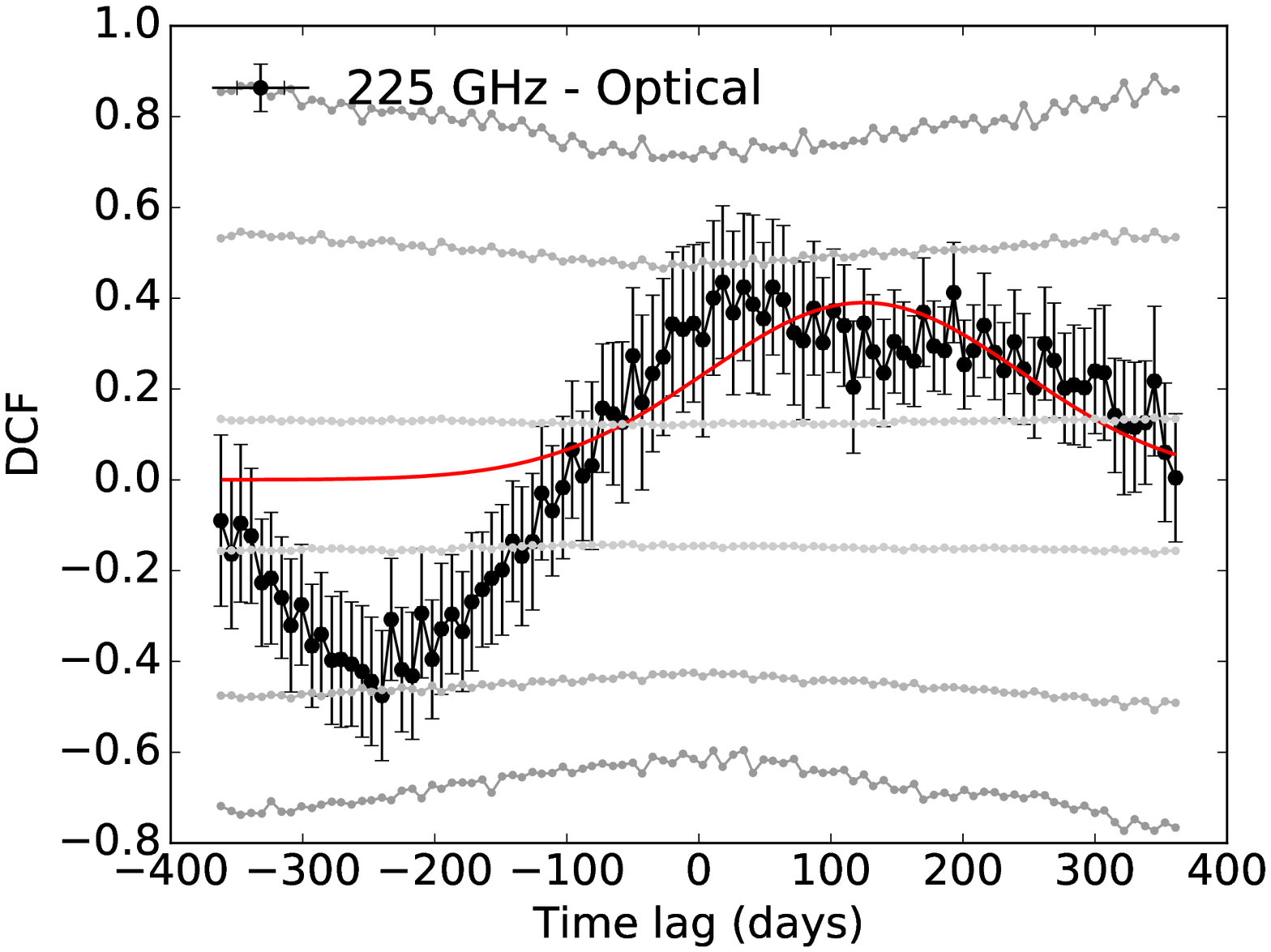}
\caption{Discrete Correlation Function of 1633+382 for radio--bands with optical data. Symbols and lines have the same meaning as in Fig. \ref{DCF}.}
\label{DCFro}
\end{figure}

\subsubsection{X--ray Correlations}
The cross-correlation analysis indicates a significant correlation ($>95\%$) between the flux variations 
at X-ray energies with those at optical frequencies and a poorer correlation between X-rays and radio bands.  
In Fig.\ \ref{DCFx}, we plot the cross-correlation analysis of X--rays with 15, 43 and 225~GHz 
radio data, and also X--rays with optical data. The estimated time lags are given in Table \ref{timelag}. 
Our analysis suggests that the flaring activity at X-rays leads the radio flux variability by $\sim$100~days. 
The values for all radio bands  appear to be compatible within the errors, without any significant trend. On 
the other hand, the estimated time lag between X-ray and optical variations is  consistent with zero within 
uncertainties, which suggests concurrent flaring activity at the two frequencies.

\begin{figure}
\includegraphics[scale=0.21,trim={0cm 0cm 0cm 0cm},clip]{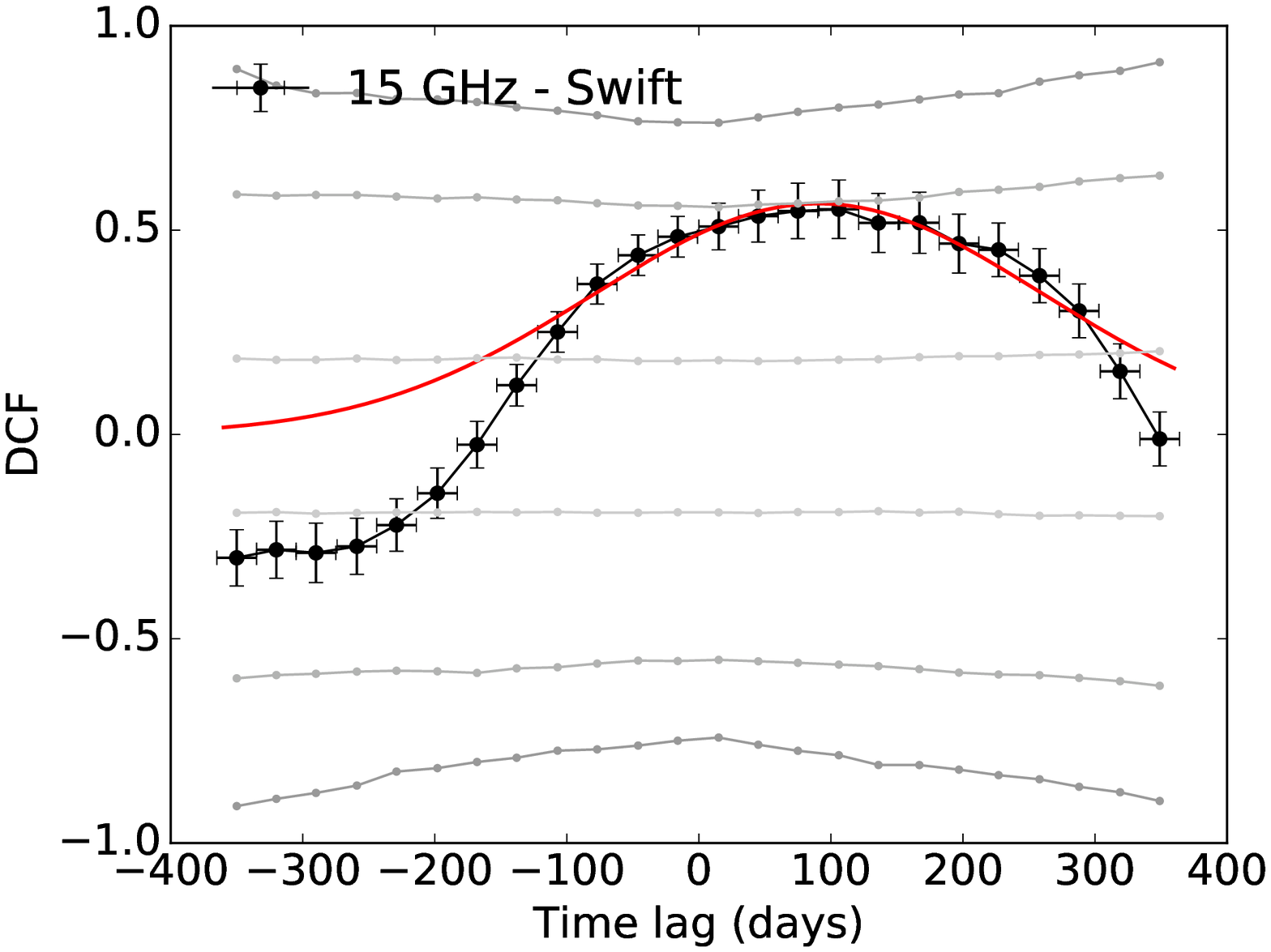}
\includegraphics[scale=0.21,trim={0cm 0cm 0cm 0cm},clip]{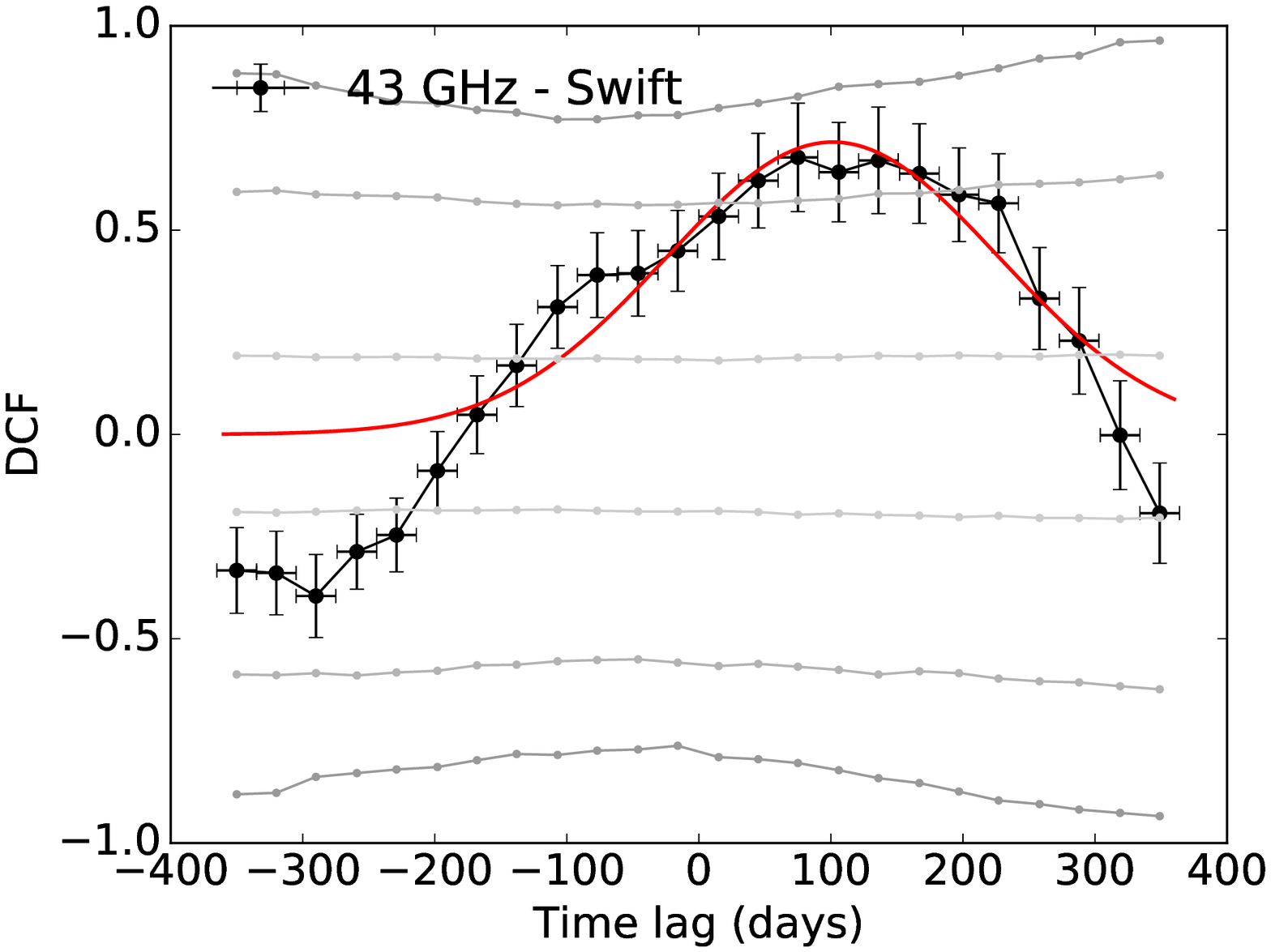}\\
\includegraphics[scale=0.21,trim={0cm 0cm 0cm 0cm},clip]{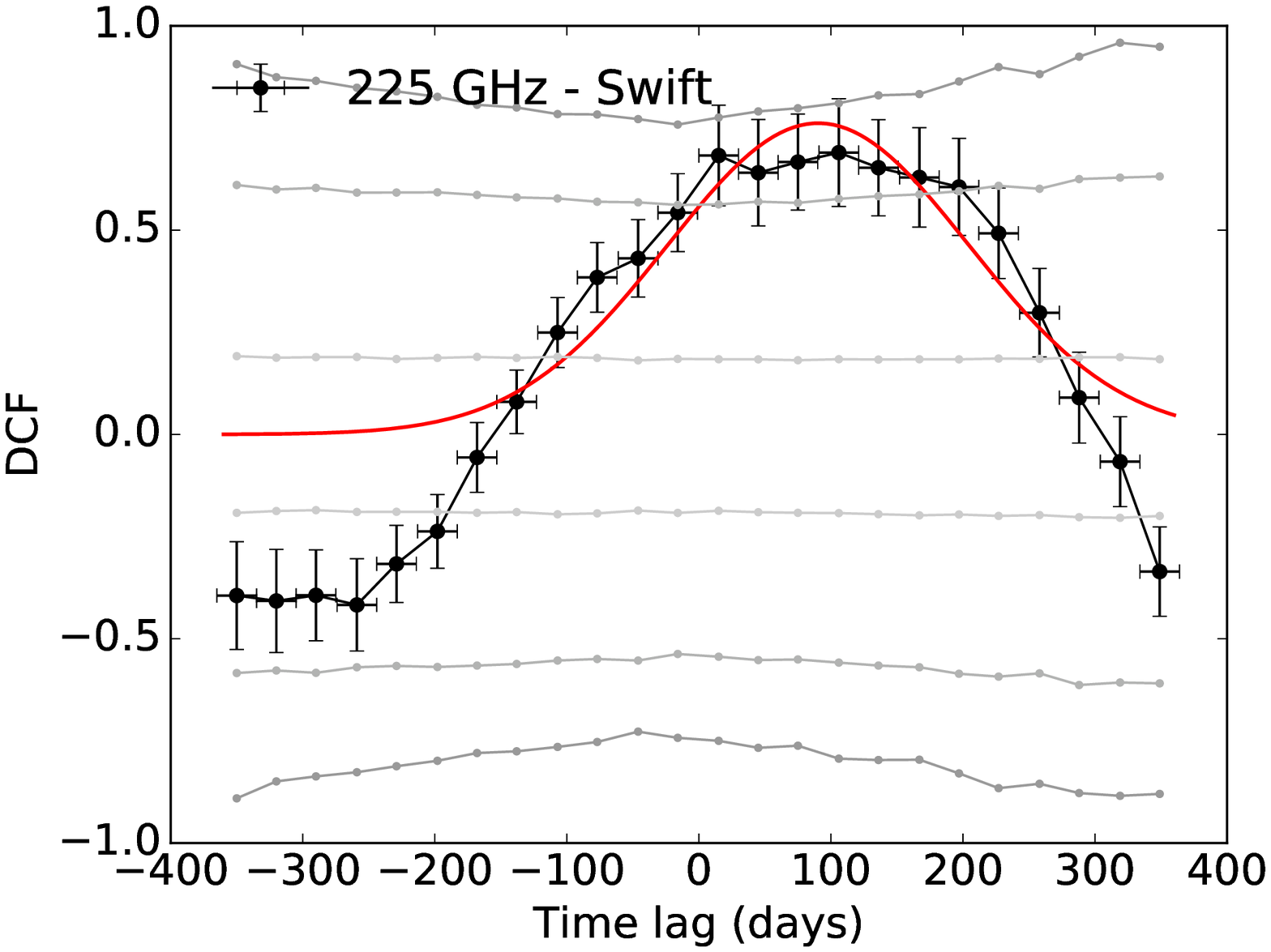}
\includegraphics[scale=0.21,trim={0cm 0cm 0cm 0cm},clip]{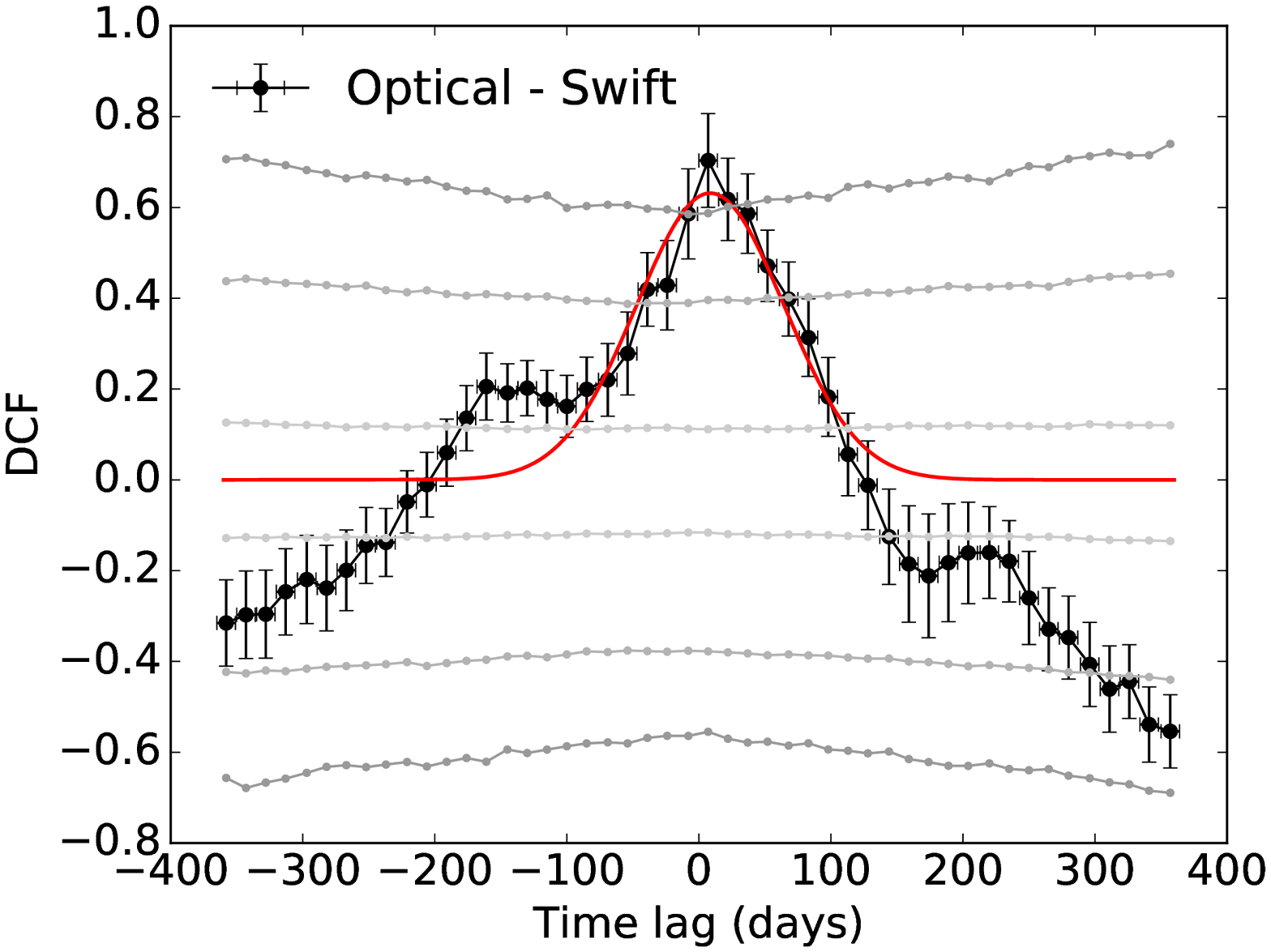}\\
\caption{Discrete Correlation Function of 1633+382 for X--rays {\it Swift}-XRT data. Symbols and lines have the same meaning as in Fig. \ref{DCF}.}
\label{DCFx}
\end{figure}

\subsubsection{$\gamma-$ray Correlations}
DCF analysis curves for $\gamma-$rays vs.\  radio--bands (15, 43, and 225~GHz), optical and X--rays are 
shown in Fig.\ \ref{DCFgamma}, and the time lags obtained are given in Table \ref{timelag}. In general the  
correlation peaks in the DCF curves are well above 95$\%$ confidence level. 
Time lags for $\gamma-$ray vs.\ radio bands suggest that the former leads the latter by $\sim$70~days.  
DCF curves  for high frequency radio bands show a small deviation from the Gaussian fit, or even a small bump on the left side of the main peak, which disappears when we investigate the truncated light curves' DCF, and thus may be attributed to the averaging of possibly different delays from the various flux density enhancements, as discussed in Section \ref{cc&td} above.

The estimated time lags between $\gamma-$ray and optical (8$\pm$17~days) and $\gamma-$ray and X--ray ($-$(6$\pm$15)~days) energies are compatible with zero. This means we are seeing concurrent flaring events at these frequencies. It should 
however be noted that for the correlation analysis, we used a weekly binned $\gamma-$ray light curve. 

\begin{figure}
\includegraphics[scale=0.21,trim={0cm 0cm 0cm 0cm},clip]{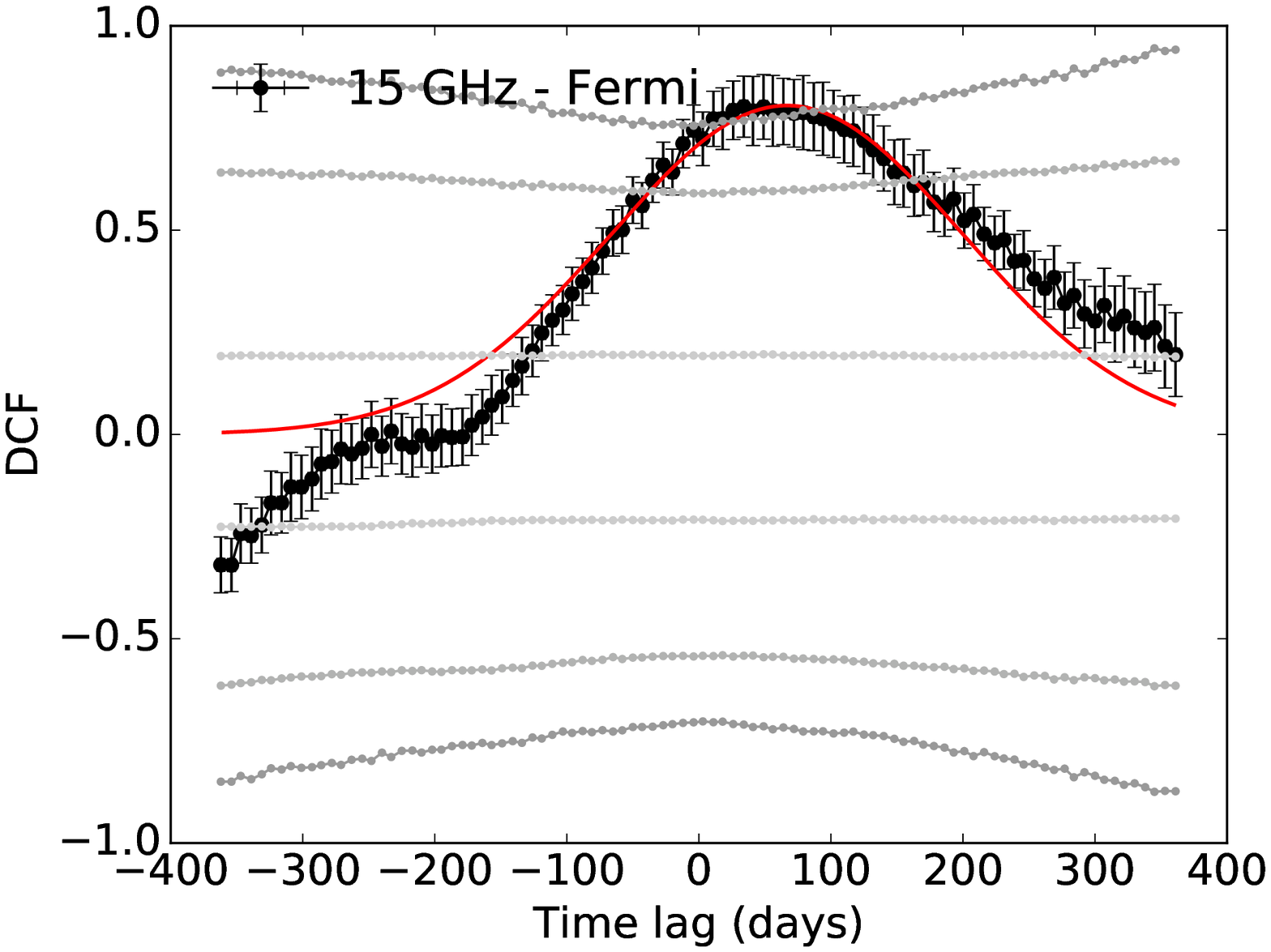}
\includegraphics[scale=0.21,trim={0cm 0cm 0cm 0cm},clip]{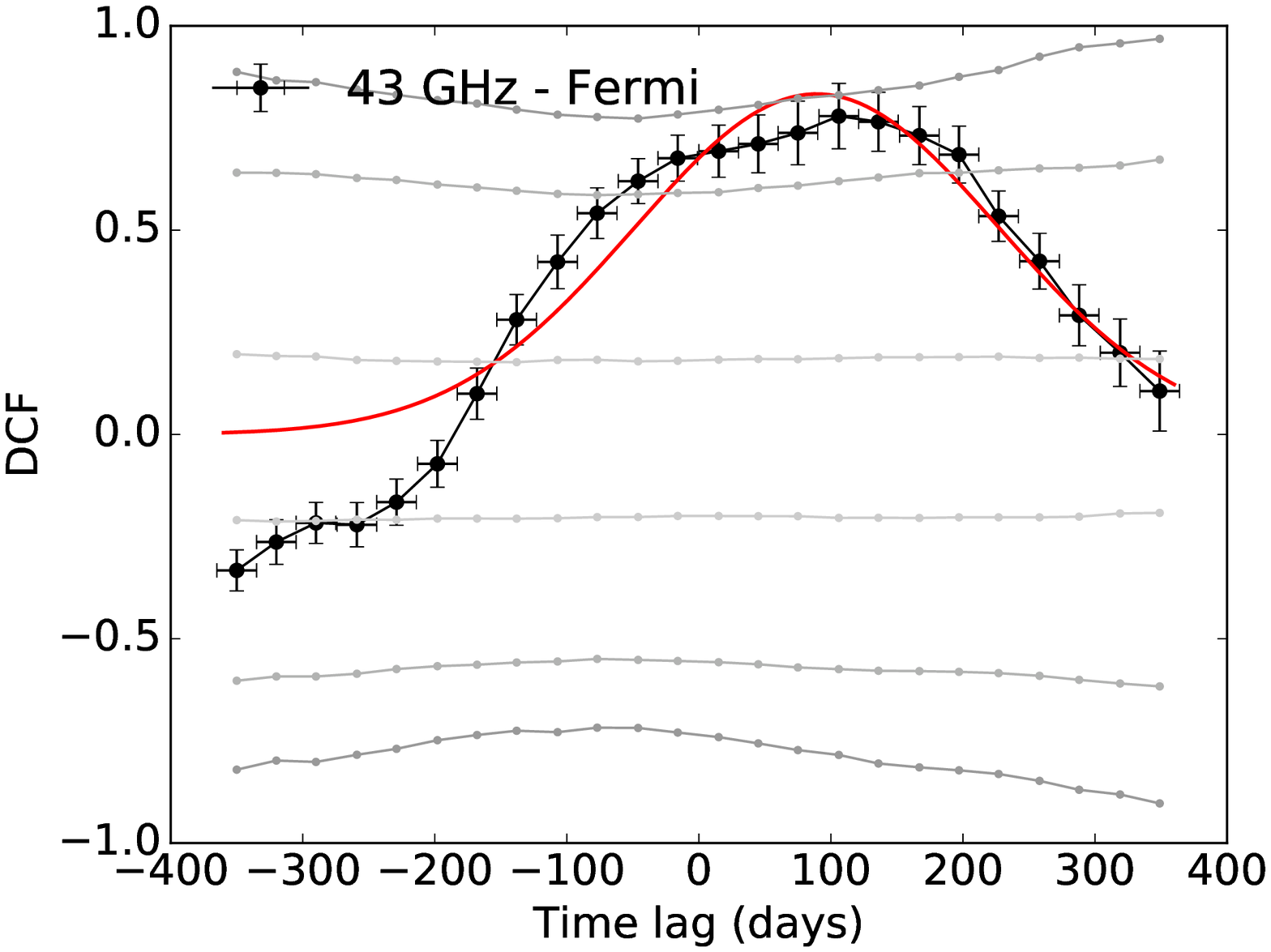}\\
\includegraphics[scale=0.21,trim={0cm 0cm 0cm 0cm},clip]{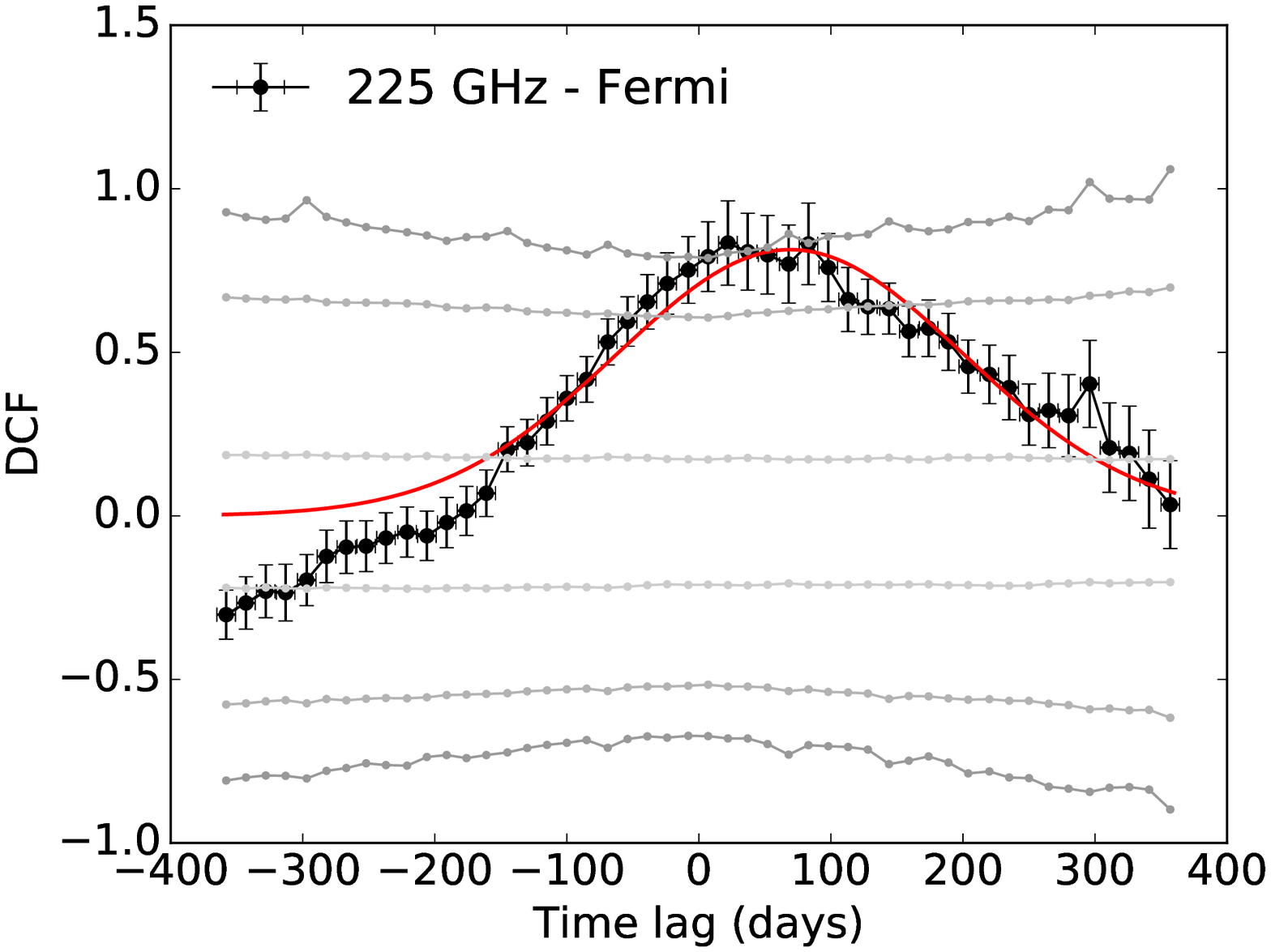}
\includegraphics[scale=0.21,trim={0cm 0cm 0cm 0cm},clip]{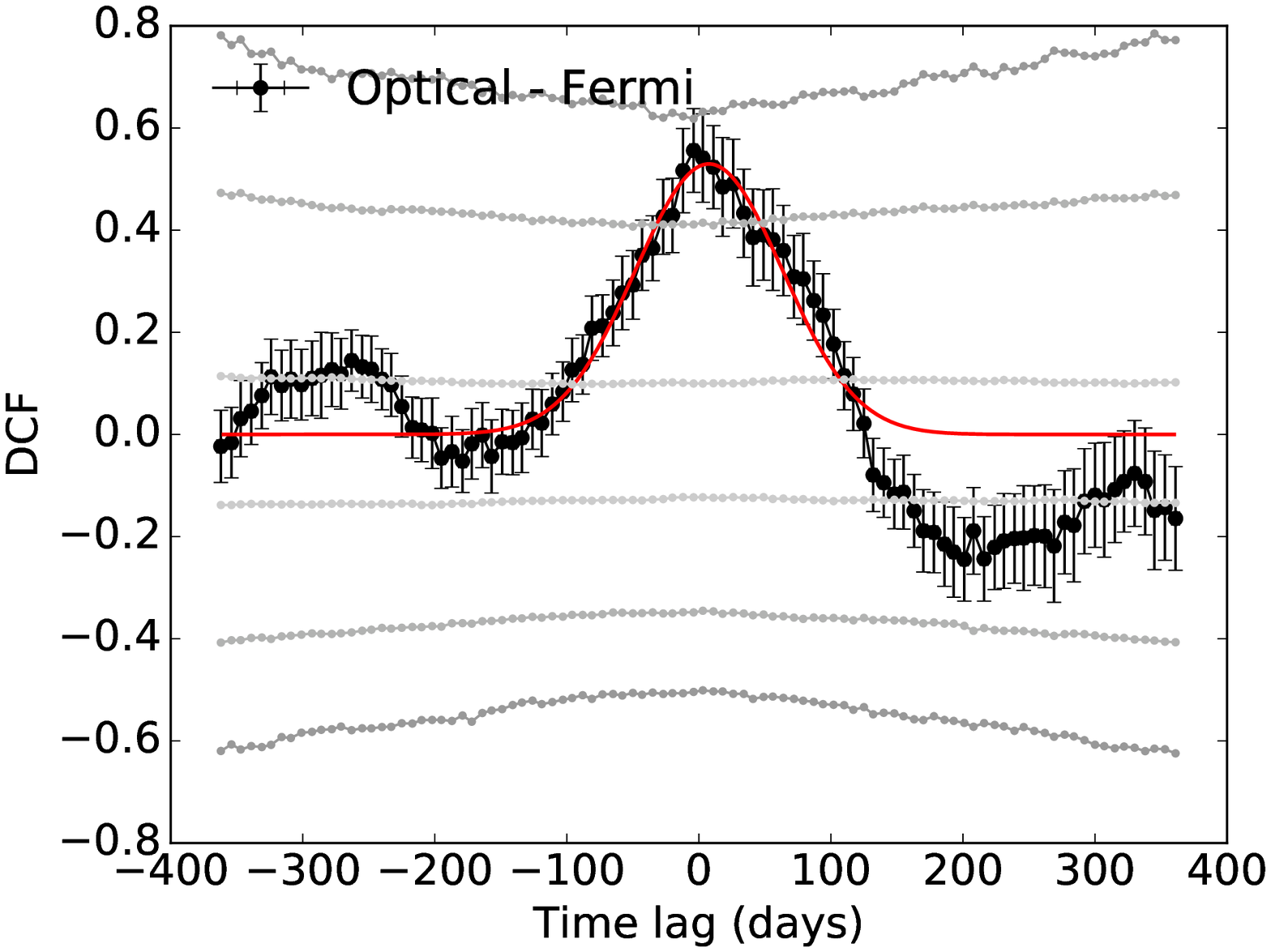}\\
\includegraphics[scale=0.21,trim={0cm 0cm 0cm 0cm},clip]{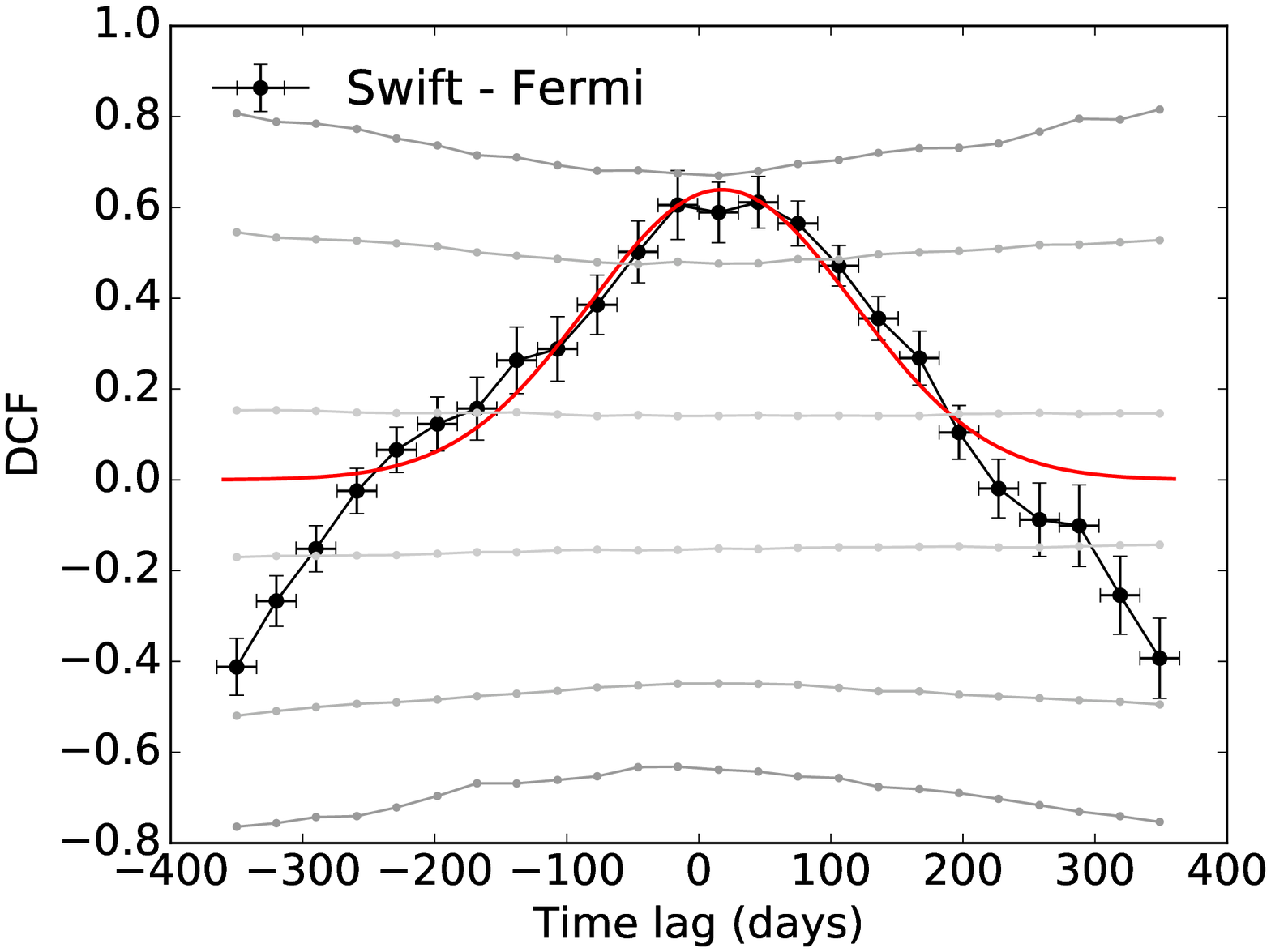}
\caption{Discrete Correlation Function of 1633+382 for $\gamma-$rays. Symbols and lines have the same meaning as in Fig. \ref{DCF}.}
\label{DCFgamma}
\end{figure}

\section{Discussion}
We used multi-wavelength observations  to pinpoint the  location of high-energy emission in the high-redshift blazar 1633+382. Visual inspection of the light curves of 1633+382 (see Figure \ref{lightcurve}) suggests that, for at least some of  the different examined frequency bands, a long--term correlation seems to exist in the observed flux densities. In shorter period scales, however, some variations or flux density enhancements are only observed in particular frequency bands and short-term correlations are not so clear. The most clear example is the orphan flux density enhancement at MJD$\sim$57050, which is only seen at optical and higher frequencies, but not in radio. This kind of phenomenology is also seen in the spectral index. While the overall spectral index evolution seems to be very similar at  various radio frequency pairs, this is not the case in optical, as the spectral index evolution between radio and optical frequency pairs seems to be different than what is seen for other bands. 

Radio spectral indices are significantly correlated with their flux densities. Larger flux densities are significantly associated with thicker opacities in the source. This effect would be natural in the context of the standard shock--in--jet model \citep{MarscherGear85} if we consider the additional flux to arise from an optically thick component. The different correlation observed at various radio bands is however quite intriguing. The fact that low--frequency spectral indices seem to be more correlated with flux than high--frequency spectral indices does not seem to be associated with bias in the data due to different scatter or spectral indices amplitudes. On the other hand, it might be possible that a connection with turnover frequency, or the particular flat radio spectra features of this source may exist. More observations and comparison with other sources with different spectral characteristics may be needed to better understand this \citep[e.g.,][]{LeeJW17}.

\begin{figure}
\includegraphics[scale=0.35,trim={0cm 0cm 0cm 0cm},clip]{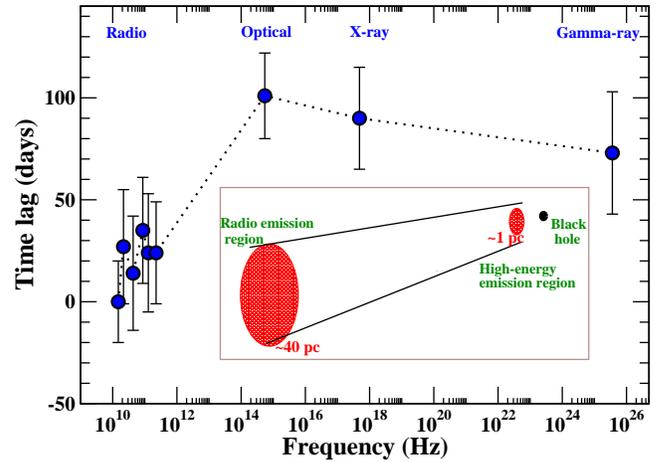}
\caption{Illustration of the resulting time lags with respect to 15~GHz using the DCF analysis. The inset shows a sketch representing the location of} the emitting regions as discussed in the text.
\label{tlag}
\end{figure}

A quantitative analysis using the DCF method indicates a significant 
correlation between the flux variations observed over twenty decades in 
energy near or above 3$\sigma$ level, excluding the optical band, for most of the frequency pairs.
Variations at different radio bands are found to be significantly 
correlated within each other with no time lag, which suggests 
concurrent flaring activity at radio bands. However we can not rule out 
the presence of time lags shorter than our sampling rate ($\sim$ 1 month) 
and the error estimated for the DCF peak, which is about 40 days.
Zero time lag between radio light curves indicates co-spatial origin 
of the radio emission regions, which is usually not what we observe 
for blazars. It has been observed for several blazars that the high 
radio frequency flux variations lead those at low frequencies 
\citep{Fuhrmann14, Chidiac16, Rani13,Karamanavis16}. Moreover, 
the time lag vs. frequency plot for single-dish observations follows 
a similar trend as is seen for VLBI core-shift measurements. 
Indications of a small core-shift in this source have been observed in 
\cite{Algaba12}, suggesting that a short time lag, smaller than our 
sampling rate, may be the case.

Flux variations at $\gamma-$ray, X-ray, and optical bands seem to be correlated at 2$\sigma$ level or higher, and the estimated time lags are consistent with zero, which suggests that any possible lag is shorter than the uncertainty in the peak, about 20 to 40 days depending on the band. Moreover, radio band light curves are significantly correlated with those at X-ray and $\gamma-$ray frequencies such that the X-ray and $\gamma-$ray flux variations lead those at radio bands by $\sim$90 and 70 days, respectively. We noticed similar time lags between the optical and radio band light curves; however the significance of correlation is below $2\sigma$. \cite{Raiteri12} found similar results to the ones here, with no correlation between radio and optical bands during a flare between 2001--2003, although their analysis suggested radio emission of increasing wavelength being emitted at larger and more external jet regions. On the other hand, they also found strong correlation for $\gamma-$rays and optical fluxes during 2008--2012, which they explain as these high-energy bands are all being located above the synchrotron bump in the SED.

The estimated time lags between radio and high-energy bands suggest that the high-energy emission may actually arise in a different region, closer to the base of the jet than that of the radio bands, as already suggested in \cite{Raiteri12}. One possibility \citep{Chidiac16} is that the high-energy emission may arise from a component emitted from a hot corona or reflected off an accretion disk.  We note however that iron emission lines associated with the disk or corona have not yet been detected in 1633+382 \citep[see e.g.][]{ReevesTurner00}. Another explanation lies in the standard shock--in--jet model \citep{MarscherGear85}, where flux variations at higher frequencies leading over the lower ones are naturally expected. Flux variations and radio/high-energy time lag correlations discussed above are well described under this model. \cite{Raiteri12} suggest that flux enhancements in different bands may result as the corresponding emitting regions are respectively more aligned to our line of sight.

\cite{Pushkarev12} suggested that the location of the VLBI core at 15~GHz is at a distance of $d\sim41$~pc from the jet apex. Although no uncertainty is provided for the VLBI core location, they estimated their typical accuracy level to be about 50 $\mu as$. This translates in an uncertainty for the 15~GHz VLBI core of 1633+382 of about 12~pc. We can estimate the location of the $\gamma-$ray flux by simply considering the separation between the 15~GHz and $\gamma-$ray emitting regions. This distance can be obtained from the time lag using $d=\beta_{app}c\Delta t/\sin\theta$, where $\beta_{app}$ is the apparent speed, $c$ is the speed of light, $\theta$ is the viewing angle, and $\Delta t$ is the obtained time lag \citep{Fuhrmann14}. Considering $\beta_{app}=29.2\pm1.3$ \citep{Lister13} and $\theta=1-3\degr$ \citep{Hovatta09,Liu10}, we infer a separation between 15~GHz and $\gamma-$ray to be $40\pm13$~pc, where we estimated the error based on the standard deviation of 10000 calculations of the distance with one standard deviation normal distributions of $\beta_{app}$, $\Delta t$ and $\theta$. This leads to a location of the $\gamma-$rays at a distance of $1\pm13$~pc from the jet apex. A schematic figure of the emitting regions of the base of the jet in 1633+382 is represented in the inset of Figure \ref{tlag}, where the high-energy 
(optical to $\gamma-$rays) emission dissipates in a parsec from the jet apex  which 
is optically thick to radio frequencies. Dissipation at radio frequencies takes 
place at a distance of $\sim$40~parsecs from the jet apex.

Energy dissipation on scales closer to the black hole could 
be due to shocks \citep{MarscherGear85}, which could also enlighten the 
radio core further downstream in the jet. Dissipation via magnetic 
reconnection \citep{kagan2015} or magnetoluminescence and/or electromagnetic 
detonation \citep{blandford2015} are equally probable in high magnetization 
regions.  Moreover, in several cases, $\gamma-$ray 
and radio flux enhancements on some sources are associated with the emergence of a new component from the VLBI core or recollimation shocks \citep[see e.g.][]{Savolainen03,Mattox01}. A detailed analysis of jet kinematics and their 
comparison with broadband flaring activity will be presented in a future 
paper.

\section{Conclusions}
We have monitored the flat spectrum radio quasar 1633+386 with simultaneous multi--frequency VLBI observations at 22, 43, 86 and 129~GHz with the KVN under the iMOGABA program. Combining these observations with additional multi--band monitoring at millimetre wavelengths, optical, X--rays and $\gamma-$rays, we are able to study the morphology of the light curves between 2012 March and 2015 August and, in particular, their variability and connection with various $\gamma-$ray flux enhancements observed during such period.

We find 1633+386 to present high flux densities, more than two times larger than usual, in radio bands between approximately MJD 56200 to 56700 which may be associated with at least three different interleaved but independent events. The associated optical, X--ray and $\gamma-$ray fluxes seem to follow a similar trend, although for optical and X-ray, the poor sampling complicates the comparison. Moreover, a very clear flux peak seen in these bands at MJD~57050, is not visible in any radio band. 

Radio spectral indices show a tendency to follow the flux variations, with flatter spectrum when higher fluxes are observed. A significant correlation of radio spectral index with their respective fluxes, with correlation coefficient $r>0.65$ for $\alpha^{22}_{43}$ and $\alpha^{22}_{86}$ is measured. In all cases, the correlation coefficient is noticeably larger when compared with the highest flux used to calculate the respective spectral index. 

Cross--correlation analysis shows the flux at the different bands to be significantly correlated, with the possible exception of optical bands, where the correlation, while still present, is not statistically significant ($<95\%$). Analysis of the DCF suggests time lags smaller than the uncertainty in the peak of the DCF of about 40 days among radio frequencies, as well as among high energies (optical, X-rays, and $\gamma-$rays), whereas a time lag of $\sim$70--90~days is found between radio and high-energy bands, suggesting the emission at high energies and radio are produced in two different jet regions, with the $\gamma-$rays located at $1\pm13$~pc and radio emission at $41\pm12$~pc from the jet apex.

\acknowledgments
\footnotesize{Acknowledgements. We are grateful to all staff members in KVN who helped to operate the array and to correlate the data. The KVN is a facility operated by the Korea Astronomy and Space Science Institute. The KVN operations are supported by KREONET (Korea Research Environment Open NETwork) which is managed and operated by KISTI (Korea Institute of Science and Technology Information). Data from the Steward Observatory spectropolarimetric monitoring project were used. This program is supported by Fermi Guest Investigator grants NNX08AW56G, NNX09AU10G, NNX12AO93G, and NNX15AU81G. This study makes use of 43 GHz VLBA data from the VLBA-BU Blazar Monitoring Program (VLBA-BU-BLAZAR), funded by NASA through the Fermi Guest Investigator Program. The VLBA is an instrument of the National Radio Astronomy Observatory. The National Radio Astronomy Observatory is a facility of the National Science Foundation operated by Associated Universities, Inc. This study has made use of the Swift-XRT Monitoring of Fermi-LAT Sources of Interest public data thanks to support from the Fermi GI program and the Swift Team. This research has made use of data from the OVRO 40-m monitoring program (Richards, J. L. et al. 2011, ApJS, 194, 29) which is supported in part by NASA grants NNX08AW31G, NNX11A043G, and NNX14AQ89G and NSF grants AST-0808050 and AST-1109911. This work used Submillimeter Array data. The Submillimeter Array is a joint project between the Smithsonian Astrophysical Observatory and the Academia Sinica Institute of Astronomy and Astrophysics and is funded by the Smithsonian Institution and the Academia Sinica. 
The {\it Fermi}/LAT Collaboration acknowledges the generous support of a number of agencies
and institutes that have supported the {\it Fermi}/LAT Collaboration. These include the National
Aeronautics and Space Administration and the Department of Energy in the United States, the
Commissariat \`a l'Energie Atomique and the Centre National de la Recherche Scientifique / Institut
National de Physique Nucl\'eaire et de Physique des Particules in France, the Agenzia Spaziale
Italiana and the Istituto Nazionale di Fisica Nucleare in Italy, the Ministry of Education,
Culture, Sports, Science and Technology (MEXT), High Energy Accelerator Research Organization
(KEK) and Japan Aerospace Exploration Agency (JAXA) in Japan, and the K.\ A.\ Wallenberg
Foundation, the Swedish Research Council and the Swedish National Space Board in Sweden.
Additional support for science analysis during the operations phase is gratefully acknowledged 
from the Istituto Nazionale di Astrofisica in Italy and the Centre National d'\'Etudes Spatiales 
in France.
G. Zhao is supported by Korea Research Fellowship Program through the National Research Foundation of Korea(NRF) funded by the Ministry of Science, ICT and Future Planning (NRF-2015H1D3A1066561). D.-W. Kim and S. Trippe acknowledge support from the National Research Foundation of Korea (NRF) via grant NRF-2015R1D1A1A01056807. J. Park acknowledges support from the NRF via grant 2014H1A2A1018695. S. S. Lee and S. Kang were supported by the National Research Foundation of Korea (NRF) grant funded by the Korea government (MSIP) (No. NRF-2016R1C1B2006697). This research was supported by an appointment to the NASA Postdoctoral Program at the Goddard Space Flight Center, administered by Universities Space Research Association through a contract with NASA. We would like to thank Stefan Larson, Marcello Giroletti, Sara Cutini, Teddy Cheung, Jeremy S. Perkins, Dave Thompson, Philippe Bruel, and Elisabetta Cavazzuti the internal referees from the Fermi/LAT team, for their useful suggestions and comments. 
}

\appendix

\section{Correlation functions}

\begin{figure*}
\includegraphics[scale=0.21,trim={0cm 0cm 0cm 0cm},clip]{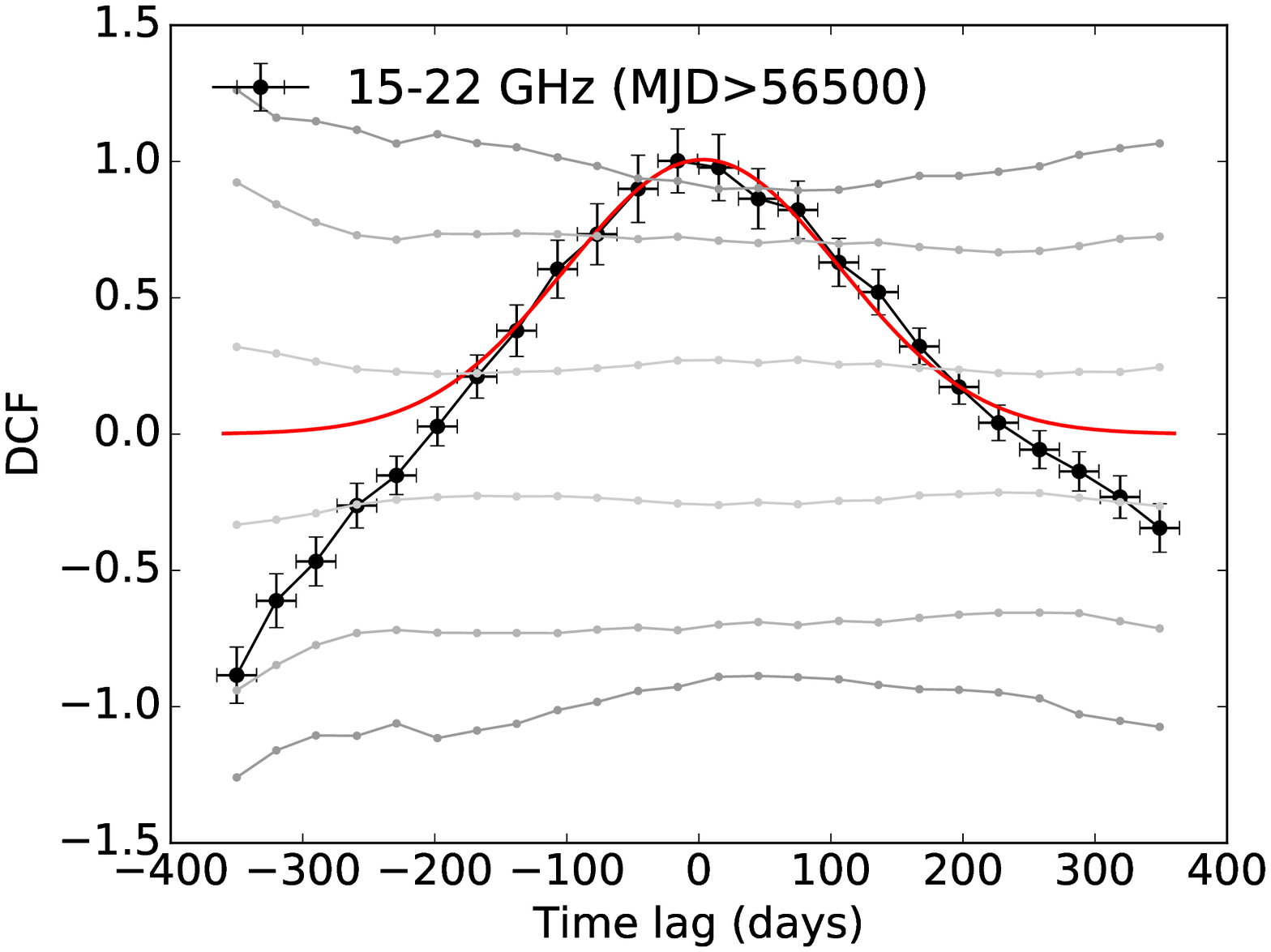}
\includegraphics[scale=0.21,trim={0cm 0cm 0cm 0cm},clip]{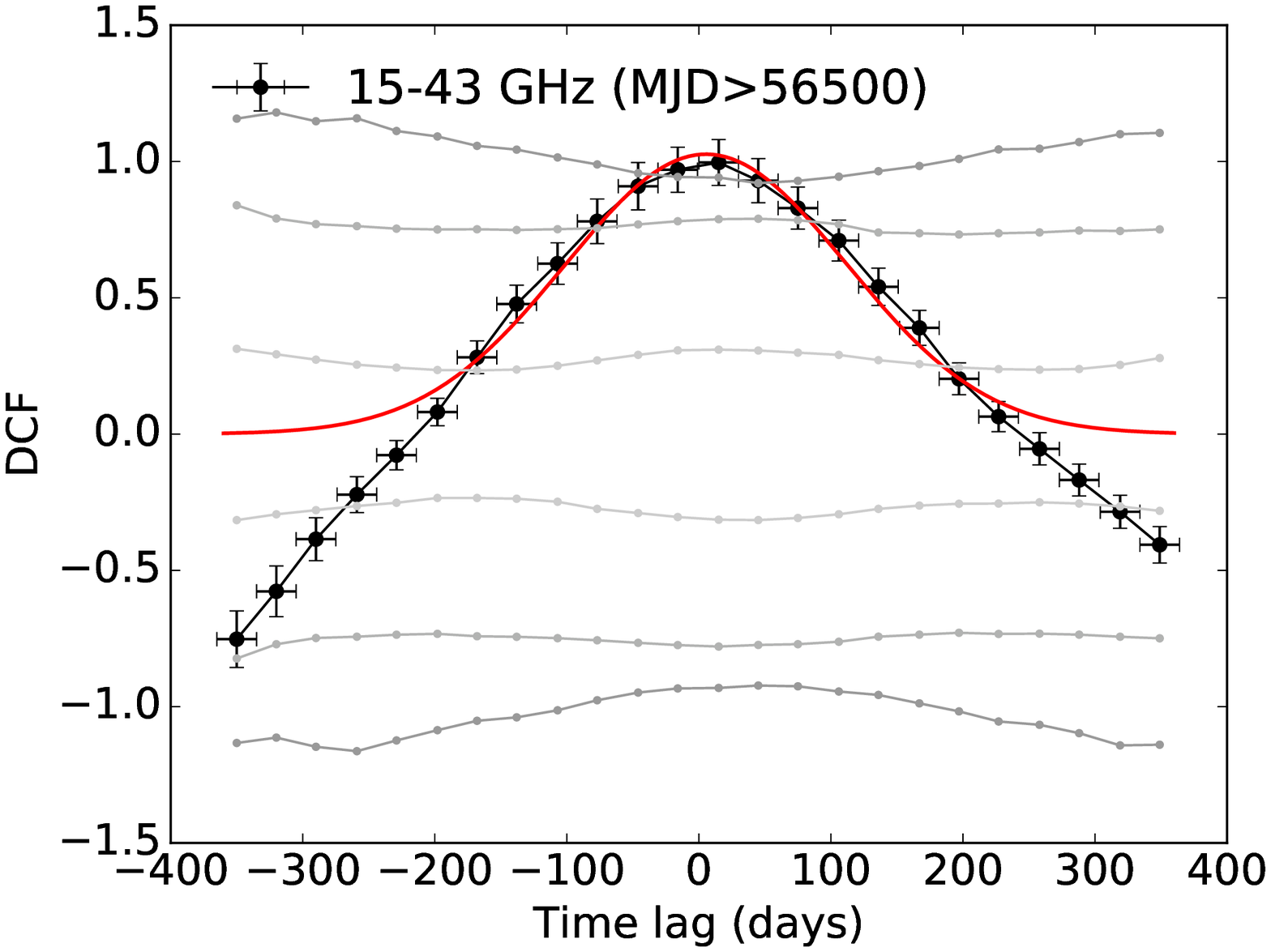}
\includegraphics[scale=0.21,trim={0cm 0cm 0cm 0cm},clip]{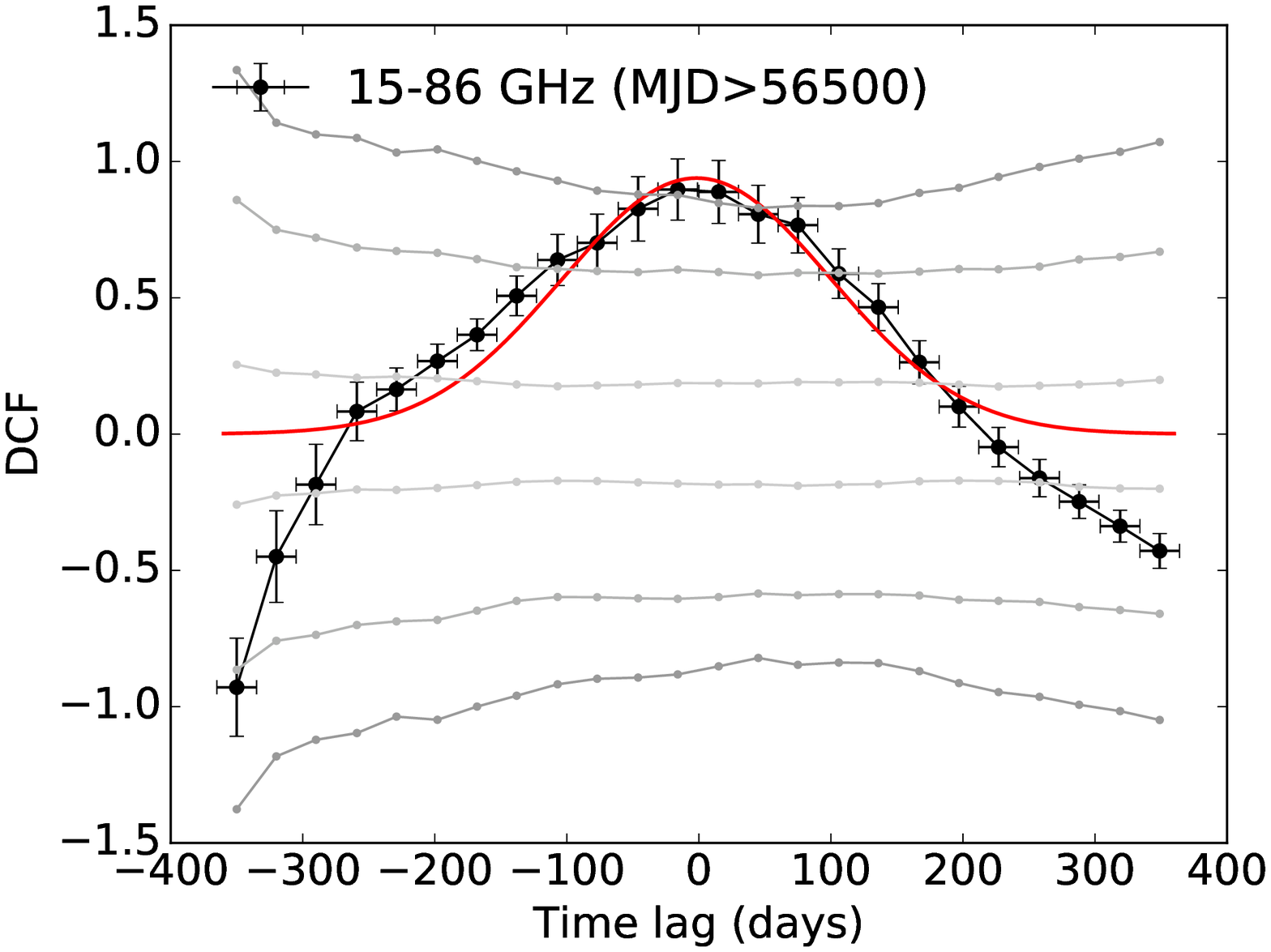}
\includegraphics[scale=0.21,trim={0cm 0cm 0cm 0cm},clip]{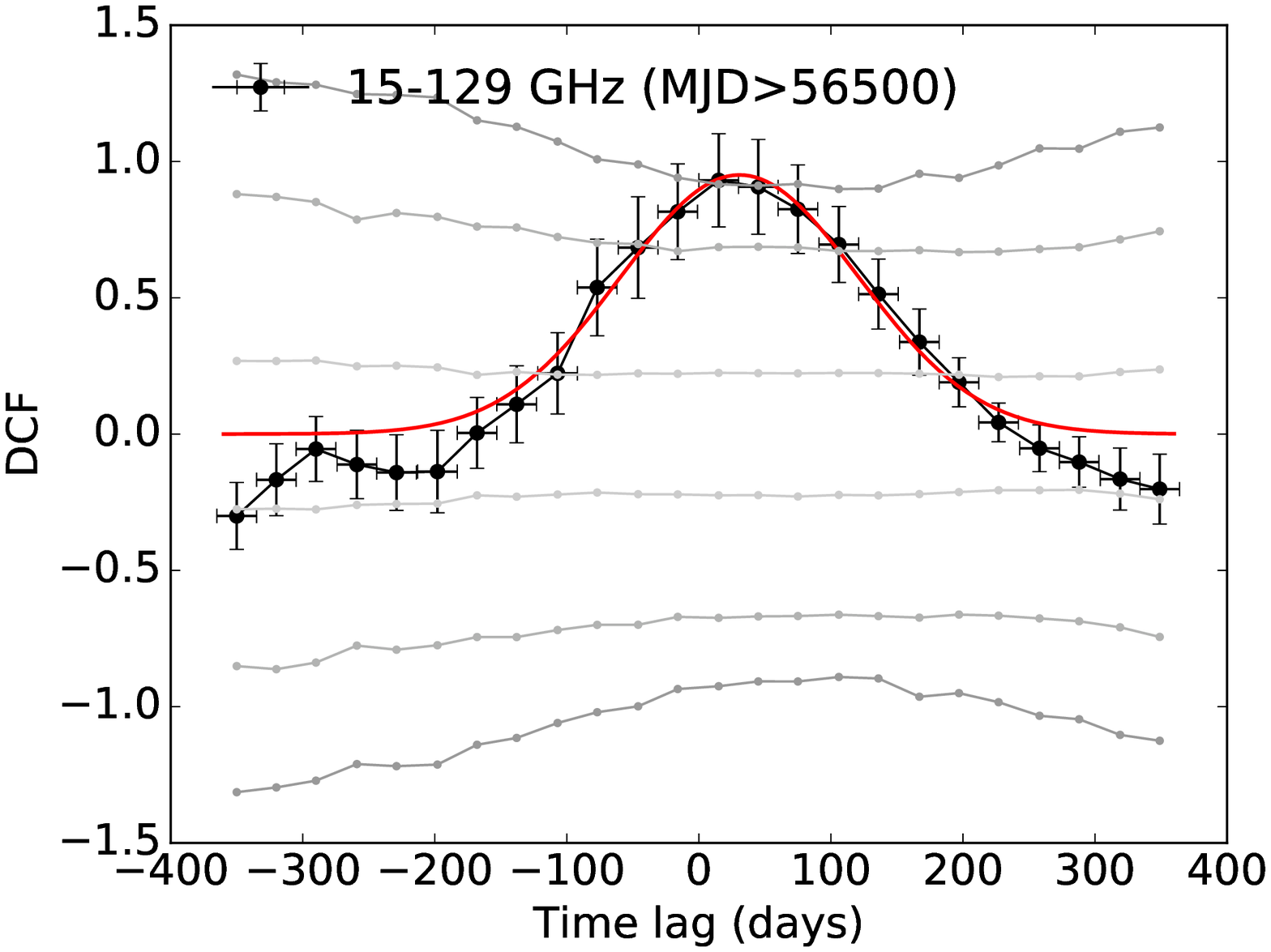}\\
\includegraphics[scale=0.21,trim={0cm 0cm 0cm 0cm},clip]{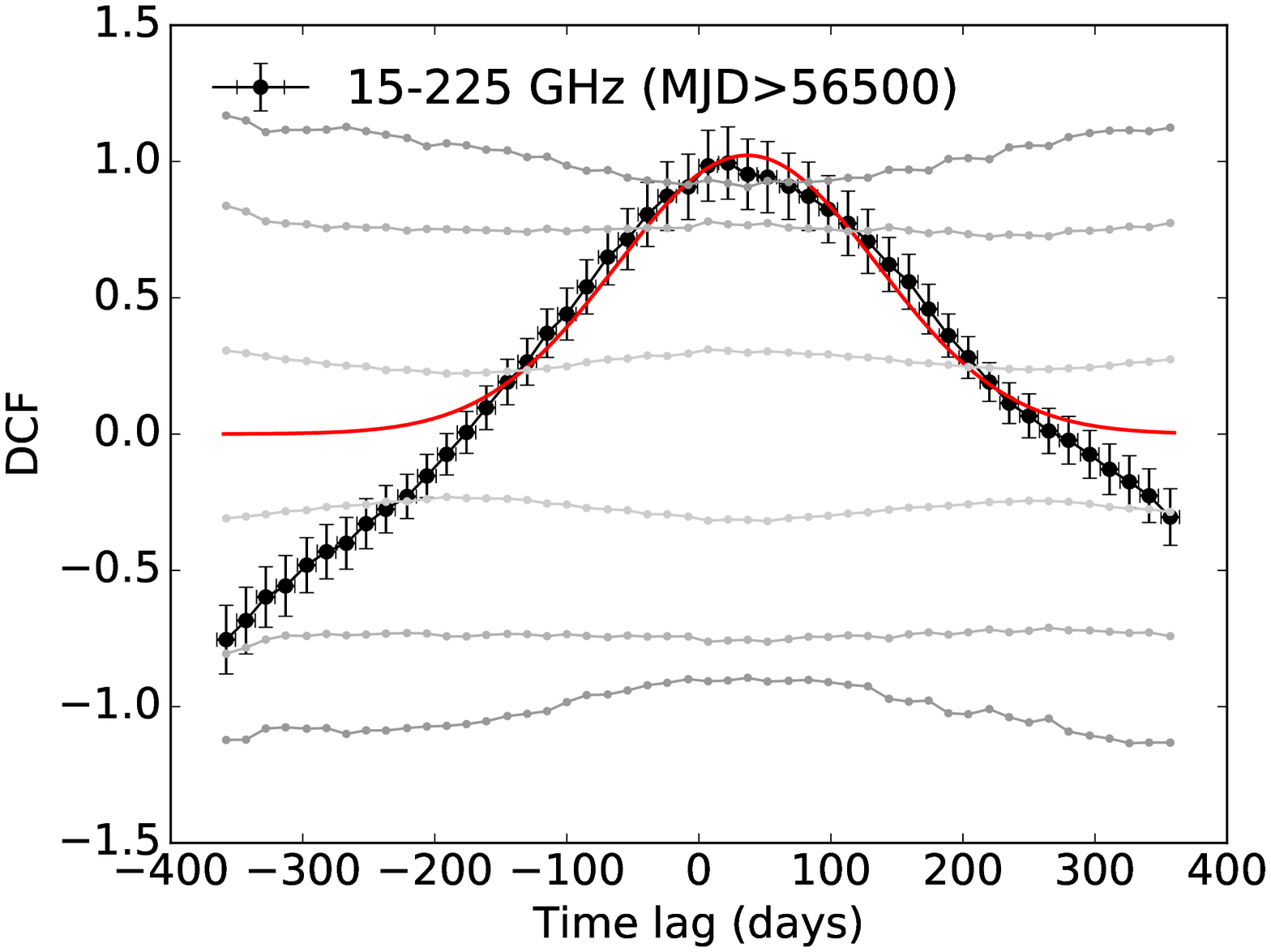}
\includegraphics[scale=0.21,trim={0cm 0cm 0cm 0cm},clip]{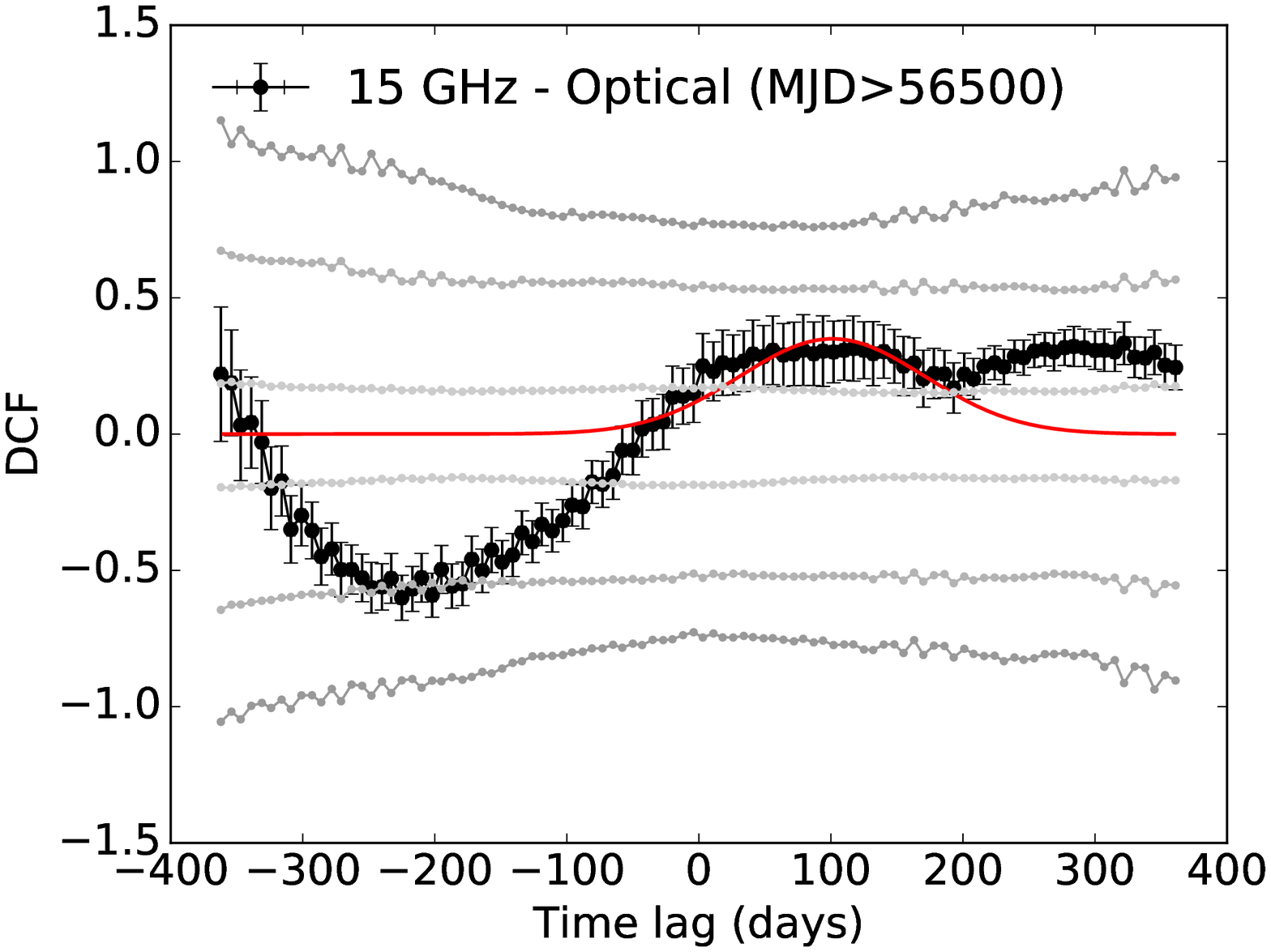}
\includegraphics[scale=0.21,trim={0cm 0cm 0cm 0cm},clip]{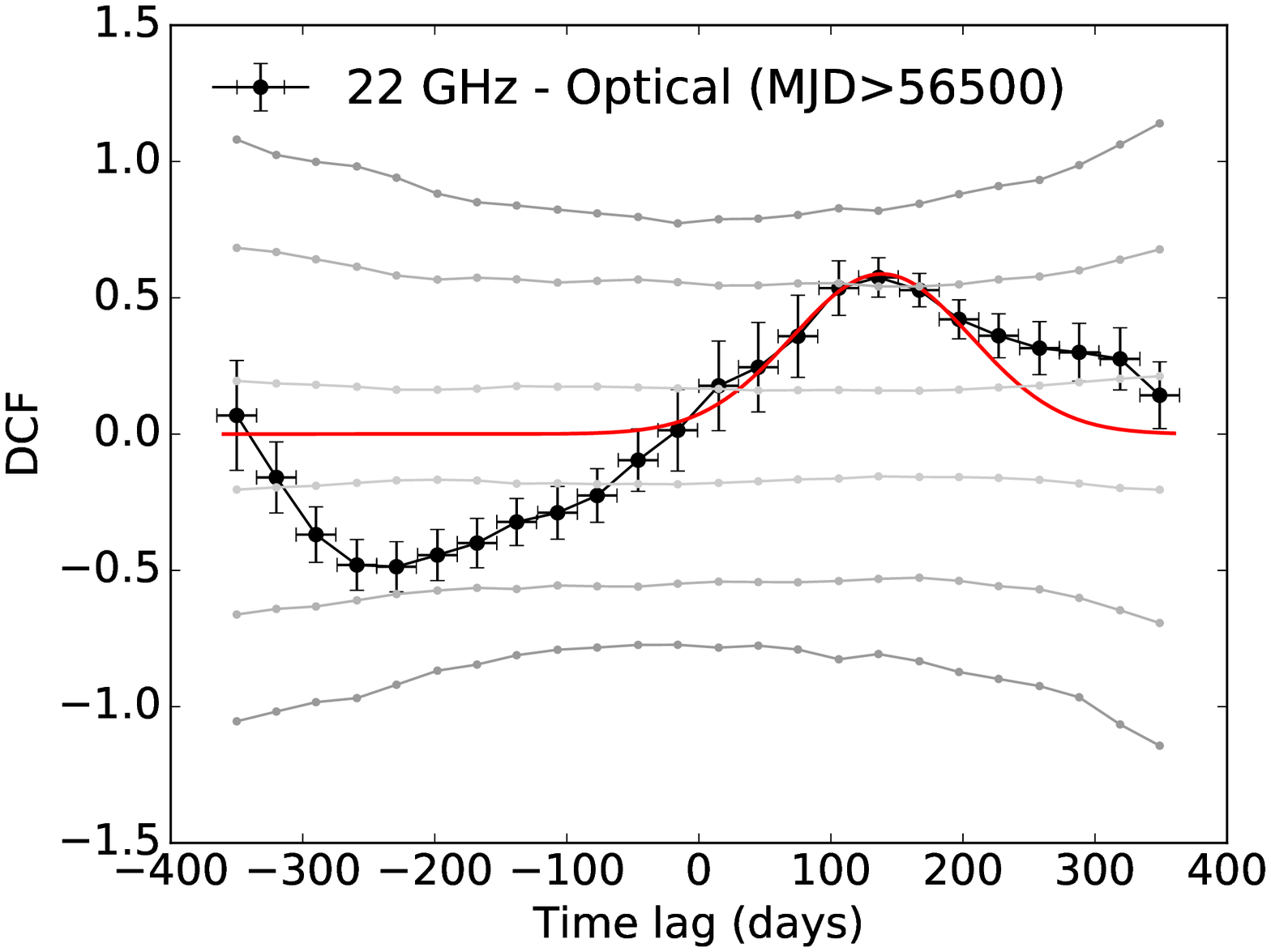}
\includegraphics[scale=0.21,trim={0cm 0cm 0cm 0cm},clip]{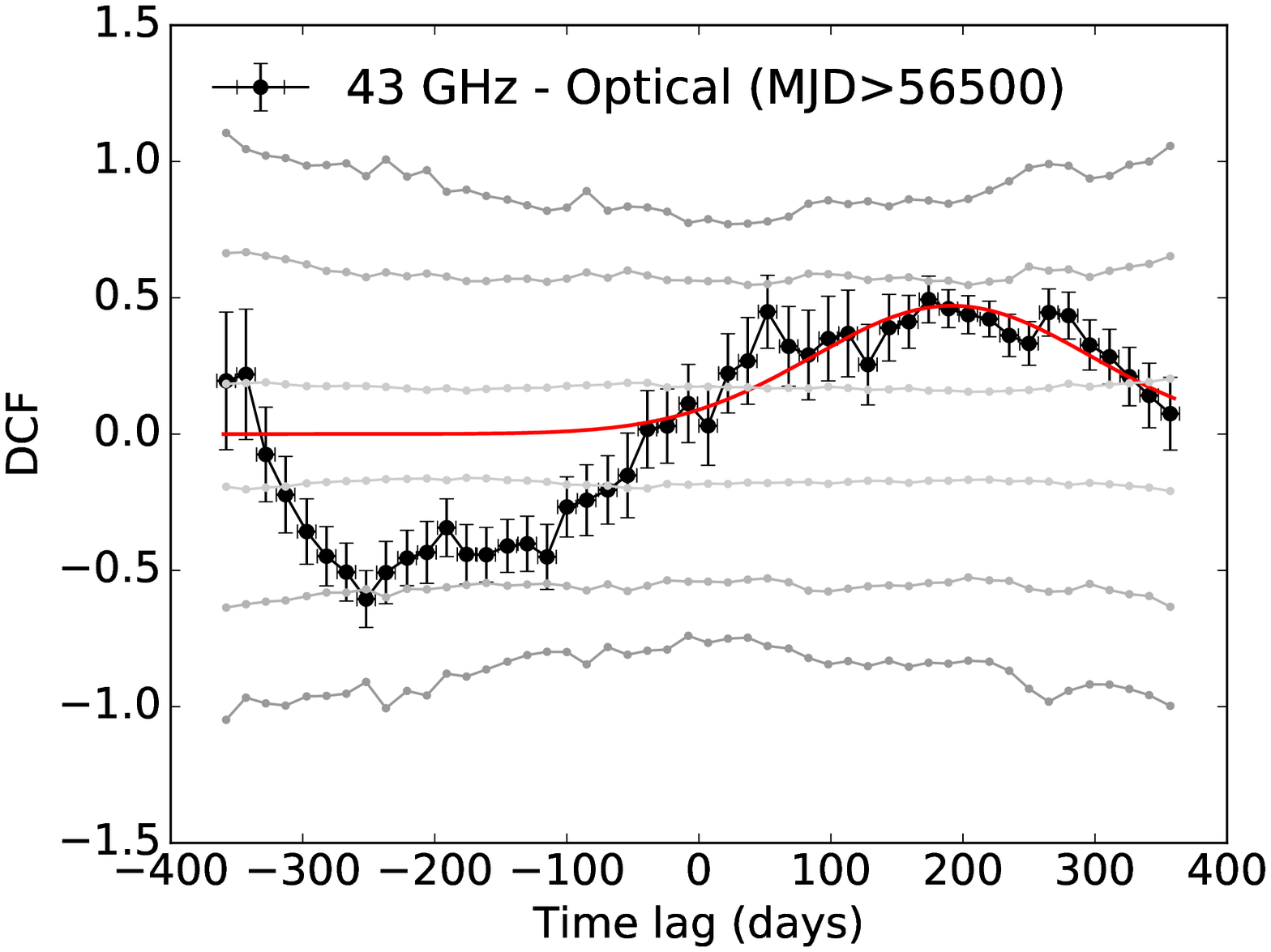}\\
\includegraphics[scale=0.21,trim={0cm 0cm 0cm 0cm},clip]{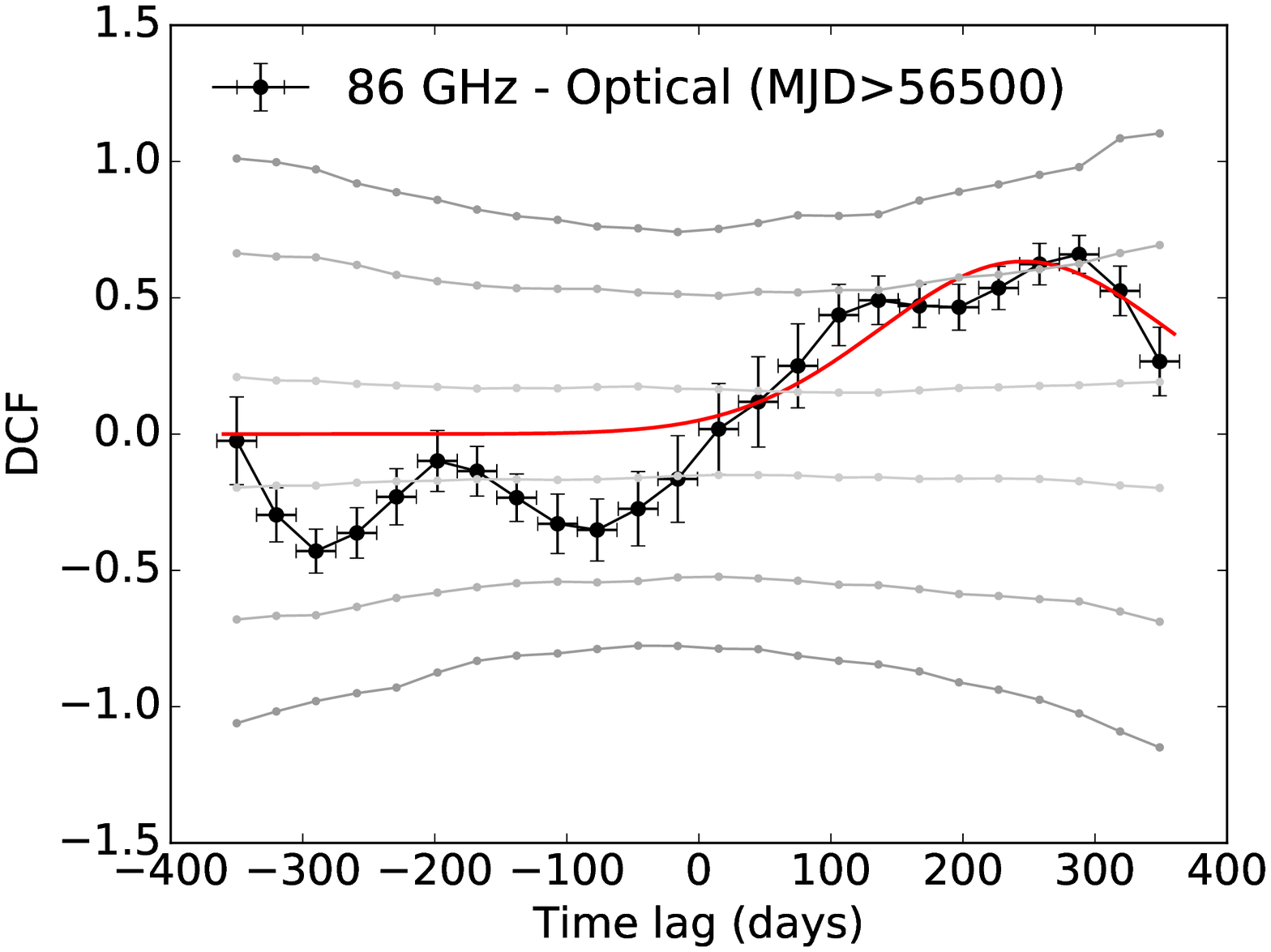}
\includegraphics[scale=0.21,trim={0cm 0cm 0cm 0cm},clip]{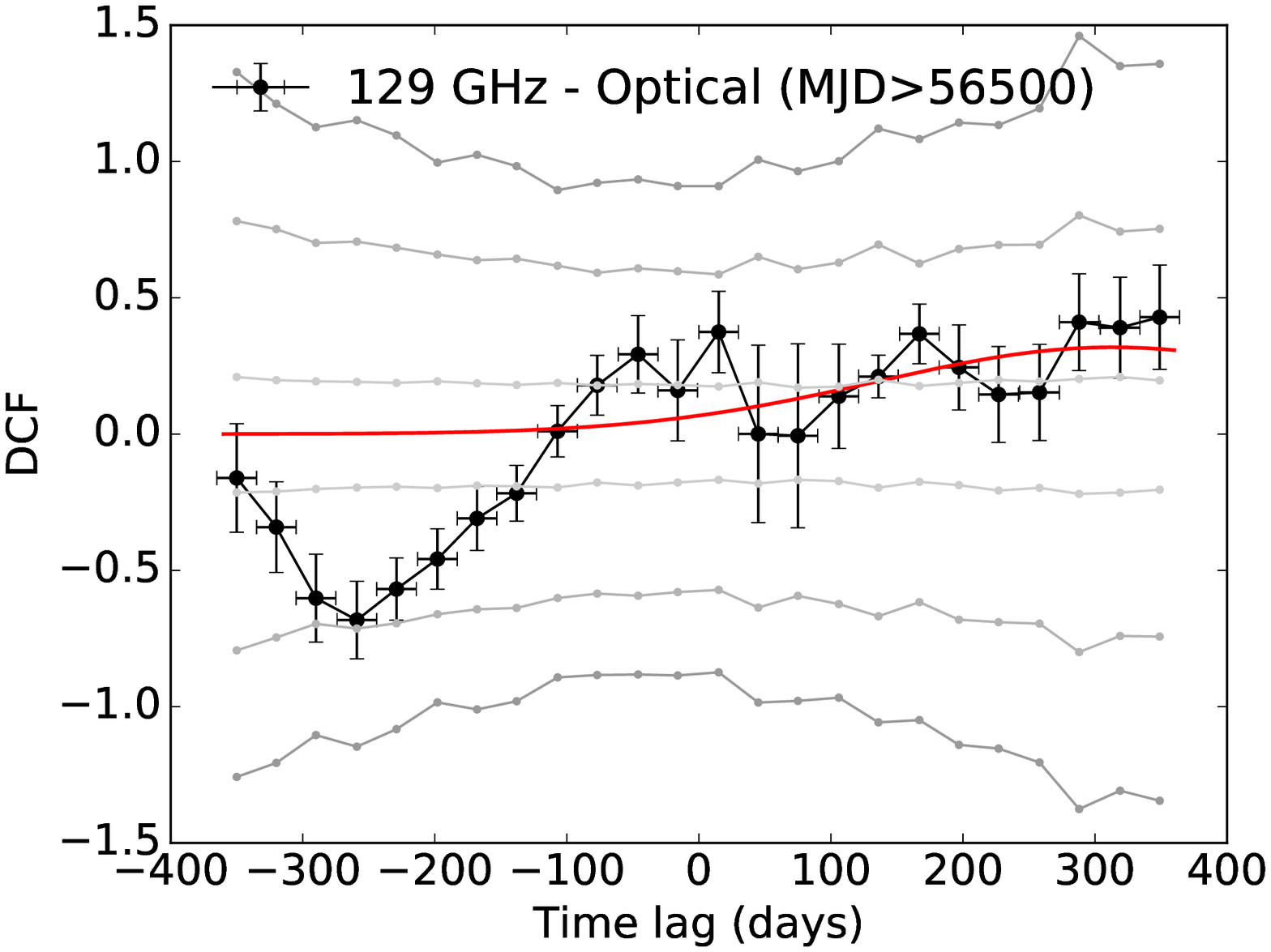}
\includegraphics[scale=0.21,trim={0cm 0cm 0cm 0cm},clip]{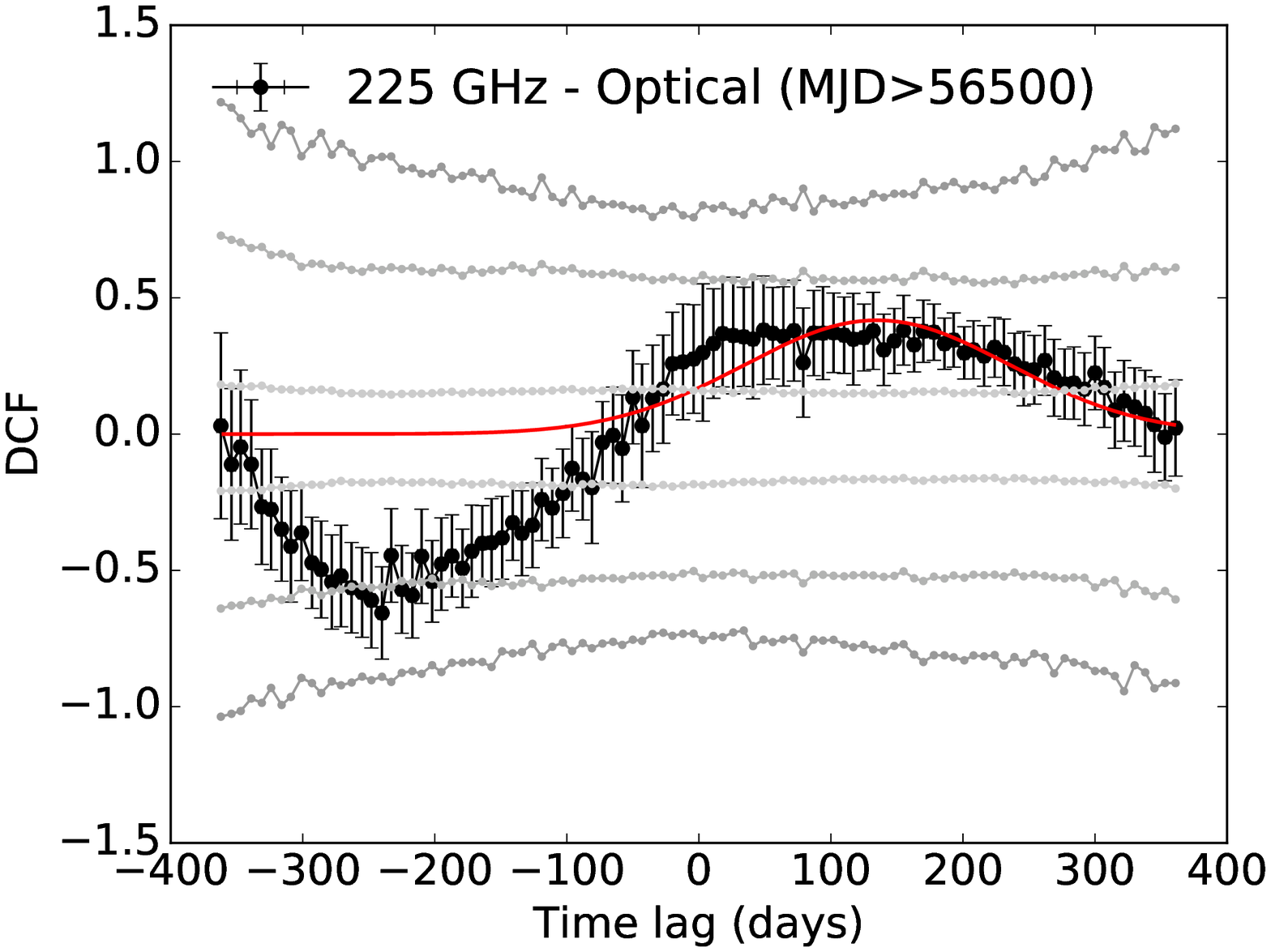}
\includegraphics[scale=0.21,trim={0cm 0cm 0cm 0cm},clip]{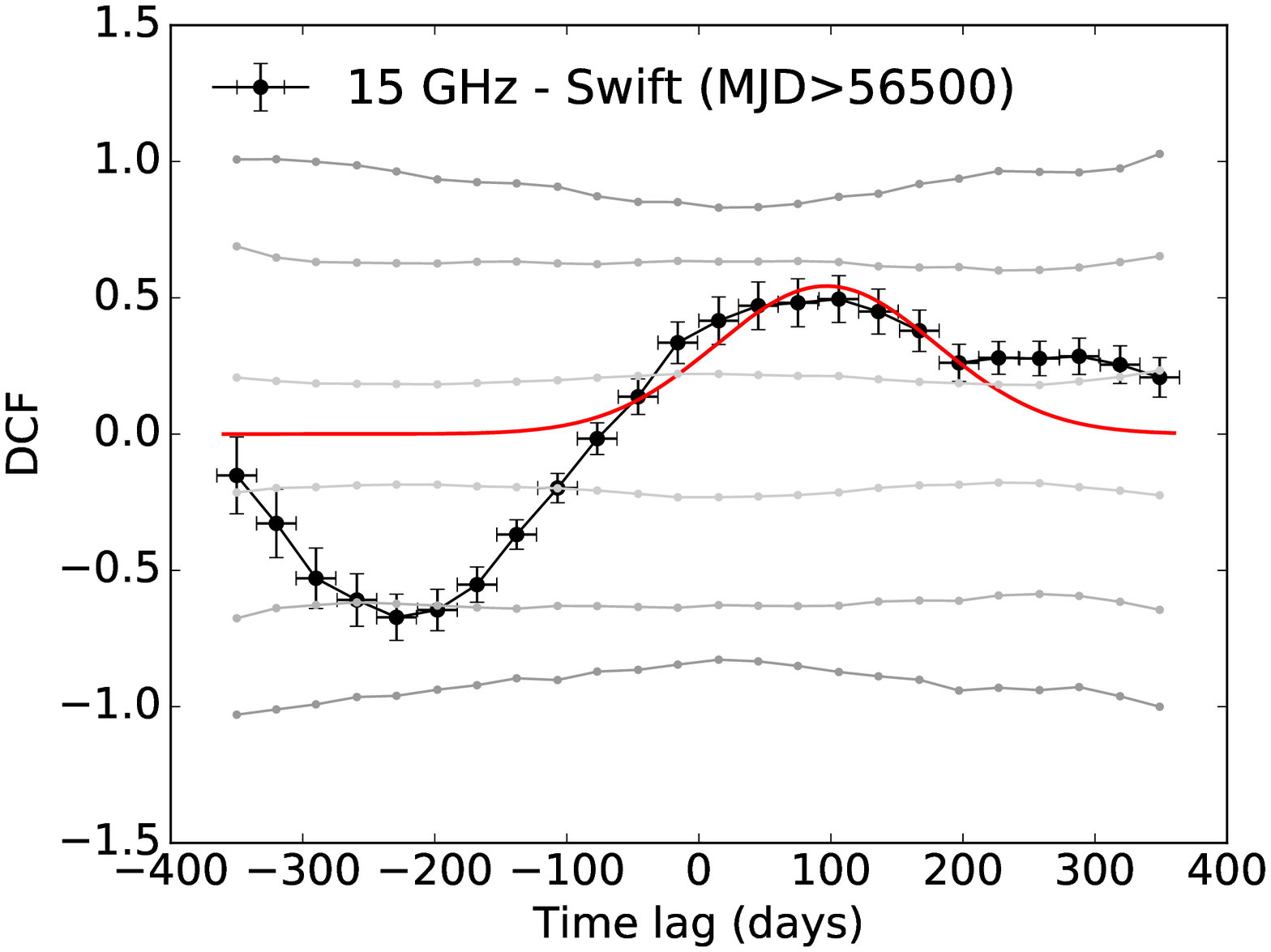}\\
\includegraphics[scale=0.21,trim={0cm 0cm 0cm 0cm},clip]{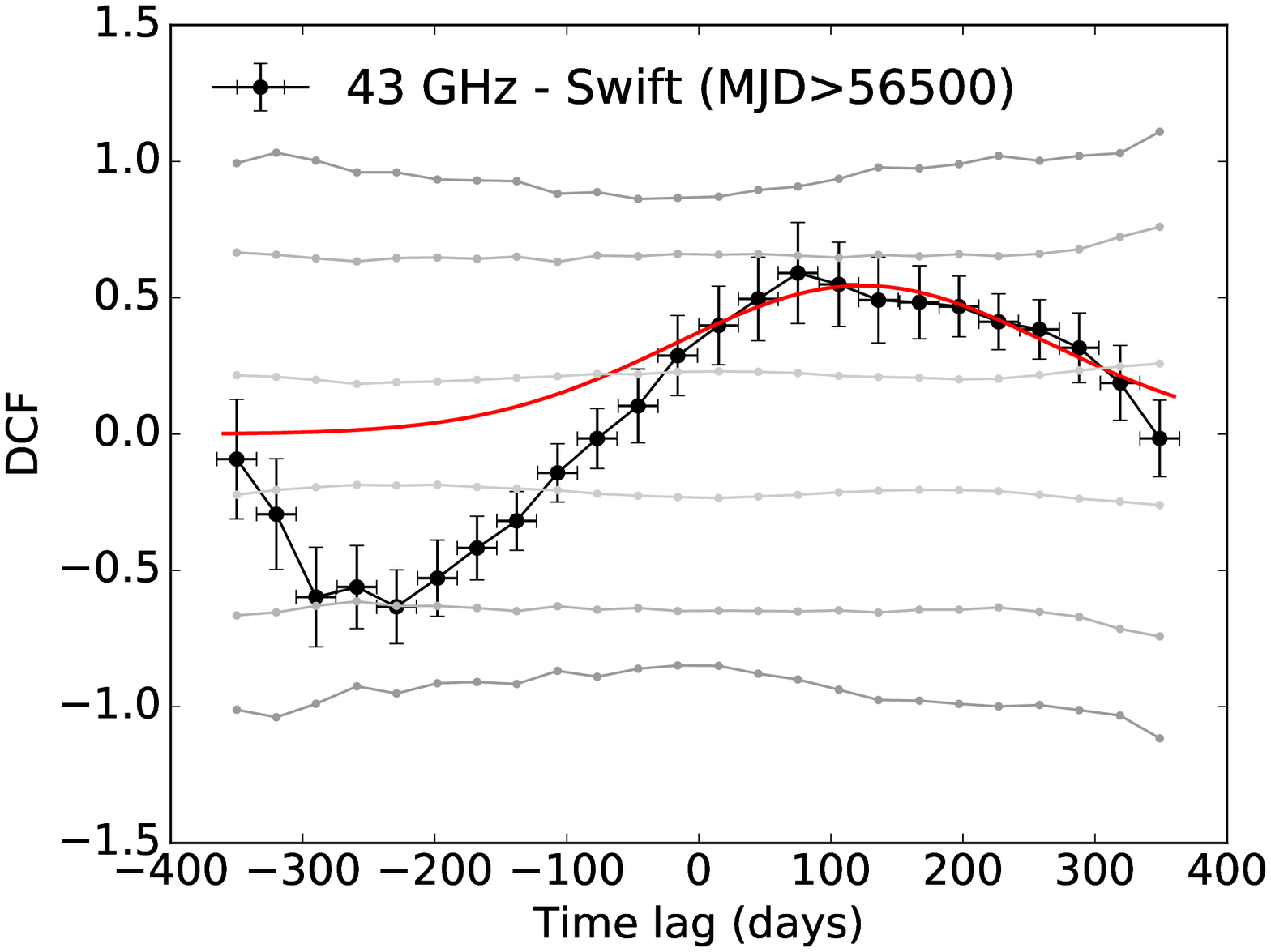}
\includegraphics[scale=0.21,trim={0cm 0cm 0cm 0cm},clip]{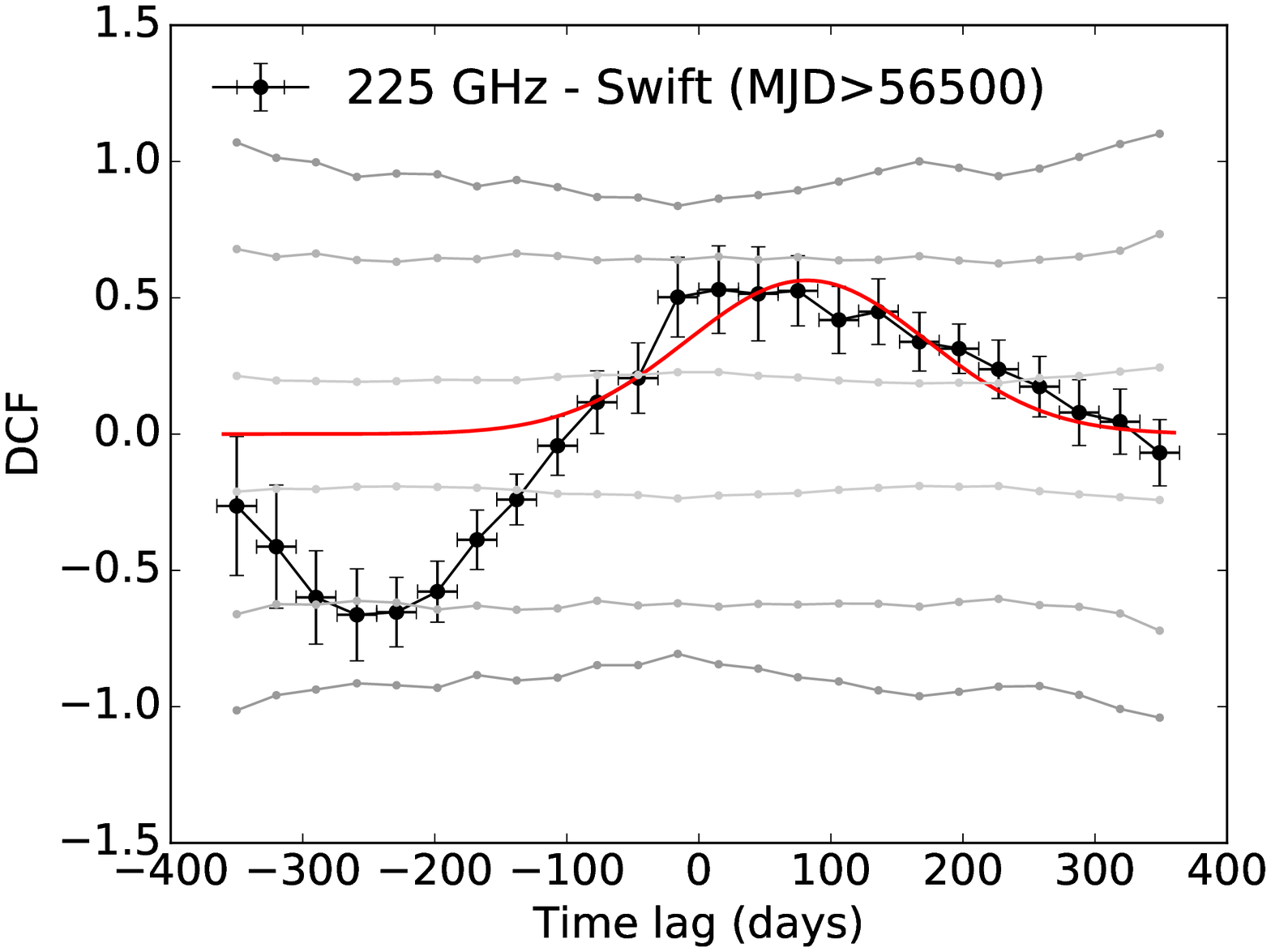}
\includegraphics[scale=0.21,trim={0cm 0cm 0cm 0cm},clip]{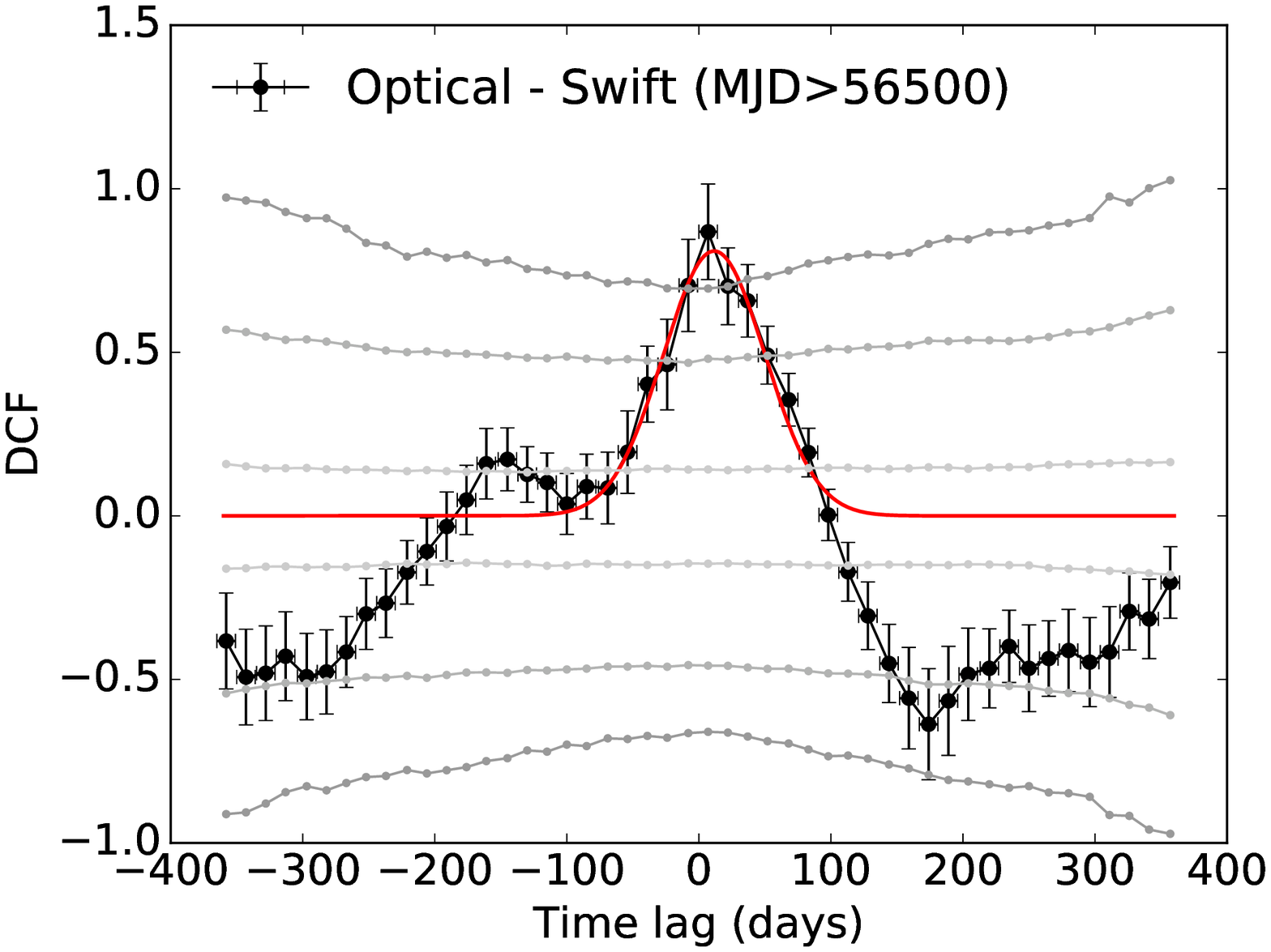}
\includegraphics[scale=0.21,trim={0cm 0cm 0cm 0cm},clip]{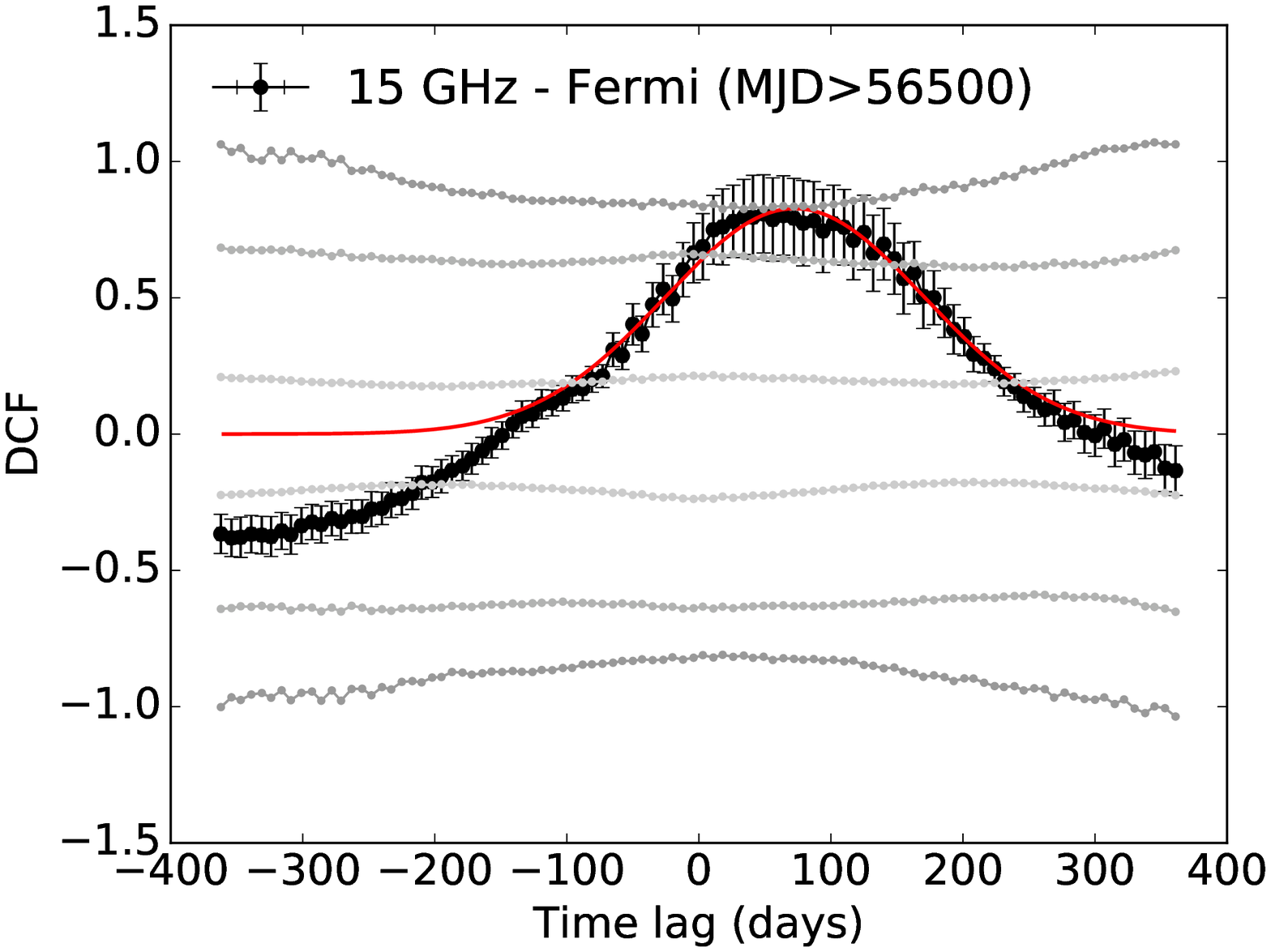}\\
\includegraphics[scale=0.21,trim={0cm 0cm 0cm 0cm},clip]{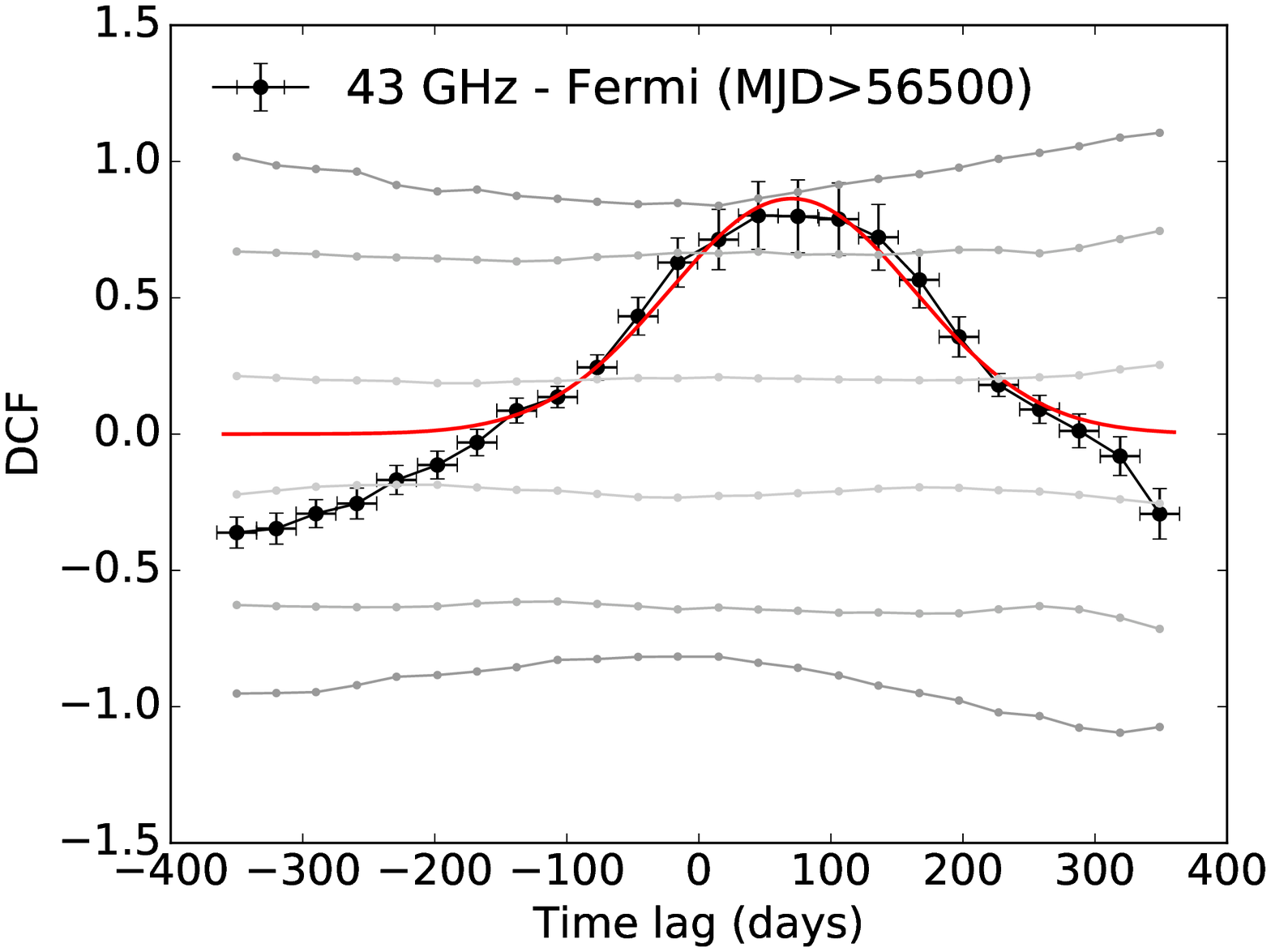}
\includegraphics[scale=0.21,trim={0cm 0cm 0cm 0cm},clip]{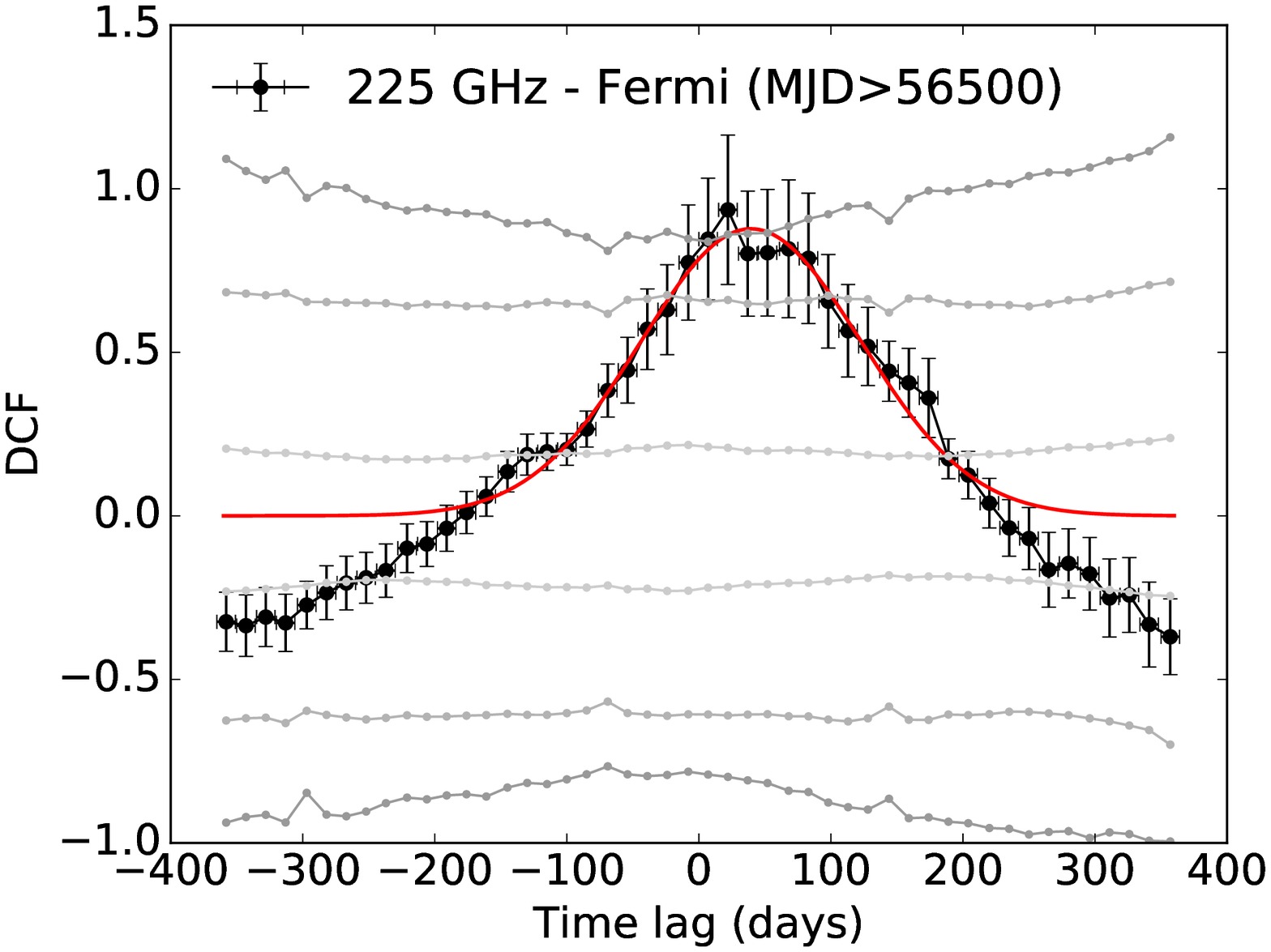}
\includegraphics[scale=0.21,trim={0cm 0cm 0cm 0cm},clip]{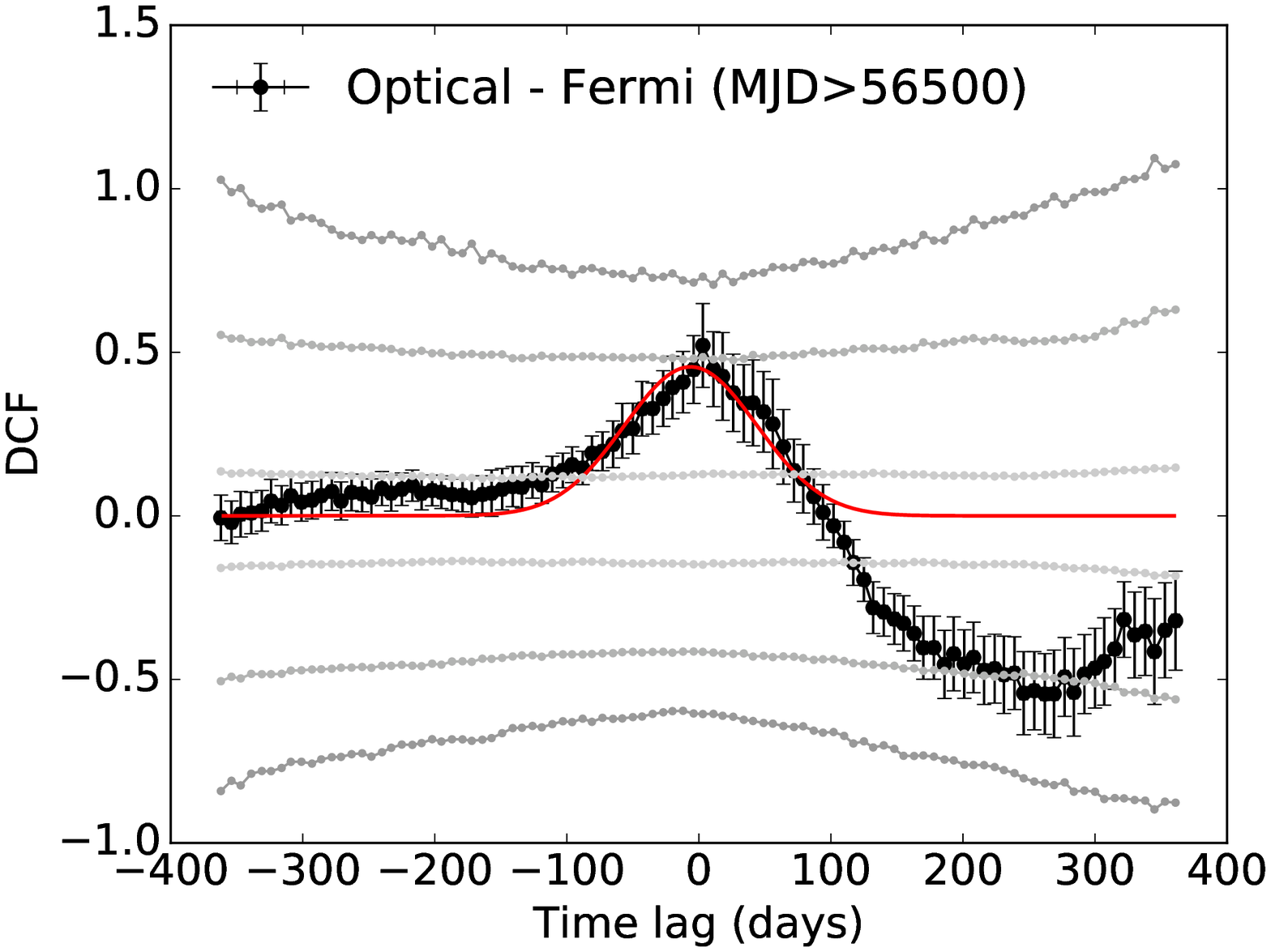}
\includegraphics[scale=0.21,trim={0cm 0cm 0cm 0cm},clip]{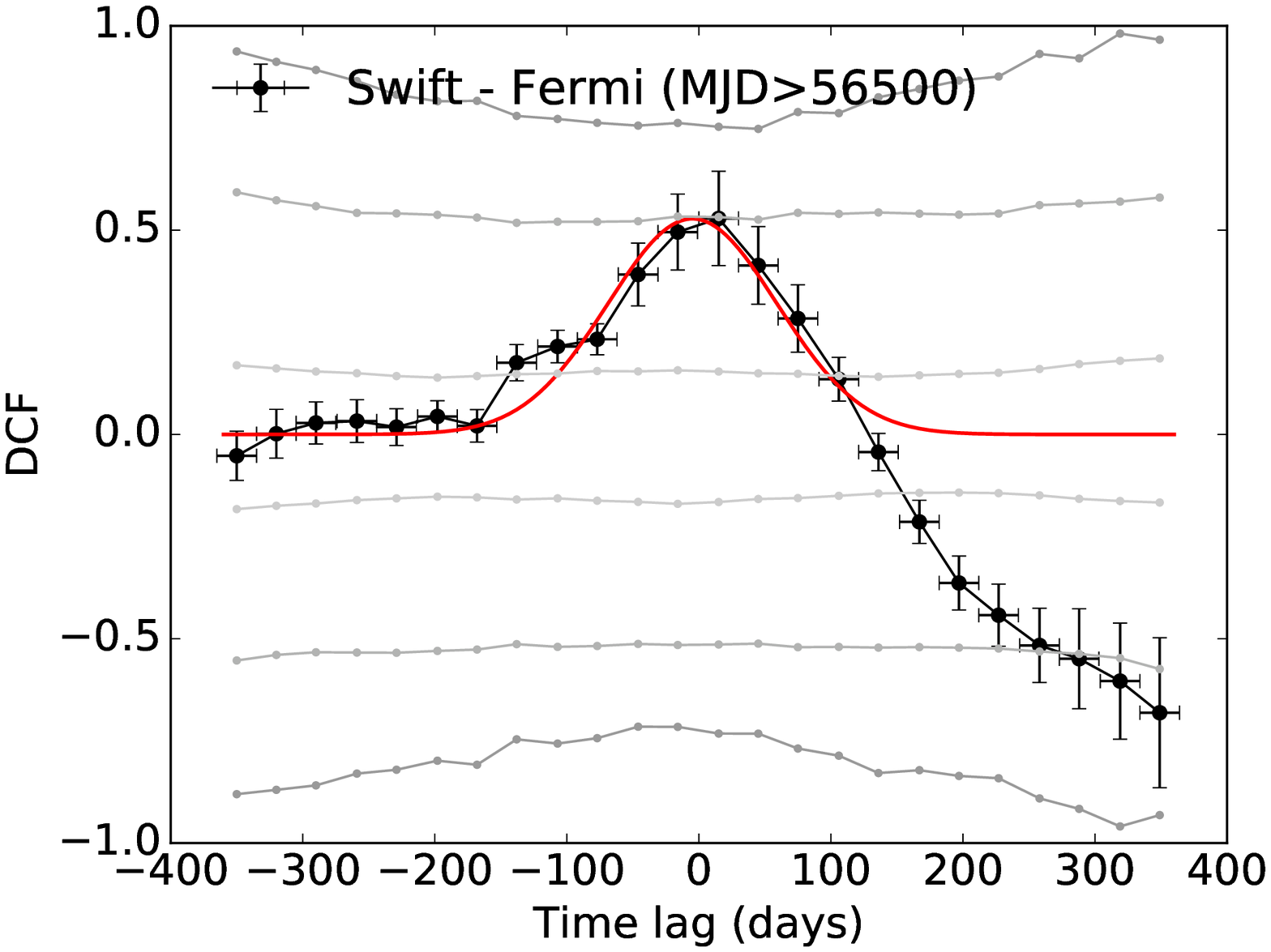}
\caption{Discrete Correlation Function of 1633+382 for MJD$>$56500. Symbols and lines have the same meaning as in Fig. \ref{DCF}.}
\label{DCFRappendix}
\end{figure*}


\end{document}